\documentclass{article}

\usepackage{todonotes}
\usepackage{arxiv}
\usepackage[utf8]{inputenc} % allow utf-8 input
\usepackage[T1]{fontenc}    % use 8-bit T1 fonts
\usepackage{hyperref}       % hyperlinks
\usepackage{url}            % simple URL typesetting
\usepackage{booktabs}       % professional-quality tables
\usepackage{amsfonts}       % blackboard math symbols
\usepackage{nicefrac}       % compact symbols for 1/2, etc.
\usepackage{microtype}      % microtypography
\usepackage{lipsum}
\usepackage{graphicx}
\graphicspath{ {./images/} }
\usepackage{xcolor}
\usepackage{url}            % simple URL typesetting
\usepackage{booktabs}       % professional-quality tables
\usepackage{amsfonts}       % blackboard math symbols
\usepackage{nicefrac}       % compact symbols for 1/2, etc.
\usepackage{microtype}      % microtypography
\usepackage{lipsum}
\usepackage{graphicx}
\usepackage{subcaption}
\usepackage{bm}
\usepackage{amsmath,amssymb}   % formule avanzate
\usepackage{bm}                % \bm per vettori in grassetto
\usepackage{mathrsfs}
\usepackage{xcolor}
\usepackage{tikz}
\usetikzlibrary{arrows.meta,positioning,shapes.geometric}
\usepackage[ruled,vlined]{algorithm2e}
\usepackage{array}    % per \newcolumntype e \arraybackslash
\usepackage{booktabs} % per \toprule \midrule \bottomrule \addlinespace
\usepackage{booktabs,siunitx,threeparttable}
\usepackage{longtable}
\usepackage[section]{placeins}
\usepackage{float}          % abilita il posizionamento [H]
\usepackage[section]{placeins} % impedisce ai float di “scappare” oltre la sezione
\usepackage{adjustbox}
\usepackage{amsthm}
\theoremstyle{remark}

\usepackage[backend=biber,
            style=apa,
            sorting=nyt,
            sortcites=true]{biblatex}
\DeclareFieldFormat{labelnumberwidth}{[#1]}

\renewcommand{\today}{}

\title{Quantifying Uncertainty in Future Event Counts at Interim Analyses of Time-to-Event Trials, with an Application to Pediatric Oncology}

\author{
Edoardo Ratti\textsuperscript{1,*},
Federico L.~Perlino\textsuperscript{2},
Stefania Galimberti\textsuperscript{1,3},
Maria G.~Valsecchi\textsuperscript{1,4}\\[0.6em]
\textsuperscript{1}Bicocca Bioinformatics, Biostatistics and Bioimaging Centre B4, \\ School of Medicine and Surgery, University of Milano--Bicocca, Monza, Italy\\ \\
\textsuperscript{2}Department of Statistics, University of Warwick, Coventry, UK\\ \\
\textsuperscript{3}Biostatistics and Clinical Epidemiology, \\ Fondazione IRCCS San Gerardo dei Tintori, Monza, Italy\\ \\
\textsuperscript{4}Tettamanti Center, \\ Fondazione IRCCS San Gerardo dei Tintori, Monza, Italy
}

\begin{document}
\raggedbottom
\maketitle

\thispagestyle{plain}

\begingroup
\renewcommand{\thefootnote}{*}
\footnotetext{Correspondence: \texttt{e.ratti9@campus.unimib.it}}
\endgroup

\begin{abstract}
Time-to-event endpoints are central to evaluating treatment efficacy across disease areas. In clinical trials with time-to-event endpoints, the information available for interim and final analyses is often determined by the number of observed events rather than by the number of enrolled patients. Interim monitoring therefore requires assessing how many additional events are expected to accrue by scheduled future analysis dates. Quantifying uncertainty around these counts is essential for assessing whether planned information levels are likely to be reached, anticipating delays or event overrunning, and supporting operational decisions while the trial is ongoing. This is especially relevant where event accrual is low as in pediatric hemato-oncology trials. Although methods for predicting time to endpoint maturation are well established, interval prediction for event counts at fixed calendar times remains less developed. We propose a patient-level framework for constructing such intervals at interim analyses of time-to-event trials. Conditionally on the interim data, the future count follows a Poisson--binomial law with patient-specific event probabilities; we estimate this law using a conditional parametric bootstrap. Under standard regularity conditions, the bootstrap is consistent and yields asymptotically calibrated prediction intervals. The framework accommodates staggered entry, patient-level covariates, administrative censoring, loss to follow-up, and possible dependence between entry dates and loss to follow-up, before conditioning on the realised interim data. We study its operating characteristics in simulation studies and illustrate the method using a real-world phase III trial in childhood acute lymphoblastic leukaemia.
\end{abstract}


\section{Introduction}

Time-to-event endpoints are widely used to evaluate treatment efficacy across disease areas. In these trials, the amount of statistical information available for interim and final analyses is driven primarily by the number, or counts, of events observed during follow-up. For this reason, sample-size calculations are usually expressed in terms of an expected number of events \parencite{Wu}. Interim analyses are also frequently incorporated to allow early stopping for efficacy or futility for ethical and economic reasons \parencite{WassmerBrannath2025GroupSequential}. In group-sequential designs, type-I error is typically controlled through alpha-spending or boundary methods \parencite{JennisonTurnbull2025GroupSequentialAdaptive}, and the planned operating characteristics depend on the information accrued at each look. Analyses should therefore occur at the information levels specified in the design; otherwise, the type-I error probability may be adversely affected \parencite{Lan1989, Proschan1992}.

Maintaining this alignment between calendar time and statistical information is difficult once a trial is underway. Planning assumptions about the survival distribution may be uncertain at the design stage and may become inaccurate as follow-up accumulates. If events accrue more slowly than anticipated, scheduled analyses may be delayed or reached with less information than expected, creating practical consequences for trial conduct. This challenge is particularly relevant in pediatric oncology trials, where limited accrual may require relatively a long follow-up after end of recruitment or even to relax conventional operating characteristics to maintain trial feasibility and generate clinically meaningful evidence \parencite{Valsecchi, RenfroAlonzo}. If events accrue either too rapidly or slowly, the trial may face over/unde-running , which requires appropriate statistical handling \parencite{Whitehead1992Overrunning}. Interim monitoring therefore benefits from methods that can forecast future event accrual and account for uncertainty to that forecast.

A well-developed literature considers the problem of predicting when a target number of events will be reached. Under a \textit{maximum-information} design, the trial ends once the planned number of events has accrued \parencite{Kim2014MaximumDurationInformation}. The corresponding prediction target is the calendar time required to reach that milestone, often called time to endpoint maturation \parencite{Ou2019, Wang2022RealTimeTTEM}. Point and interval predictions for this target have been extensively studied \parencite{Bagiella2001,YingHeitjanChen2004NonparametricPrediction,Ying2008,Ying2013,Ying2017,Chen2016,Lan2018,Machida2025DynamicPredictionTiming}. \textcite{Heitjan2015} provide a comprehensive review and emphasize the importance of prediction intervals for quantifying uncertainty around the closing date of a study. Some methods also allow prediction while accrual is still ongoing \parencite{Wang2022RealTimeTTEM,Zhang2012,Anisimov2011}.

The monitoring question addressed in this paper is the complementary one. Rather than predicting the date on which a fixed event target will be reached, we consider prediction of the number of additional events that will be observed by future scheduled analysis dates. This fixed-calendar perspective is natural in \textit{maximum-duration} design, where follow-up is limited by a prespecified calendar time \parencite{Kim2014MaximumDurationInformation}. It is also useful in maximum-information designs, since censoring may prevent the target number of events from being reached in reasonable time, e.g. due to a high beneficial effect of the experimental treatment, and operational decisions often depend on whether the planned information level is likely to be available by a given date.

Existing methods for this fixed-calendar event-count problem mainly provide point predictions under a limited set of survival models. \textcite{Fang2011} proposed a hybrid nonparametric--parametric approach for estimating the survival curve, in which change points are detected using the algorithm of \textcite{Goodman2011}; the survival curve before the final change point is estimated nonparametrically, while the tail is modelled using an exponential or Weibull distribution. This method is implemented by \textcite{eventTrack} in the \textit{eventTrack} R package. Event-count prediction has also been used in sample-size re-estimation \parencite{Whitehead2001,Hade2010,Todd2012,Friede2019,Mori2024}. These approaches are valuable for planning, but point predictions do not describe how uncertain the future number of events is.

Prediction intervals for event counts are less developed. To the best of our knowledge, only two approaches have been 
proposed to construct prediction intervals for future event 
counts at interim analyses of time-to-event trials 
\parencite{Anisimov2011,Anisimov2021AdvancedModelsArxiv}. 
Both derive closed-form prediction intervals accounting for 
newly recruited and at-risk patients under different 
recruitment scenarios in ongoing multicentre trials. These intervals use the plug-in method \parencite{MeekerHahnEscobar2017StatisticalIntervals}, replacing unknown model parameters by point estimates and then treating them as fixed. For within-sample prediction, it was demonstrated that this leads to coverage far from the nominal level because uncertainty from parameter estimation is not carried through to the predictive distribution \parencite{Tian2022}.

We take a conditional forecasting view of the interim problem. At an interim analysis, the patients still event-free form the risk set for future events, and each patient has an individual follow-up history and covariate profile. Given a fitted survival model, the future count can be built from patient-specific event probabilities. Instead of using a plug-in predictive distribution, we estimate the distribution of the future count with a conditional parametric bootstrap. To do this, direct-bootstrap prediction intervals for future failures in reliability studies \parencite{Tian2022} are extended to the clinical-trial setting. The proposed approach propagates uncertainty from the fitted event-time and loss to follow-up models while keeping the prediction target tied to the patients in the risk set at the interim analysis.

The resulting framework accommodates staggered entry, unequal follow-up, patient-level covariates, administrative censoring, and loss to follow-up. It also allows dependence between entry dates and loss to follow-up before conditioning on the realised interim follow-up windows. These features are motivated by a recently completed phase III trial in childhood acute lymphoblastic leukaemia (ALL) \parencite{Conter2024PEGAsparaginaseHRALL}, where the practical goal was to predict, based on interim data, the number of events expected to occur by the time of the planned final analysis.

The remainder of the paper is organised as follows. Section~\ref{sec:framework} introduces the interim prediction problem, the patient-level survival model, and the Poisson--binomial representation of the future event count. Section~\ref{sec:cpb} develops the conditional parametric bootstrap and the resulting prediction intervals. Section~\ref{Simulations} presents simulation studies evaluating the operating characteristics of the method. Section~\ref{Case} illustrates the approach in a phase III trial in childhood ALL. Finally, Sections~\ref{Disc} and~\ref{FR} conclude the paper by discussing the findings and outlining potential extensions. 

\section{Statistical framework}\label{sec:framework}

Consider a trial with a time-to-event endpoint and closed accrual at the interim analysis. Let $t_c$ denote the calendar time of the interim look and let $\Delta>0$ be a future prediction horizon. For patient $i=1,\ldots,n$, let $A_i$ be the calendar entry time, $T_i$ the time to event measured from entry, $C_i$ the time to loss to follow-up measured from entry, and $\bm z_i$ a vector of baseline covariates. The follow-up available at interim is $\tau_i=t_c-A_i$, and the observed data are $X_i=\min(T_i,C_i,\tau_i)$ together with an indicator $\delta_i\in\{\mathrm E,\mathrm L,\mathrm A\}$ denoting, respectively, an observed event, loss to follow-up before the interim cutoff, or administrative censoring at $t_c$. We write $\mathcal D_c=\{(X_i,\delta_i,\tau_i,\bm z_i):i=1,\ldots,n\}$ for the interim data and $\mathcal R_c=\{i:\delta_i=\mathrm A\}$  for the set of patients still at risk, i.e. event-free under observation at the interim analysis.

The prediction target is the number of additional events among patients in $\mathcal R_c$ during the future calendar window $(t_c,t_c+\Delta]$,
\begin{equation}\label{eq:YDelta}
Y_\Delta
=\sum_{i\in\mathcal R_c}
\mathbb I\{\tau_i<T_i\leq \tau_i+\Delta,\, T_i\leq C_i\}.
\end{equation}
This is a within-sample prediction problem: the future event count concerns the same patients observed at the interim analysis, rather than an independent future sample \parencite{MeekerHahnEscobar2017StatisticalIntervals}. The relevant object is therefore the conditional distribution of $Y_\Delta$ given the interim data. In the special case without loss to follow-up and without covariate heterogeneity, staggered patient entry can be viewed as a generalized type-I censoring structure \parencite{KleinMoeschberger2003SurvivalAnalysis}, closely related to the multiple-cohort reliability setting studied by \textcite{Tian2022}. Our formulation keeps the clinical-trial problem as the primary object and extends this connection to patient-level covariates and loss to follow-up.

We model the event time through a parametric survival function
$S_{\bm\theta}(t\mid\bm z_i)$ with density
$f_{\bm\theta}(t\mid\bm z_i)$. Furthermore, patients typically enter a trial on different calendar dates, so the available follow-up window $\tau_i=t_c-A_i$ at interim is determined by the accrual process. The time to loss to follow-up
may also depend on when a patient entered \parencite{Jiang2021}. For example, if retention
procedures are strengthened during recruitment, patients who enter later
may be less likely to be lost to follow-up within the same time since
entry. To allow the distribution of time to loss to follow-up to vary with entry time, we model it conditional on when each patient enters the trial. We write $G(t)$ and $g(t)$ for the corresponding conditional survival function and density. We assume that patients are independent conditional on their entry times and covariates and that, within each patient, $T_i\perp\!\!\!\perp C_i\mid(A_i,\bm z_i)$.

Let $Q_{\bm\xi}(a)$ and $q_{\bm\xi}(a)$ denote the cumulative distribution function (CDF) and density of
the entry-time distribution, respectively, from a
regular parametric family with known, parameter-independent
support $(0,a_{\max})$, where $a_{\max}$ is accrual closure date and $\bm\xi\in\Xi$. We use a Gaussian copula to flexibly link the continuous marginal distributions of entry time and time to loss to follow-up \parencite{SureshTaylorTsodikov2021, PanJoe2022, ZhangCharalambousFoster2023}. For a given accrual distribution, this construction gives closed-form expressions for $G(t)$ and $g(t)$ in terms of $\bm \xi$ and $\bm\psi=(\bm\nu,\rho)$, where $\bm\nu$ indexes the marginal loss-to-follow-up distribution and $\rho\in(-1,1)$ is the copula correlation parameter. In what follows, the dependence of \(G(t)\) on entry time and model parameters is left implicit and made explicit whenever necessary. The accrual distribution can be chosen from any continuous family whose cumulative distribution function can be inverted analytically or numerically. Supplementary materials~\ref{app:gaussian-dropout} gives the details of the construction. We assume that $\bm\theta$, $\bm\psi$ and $\bm\xi$ are
variation-independent, with joint parameter space
$\Theta\times\Psi\times\Xi$.
Accordingly, the predictive parameter is
$\bm\eta=(\bm\theta,\bm\psi,\bm\xi)$.

For each patient event-free at $t_c$, the conditional probability of contributing an event by $t_c+\Delta$ is
\begin{equation}\label{eq:piDelta}
\begin{aligned}
\pi_{i,\Delta}(\bm\eta)
&=\Pr\{T_i\leq \min(\tau_i+\Delta,C_i)\mid T_i>\tau_i,\,C_i>\tau_i,\, a_i, \, \bm z_i\} \\
&=\frac{\int_{\tau_i}^{\tau_i+\Delta} f_{\bm\theta}(u\mid\bm z_i)G_{\bm\psi}(u)\,\mathrm du}
{S_{\bm\theta}(\tau_i\mid\bm z_i)G_{\bm\psi}(\tau_i)}.    
\end{aligned}
\end{equation}

Thus, conditional on $\mathcal D_c$ and on $\bm\eta$, the future count follows a Poisson--binomial law,
\begin{equation}\label{eq:PoiBinDelta}
Y_\Delta\mid \mathcal D_c,\bm\eta
\sim
\mathrm{PoiBin}\{\bm\pi_\Delta(\bm\eta)\},
\qquad
\bm\pi_\Delta(\bm\eta)=\{\pi_{i,\Delta}(\bm\eta):i\in\mathcal R_c\},
\end{equation}

where $\mathrm{PoiBin}(\bm p)$ denotes the distribution of a sum of independent Bernoulli random variables with possibly unequal success probabilities \parencite{Hong2013,Tang2022}. See Section A of the Supplementary Materials for further details on the Poisson--binomial distribution. This formulation accommodates patient-specific follow-up times, patient-level covariates, and loss to follow-up through the individual probabilities in \eqref{eq:piDelta}. In the special case with no loss to follow-up, $G_{\bm\psi}\equiv1$ and $\pi_{i,\Delta}(\bm\theta)=1-S_{\bm\theta}(\tau_i+\Delta\mid\bm z_i)/S_{\bm\theta}(\tau_i\mid\bm z_i)$. Derivations and additional special cases are given in Section B of the Supplementary Materials.

Under the assumptions above, the predictive parameter can be estimated as follows. The accrual parameter is first estimated from its
marginal likelihood,
$\mathcal{L}_A(\bm\xi)=\prod_{i=1}^n q_{\bm\xi}(a_i)$.
The conditional loss-to-follow-up likelihood is then maximized with $\bm{\xi}$ fixed at $\widehat{\bm{\xi}}$, while $\bm{\theta}$ is estimated separately. Both estimators are obtained by maximizing the corresponding right-censored likelihoods:
\begin{equation}\label{eq:lik}
\begin{aligned}
&\mathcal L_T(\bm\theta)
\propto
\prod_{i=1}^{n}
 f_{\bm\theta}(X_i\mid\bm z_i)^{\mathbb I\{\delta_i=\mathrm E\}}
 S_{\bm\theta}(X_i\mid\bm z_i)^{1-\mathbb I\{\delta_i=\mathrm E\}},
\\
&\mathcal L_C(\bm\psi)
\propto
\prod_{i=1}^{n}
 g_{\bm\psi}(X_i ; {\widehat{\bm \xi}})^{\mathbb I\{\delta_i=\mathrm L\}}
 G_{\bm\psi}(X_i; {\widehat{\bm \xi}})^{1-\mathbb I\{\delta_i=\mathrm L\}}.    
\end{aligned}
\end{equation}

Thus
$\bm{\widehat\eta}=
(\bm{\widehat\theta},\bm{\widehat\psi},\bm{\widehat\xi})$. The derivation of the factorisation, including the role of the entry-time process, is given in Section \ref{App:Lik} of the Supplementary Materials. The framework does not require a specific parametric family for the event or loss-to-follow-up distributions beyond identifiability and regularity conditions. Standard parametric and flexible parametric survival specifications, including Royston--Parmar models \parencite{Royston2002} and generalized gamma models \parencite{Stacy1962}, can therefore be used.

A level $1-\alpha$ prediction interval for $Y_\Delta$ is a random interval $\mathrm{PI}_{1-\alpha}(\mathcal D_c)=[L(\mathcal D_c),U(\mathcal D_c)]$ whose repeated-sampling coverage is evaluated for the future count. Since $Y_\Delta$ is discrete, exact equality between coverage and the nominal level is generally not attainable, and the natural target is asymptotic. A simple plug-in interval is obtained by replacing $\bm\eta$ with $\widehat{\bm\eta}$ in \eqref{eq:PoiBinDelta} and taking the corresponding lower and upper quantiles. Although appealing, this treats the fitted parameters as known. For within-sample prediction, such intervals can undercover because they do not propagate the sampling variability of $\widehat{\bm\eta}$ into the predictive distribution \parencite{Beran1990,Tian2022}. This limitation motivates the conditional parametric bootstrap developed next.
\section{Conditional parametric bootstrap}\label{sec:cpb}

While the plug-in interval described above uses the fitted parameter value
$\widehat{\bm \eta}$ as if it were known, we account here for parameter-estimation
uncertainty while preserving the conditional nature of the prediction problem.
We construct a bootstrap predictive distribution for $Y_\Delta$ around the
observed interim data. The construction is related to the direct-bootstrap prediction intervals of \textcite{Harris1989}, \textcite{Shen2018}, and \textcite{Tian2022}. In particular, \textcite{Tian2022} compare direct bootstrap methods with calibration-bootstrap ideas from \textcite{Beran1990} and generalized pivotal-quantity constructions related to \textcite{Hannig2006} and \textcite{Wang2012}. We use the direct-bootstrap logic because it is naturally aligned with the conditional predictive distribution in \eqref{eq:PoiBinDelta} and can be adapted to the patient-level censoring and covariate structure of interim monitoring.

Given $\widehat{\bm\eta}=(\widehat{\bm\theta},\widehat{\bm\psi}, \widehat{\bm \xi})$, bootstrap datasets are generated conditionally on the realised follow-up windows $\tau_i$ and covariates $\bm z_i$. For each patient, an event time and a loss-to-follow-up time are sampled independently from $f_{\widehat{\theta}}(t\mid\bm z_i)$ and $g_{\widehat{\psi}}(t; \widehat{\bm \xi})$, respectively. For each patient, the bootstrap observed time is set to the minimum of the sampled event time, the sampled loss to follow-up time, and the original follow-up window $\tau_i$. For patients administratively censored in the bootstrap sample, the observed time is therefore equal to $\tau_i$. The corresponding indicator records whether an event, loss to follow-up, or administrative censoring occurs first.  Thus, each bootstrap dataset retains the original follow-up windows $\tau_i$ and covariates $\bm z_i$, while event and loss-to-follow-up times are resampled from the fitted models and the corresponding event/censoring indicators are reconstructed, as in classical bootstrap schemes for censored survival data \parencite{Efron1981CensoredBootstrap,Hjort1985BootstrappingCox,Akritas1986BootstrappingKM,GrossLai1996BootstrapTruncatedCensored,Bilker1997}. 

Although inference is conditional on the observed entry times, sampling variability in \(\bm\xi\) should also be propagated when relevant, as it also affects the estimation of $\bm{\psi}$. We therefore generate auxiliary entry times from $Q_{\widehat{\bm \xi}}(a)$ solely to propagate uncertainty in $\widehat{\bm{\xi}}$. In each bootstrap replicate, the conditional loss-to-follow-up likelihood is evaluated using the original entry times, with $\bm{\xi}$ fixed at its bootstrap estimate $\widehat{\bm{\xi}}^{*(b)}$. Re-fitting the event and loss-to-follow-up models to the $b$th bootstrap dataset gives $\widehat{\bm\eta}^{*(b)}$.

For each bootstrap estimate, the conditional event probabilities in \eqref{eq:piDelta} are computed only for patients in the original interim risk set $\mathcal R_c$ and used to construct the conditional future-count distribution through the Poisson--binomial law in \eqref{eq:PoiBinDelta}. Averaging these distribution functions gives
\begin{equation}\label{eq:FhatStar}
\widehat F^*_{\Delta}(y\mid\mathcal D_c)
=
\frac{1}{B}\sum_{b=1}^B
F_{\mathrm{PB}}\{y;\bm\pi_\Delta(\widehat{\bm\eta}^{*(b)})\},
\end{equation}
where $F_{\mathrm{PB}}(\cdot;\bm p)$ denotes the distribution function of a Poisson--binomial random variable with success probabilities $\bm p$. The equal-tailed level $1-\alpha$ prediction interval is then obtained from the empirical predictive distribution as
\begin{equation}\label{eq:PIstar}
\mathrm{PI}_{1-\alpha}(\mathcal D_c)
=
\left[\widehat q^*_{\alpha/2},\widehat q^*_{1-\alpha/2}\right],
\qquad
\widehat q^*_p=\inf\{y:\widehat F^*_{\Delta}(y\mid\mathcal D_c)\geq p\}.
\end{equation}

Evaluating $\bm\pi_\Delta(\widehat{\bm\eta}^{*(b)})$ on the original risk set, kept fixed across bootstrap replicates, is essential. If the conditional probabilities were evaluated on the risk sets defined by the regenerated event/censoring indicators, the set of patients contributing to the prediction would vary across bootstrap samples. A predictive distribution evaluated on these sets would average over different risk sets and would no longer target the within-sample forecasting problem defined by the observed interim data. We therefore regenerate the indicators, which carry information on $\bm \eta$, only to re-estimate $\bm \eta$, and always evaluate the predictive distribution on the original risk set $\mathcal{R}_c$. Sampling the event and loss-to-follow-up times can be implemented by inverse-transform methods for a wide range of parametric survival models \parencite{BenderAugustinBlettner2005GeneratingSurvivalTimes,CrowtherLambert2013SimulatingComplexSurvival}.

Under standard regularity conditions for the fitted parametric models, consistency of the parametric bootstrap for the parameter estimator, and smoothness and non-degeneracy conditions on the patient-specific probabilities in \eqref{eq:piDelta}, the bootstrap distribution of the standardised prediction error induced by \eqref{eq:FhatStar} consistently approximates its repeated-sampling distribution. Consequently, the quantiles in \eqref{eq:PIstar} yield asymptotically calibrated prediction intervals. The argument follows the direct-bootstrap theory of \textcite{Tian2022}, with the additional technical conditions required by subject-specific follow-up windows, covariates, loss to follow-up, and the fixed interim risk set. The resampling scheme of the conditional parametric bootstrap (CPB) is reported in Section \ref{App:Algo} of the Supplementary materials. The formal conditions and their connection to Theorems~1--3 and Lemmas~1--3 of \textcite{Tian2022} are given in Section \ref{App:Cov} of the Supplementary materials.

\section{Simulation study}\label{Simulations}

We conducted two simulation studies to evaluate the operating characteristics of the proposed prediction intervals in settings motivated by interim monitoring of time-to-event trials. Study $\mathcal S_1$ assesses the full proposed framework under correct model specification, including loss to follow-up and dependence between entry time and loss to follow-up before conditioning on the interim data. Study $\mathcal S_2$ considers administrative censoring only and evaluates the sensitivity of the prediction intervals to the parametric survival model used for extrapolation.

In both studies, the estimand is the future event count $Y_\Delta$ among patients still at risk at the interim analysis, and the performance of the interval procedure is evaluated over repeated simulated interim datasets. We use the same notation as in Sections~\ref{sec:framework}--\ref{sec:cpb}: $\mathcal D_c^{(m)}$ denotes the $m$th Monte Carlo interim dataset, while $\mathcal D_c^{*(b)}$ denotes the $b$th bootstrap replicate generated from a fixed dataset.

\subsection{Simulation design}

In $\mathcal S_1$, event times were generated from a Weibull proportional hazards model with treatment as a binary covariate and 1:1 randomisation. Entry times were uniformly distributed over a three-year accrual period. Loss-to-follow-up times followed an exponential distribution and were allowed to be dependent on entry times, so that patients enrolled earlier were more likely to be lost to follow-up before the interim analysis. This setting directly targets the extension introduced in Section~\ref{sec:framework}, where inference is conditional on the realised follow-up windows and does not require independence between entry time and loss to follow-up before conditioning. Implementation details and resulting parameter values are reported in Section F of the Supplementary Materials.

In $\mathcal S_2$, the data-generating mechanism was analogous, but only administrative censoring at the interim cutoff was imposed. This study isolates the role of survival extrapolation by comparing prediction intervals obtained under several fitted parametric models: the correctly specified Weibull model, generalized gamma models \parencite{Stacy1962,Prentice1975,Cox2007}, and Royston--Parmar models with increasing numbers of knots (1 to 3) chosen as equally-spaced quantiles of the log events times \parencite{Royston2002}. Following \textcite{Mori2024}, Royston--Parmar models were included as flexible parametric alternatives to standard survival regression models. 

The factors that varied across scenarios were follow-up maturity at interim prediction, treatment effect size, future prediction horizon, and censoring intensity. Follow-up maturity was defined by the timing of the interim analysis relative to the end of accrual, with later interim analyses allowing longer patient follow-up and more precise estimation of the event-time distribution. In $\mathcal S_1$, we additionally varied the strength of dependence between entry time and loss to follow-up and the relative rate of loss to follow-up. In $\mathcal S_2$, we varied the proportion of patients administratively censored at the interim analysis. Full scenario definitions are given in Section F of the Supplementary Materials.

For each scenario, we generated $N=1000$ Monte Carlo datasets of size $n=1000$. For each dataset, two-sided 95\% prediction intervals were constructed using $B=1000$ bootstrap replicates. This number of bootstrap replicates was chosen to keep the computational burden manageable across the full factorial simulation design. In $\mathcal{S}_1$, prediction intervals based on the conditional parametric bootstrap were additionally compared with those obtained using the corresponding plug-in approach. All simulations were implemented in R; survival models were fitted using \texttt{survival} and \texttt{flexsurv}, and Poisson--binomial probabilities were evaluated using \texttt{poibin}. Additional package details are reported in Section F of the Supplementary Materials. 
%R code for implementing the parametric-bootstrap procedure used in the simulation study and case study is provided in the Supplementary Materials. 

\subsection{Performance measure}

In both simulation study, the performance measure is unconditional coverage probability, defined as the repeated-sampling probability that the interval constructed from the interim data contains the future count. Since the conditional distribution of $Y_\Delta$ is available in each simulated dataset when evaluated at the true data-generating parameters, we estimate coverage by averaging the conditional coverage over Monte Carlo replications:
\[
\widehat{\mathrm{UCP}}
=
\frac{1}{N}\sum_{m=1}^{N}
\Pr_{\bm\eta_0}\!\left\{
Y_\Delta\in
\mathrm{PI}_{1-\alpha}(\mathcal D_c^{(m)})
\,\middle|\,
\mathcal D_c^{(m)}
\right\},
\]
where $\bm\eta_0$ denotes the true simulation parameter and the probability is computed from the Poisson--binomial distribution with success probabilities evaluated at $\bm\eta_0$.

\subsection{Results}

In $\mathcal S_1$, the coverage of the conditional parametric bootstrap intervals was generally close to the nominal level and met or exceeded it in every scenario. Coverage varied most with the prediction horizon ($\Delta t$) and was slightly lower for longer-term predictions. The other factors considered showed no systematic association with coverage across the simulated scenarios. (Figure~\ref{fig:cov_drop}, panel A; Monte Carlo standard errors are reported in Figure~2 of the Supplementary Materials). The plug-in intervals consistently had lower coverage than the bootstrap intervals. Plug-in coverage was more sensitive to follow-up maturity, and this sensitivity varied with the prediction horizon improving with more mature follow-up at the interim analysis for both short- and long-term predictions. Nevertheless, at longer prediction horizons, the coverage remained consistently below the nominal level.

In $\mathcal S_2$, the coverage was generally close to the nominal level across fitted models and exceeded it in most scenarios (Figure~\ref{fig:cov_drop}, panels B and C; Monte Carlo standard errors are reported in Figure~3 of the Supplementary Materials). Coverage improved with more mature follow-up at the interim analysis, with only small differences across prediction horizons. The modest reductions in coverage at long-term prediction horizon were more apparent when a greater proportion of patients were administratively censored at interim. This proportion varied with both the treatment effect and the baseline hazard. The more flexible models generally achieved coverage comparable to that of the simpler Weibull model, with slightly lower coverage in a few scenarios. The generalized gamma and Royston--Parmar models yielded similar coverage across scenarios, with the closest agreement between the generalized gamma model and the three-knot Royston--Parmar model. The one- and two-knot Royston--Parmar models yielded coverage similar to that of the Weibull model, and neither spline specification consistently outperformed the other.

\begin{figure}[t]
  \centering
          \makebox[\textwidth][c]{%
            \includegraphics[width=1.15\textwidth]{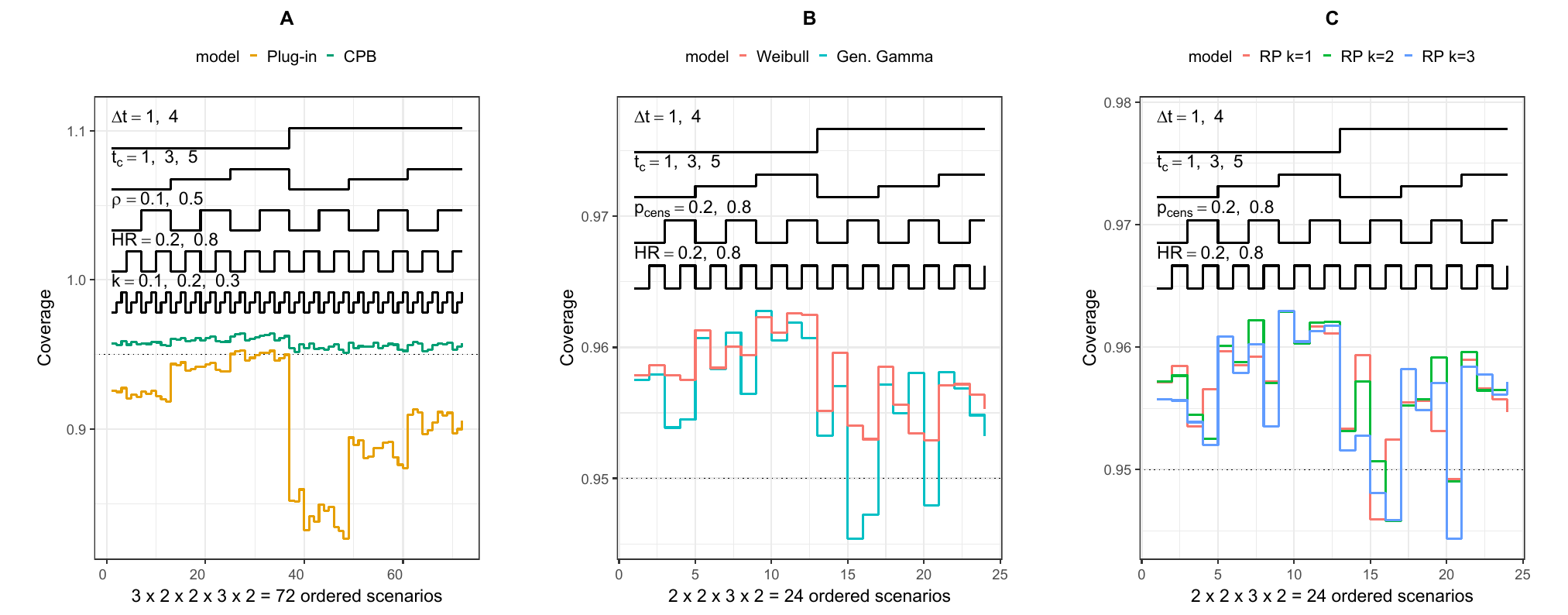}
        }
\caption{Coverage of the two-sided 95\%  prediction intervals across the investigated simulation scenarios.
Panel~A reports results from Study $\mathcal{S}_1$, comparing the conditional parametric bootstrap (CPB) with the plug-in approach.
Panels~B and~C report results from Study $\mathcal{S}_2$: Panel~B compares the Weibull and generalized gamma models, whereas Panel~C compares Royston--Parmar (RP) models with different numbers of knots ($k=1,2,3$).
$\Delta t$ denotes the prediction horizon, $t_c$ the time of the interim analysis from accrual end, $\rho$ the marginal correlation between loss-to-follow-up times and entry dates, and $\mathrm{HR}$ the hazard ratio.
In Study $\mathcal{S}_1$, $k$ denotes the ratio between the loss-to-follow-up rate and the baseline scale parameter of the event-time distribution.
In Study $\mathcal{S}_2$, $p_{\text{cens}}$ denotes the proportion of patients administratively censored at the interim analysis.
The dotted horizontal line indicates the nominal coverage level of 0.95.}
  \label{fig:cov_drop}
\end{figure}

\section{Case study in Pediatric ALL}
\label{Case}
To illustrate the proposed method, we considered data from the AIEOP-BFM ALL 2009 protocol,  an international multicenter study on newly diagnosed children (<18 years of age) with acute lymphoblasticleukaemia (ALL). 
The protocol included three phase III studies; in particular, one open-label randomised study aimed at assessing whether intensifying asparaginase exposure during early consolidation (phase IB) could reduce minimal residual disease (MRD) and improve outcome in children and adolescents with high-risk ALL. In the experimental arm, patients received four additional weekly doses of pegylated L-asparaginase (PEG-ASNase) on top of standard chemotherapy, while the control arm followed the standard regimen  \parencite{Conter2024PEGAsparaginaseHRALL}. Between June 2010 and February 2017, 6136 children (aged 1–17 years) with Philadelphia-negative ALL were enrolled; 1097 met high-risk criteria (defined at the end of induction as poor response and unfavourable genetic features) and 809 were randomised (404 experimental vs 405 control arm). 

For the application of our method, we excluded five
patients from the analysis: two misclassified as
high-risk and three classified as resistant according
to the protocol (two in the experimental arm and one
in the control arm), leaving 804 patients. We focused on event-free survival (EFS) and constructed
two-sided prediction intervals with a nominal coverage
of 95\% for the cumulative number of additional EFS events expected every six months from June 1, 2017 (accrual closure) to June 1, 2021, when data were frozen for final analysis. The additional events actually observed in randomized patients after June 1, 2017 are then reported by semester to evaluate if the observed events are included by the prediction intervals. EFS is defined as the time from random assignment to the date of the first event (among: resistance to treatment, relapse, death, or second malignant neoplasm) or, in the absence of an event, to the last follow-up (censored observation). Since no loss to follow-up occurred at the time of the final analysis, only administrative censoring was considered, leading to a simplified formulation of \eqref{eq:PoiBinDelta} and of the sampling scheme, as detailed in Sections B.1
and D.1 of the Supplementary Materials. By June 1, 2017, 164 events had occurred and, by the final analysis, 62 additional events had been observed, giving a total of 226 events over the 804 analyzed patients. 

We obtained prediction intervals under various regression models with treatment assignment as the only covariate. The baseline time-to-event distribution was specified as Weibull, Log-logistic, Log-normal, their Royston–Parmar counterparts --proportional hazards (PH), proportional odds (PO), and linear probit (LP) with 1, 2, or 3 internal knots chosen as equally-spaced quantiles of the log events times-- and the Generalized Gamma distribution. Model fit at interim was assessed both graphically  and with information criteria. For a fitted model with loglikelihood $\ell=\ell(\hat{\bm \theta})$, $q$ parameters and sample size $n$, we computed $\mathrm{AIC} \;=\; -2\,\ell + 2q$ and $\mathrm{BIC} \;=\; -2\,\ell + q\log n$. We used 5000 bootstrap samples to construct the prediction intervals. 

Model-fit comparisons based on information criteria (Table~\ref{tab:modelfit}, ordered by increasing BIC) highlight the trade-off between parsimony and goodness of fit. Maximized log-likelihoods were broadly similar among the Generalized Gamma and Royston–Parmar models, whereas the Weibull and Log-logistic models showed poorer fit. BIC tended to favor more parsimonious specifications, penalizing models with a greater number of spline knots, whereas AIC favored some of the more flexible Royston–Parmar specifications over their simpler counterparts. The Generalized Gamma model achieved the lowest values of both AIC and BIC, despite some Royston–Parmar specifications attaining slightly higher maximized log-likelihoods.

The smoothed arm-specific hazard estimates suggested possible departures from proportional hazards, particularly during later follow-up (Figure~4, Supplementary Materials). The nonlinear relationships between transformations of the survival function and log time further supported the use of flexible Royston–Parmar models (Figure~5, Supplementary Materials).

\begin{figure}[H]
  \centering
\includegraphics[width=1.0\textwidth]{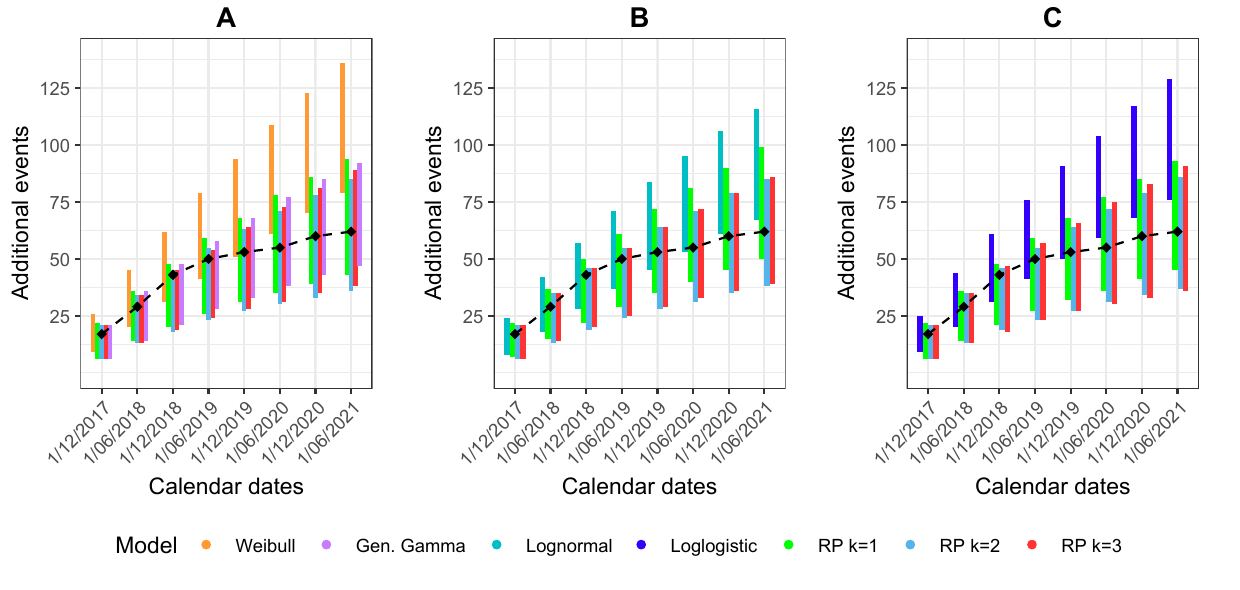}
\caption{Two-sided 95\% prediction intervals for the cumulative number of additional EFS events from June~1$^{\mathrm{st}}$ 2017 (interim analysis) to June~1$^{\mathrm{st}}$ 2021 by each semester. Black diamonds indicate the observed counts. Panel~A shows intervals under the Weibull model and its flexible specifications; Panel~B shows intervals under the log-normal model and its flexible specifications; Panel~C shows intervals under the log-logistic model and its flexible specifications. RP = Royston--Parmar; $k$ denotes the number of knots.}
  \label{figCNE}
\end{figure}

Prediction intervals for the number of additional EFS events obtained from simple parametric regression models (Weibull, log-logistic, and log-normal) were broadly similar across the prediction window and were consistently shifted upwards relative to the flexible counterparts (Figure~\ref{figCNE}). Within the Royston–Parmar models, different link functions (PH, PO, LP) produced comparable intervals and displayed a coherent pattern as the number of internal knots increased: specifications with 2–3 knots yielded more downward-shifting prediction intervals as the prediction horizon lengthened, compared with the 1-knot models. At the final analysis, 62 additional EFS events had been observed. This count fell below the lower prediction limits obtained under all three standard parametric models, whereas it was contained within the intervals obtained under the generalized gamma and all Royston–Parmar specifications.

Model discrimination was also examined in relation to extrapolative performance. 
Visual inspection of the fitted hazard functions (Figures~6 to Figures~12, Supplementary Materials) showed that the models produced different hazard trajectories. 
These differences were less pronounced in the cumulative hazard functions (Figures~13 to Figures~15, Supplementary Materials). 
Although the baseline hazard shapes and covariate-effect parameterizations differed across models, the Weibull, log-logistic, and log-normal models produced broadly similar predicted event counts over the prediction horizon. Similar results were observed across the different Royston--Parmar specifications, which also yielded comparable predictions despite differences in model flexibility and baseline hazard representation.

\begin{table}[H]
\centering
\small
\caption{Model fit measures on data at June 1, 2017 (end of accrual)}
\vspace{2mm}
\label{tab:modelfit}
\begin{tabular}{lrrrr}
\toprule
\textit{Model}      & $q$ & \textit{logLik}   & \textit{AIC}  & \textit{BIC}  \\
\midrule
Generalized Gamma   & 4   & $-562.9134$ & 1133.827 & 1152.585 \\
RP -- PO(1)         & 4   & $-564.2135$ & 1136.427 & 1155.185 \\
Log-normal          & 3   & $-567.7988$ & 1141.598 & 1155.666 \\
RP -- PH(1)         & 4   & $-564.4817$ & 1136.963 & 1155.722 \\
RP -- LP(1)         & 4   & $-564.6479$ & 1137.296 & 1156.054 \\
RP -- PH(2)         & 5   & $-562.1776$ & 1134.355 & 1157.803 \\
RP -- PO(2)         & 5   & $-563.1182$ & 1136.236 & 1159.684 \\
RP -- LP(2)         & 5   & $-563.4079$ & 1136.816 & 1160.264 \\
RP -- PH(3)         & 6   & $-561.4914$ & 1134.983 & 1163.120 \\
RP -- PO(3)         & 6   & $-562.4590$ & 1136.918 & 1165.056 \\
RP -- LP(3)         & 6   & $-562.7311$ & 1137.462 & 1165.600 \\
Log-logistic        & 3   & $-573.9548$ & 1153.910 & 1167.978 \\
Weibull             & 3   & $-576.7842$ & 1159.568 & 1173.637 \\
\bottomrule
\end{tabular}

\vspace{1mm}
\begin{minipage}{0.9\textwidth}
\footnotesize\emph{Note:} Models are ordered by increasing BIC. $q$ denotes the number of parameters. RP = Royston--Parmar model,
PH($k$) = Proportional Hazards, PO($k$) = Proportional Odds, LP($k$) = Linear Probit,
where $k$ denotes the number of knots.
\end{minipage}
\end{table}
  
%--------------------------------------------------------
\section{Discussion}\label{Disc}

In this work, we adopted a conditional forecasting framework for predicting future event counts at interim analyses in time-to-event trials. Conditional on the patients observed at interim, fitted survival and loss-to-follow-up models were used to compute patient-specific future event probabilities, and prediction intervals were obtained through a conditional parametric bootstrap extending the direct-bootstrap approach of \textcite{Tian2022} to the clinical-trial setting. The proposed framework preserves their formal conditions for asymptotic nominal coverage while accommodating staggered entry, unequal follow-up, covariates, administrative censoring, and loss to follow-up. The approach is illustrated in a phase III randomized trial on childhood ALL, where the objective was to support planning of the final analysis by forecasting the number of events expected to have accrued by that time.

The Poisson–binomial representation of the future event count provides a flexible framework to model heterogeneous, subject-specific conditional probabilities of experiencing the event within the future time window, driven by staggered entry and, more generally, by covariate profiles. This formulation also naturally allows different parametric models for event times and censoring times. Embedding such models within the conditional parametric bootstrap scheme proposed requires simulating event and loss to follow-up times from the chosen distributions. This provides substantial flexibility and allows sensitivity analyses to be carried out straightforwardly across different distributional assumptions by leveraging a broad class of parametric survival models available in the literature \parencite{CrowtherLambert2014,LiuPawitanClements2018}. In this work we assumed a common loss to follow-up mechanism across covariate profiles, but the approach readily extends to settings where dropout also varies by covariates (e.g., treatment arm). A further assumption of the proposed framework is that the trial population and the main event and censoring mechanisms remain reasonably stable over time, a generally reasonable point in clinical trials, even in the context of adaptive designs \parencite{Friede2009}. Under this perspective, the interim dataset provides a relevant basis for predicting the future course of the same ongoing trial, and the proposed conditional bootstrap scheme is designed to reflect this conditional prediction setting.

%%%%%%%%%%%%%%
We examined the operating characteristics of the proposed method across a range of settings motivated by interim monitoring of time-to-event trials and by the case study on pediatric ALL. The prediction horizon emerged as the main determinant of bootstrap interval performance. Coverage tended to decline modestly at longer
prediction horizons but met or exceeded the nominal
level in all scenarios of Study $\mathcal{S}_1$ and in most
scenarios of Study $\mathcal{S}_2$. The plug-in intervals showed systematically lower coverage than their bootstrap counterparts and were more sensitive to follow-up maturity. Coverage improved with more mature follow-up but declined markedly at longer prediction horizons. At those horizons, coverage remained well below the nominal level even at later interim analyses.

Across the simulated scenarios, more flexible parametric extensions generally maintained the predictive performance of the simpler models, with no consistent improvement. Under more mature follow-up, coverage remained above the nominal level across the fitted models, with only modest deterioration as the prediction horizon increased, particularly when a substantial proportion of patients administratively censored occured at early interim analyses. This pattern was especially evident for the Royston--Parmar models, for which increasing the number of spline knots did not translate into systematically better performance. These findings suggest that more flexible parametric models may represent a reasonable choice when assumptions about the underlying survival distribution are difficult to ascertain.

The analysis of the real case suggests that model choice should be guided not only by information criteria but also by extrapolative behaviour beyond the maximum follow-up observed at the interim through  graphical inspection. Models with different hazard shapes and different covariate-effect scales may nevertheless yield similar extrapolations when they imply similar behaviour on the survival, or equivalently, on the cumulative hazard scale over the prediction horizon. In this sense, apparent differences in the fitted hazard functions do not necessarily translate into materially different predicted event probabilities.

\section{Further research}\label{FR}
This work opens several promising avenues for future research aimed at addressing the limitations of the proposed framework and connect it to a set of active areas. First, we focus on settings in which accrual is complete at the prediction time. In many trials, interim analyses occur while recruitment is ongoing, so a practically relevant extension is joint prediction of both future accrual and future events as often considered in literature on time to endpoint maturation. Second, we assume that the data at the interim analysis is free of reporting delays. In multicenter trials, however, delays in data entry and event reporting are common and can materially affect interim monitoring and prediction \parencite{TsiatisDavidian2022,WangKeJiangZhangSnapinn2012,AubelAntignyFougerayDuboisSaintHilary2021}. Extensions that explicitly account for delayed reporting would therefore be valuable. Third, we assume access to unblinded, patient-level information. In practice, interim analyses are frequently conducted under blinding constraints to preserve trial integrity, limiting the availability of individual-level covariates or even treatment indicators. Related work on time to endpoint maturation prediction under blinded data \parencite{Donovan2006,Fu2025BayesianPredictionEventTimes,ZhangPuDengRoychoudhuryChuRobinson2025StudyDuration} suggests promising directions to adapt our framework to blinded settings. Regarding the survival model, we restrict attention to regression-based parametric specifications without explicitly modeling potential violations of the covariates effect scale (e.g., departures from proportional hazards, proportional odds, or accelerated failure time assumptions). While in the motivating application ignoring such violations did not lead to marked differences in prediction intervals, this warrants further investigation. Furthermore, our work may benefit from the extensive literature on parametric survival extrapolation from clinical trials for economic evaluation \parencite{Gray2021,Cooper2022,Latimer2022,Sweeting2023,Jackson2023survextrap,Palmer2023GuideFlexibleSurvival,Chen2024}, where model choice and extrapolative performance are central concerns. 

Motivated by the key role of interim monitoring in event-driven clinical trials, our approach involved a parametric bootstrap framework. This choice makes the resulting prediction intervals directly aligned with trial protocols in which interim decisions at late phases are typically based on frequentist error-control criteria. Within this framework, our approach provides a practical way to propagate parameter-estimation uncertainty while preserving the observed interim risk set and the conditional nature of the prediction target.
A full Bayesian formulation of the same prediction problem would provide a complementary inferential route. In such a formulation, uncertainty about the event-time, accrual, and loss-to-follow-up processes could be encoded through prior distributions, and prediction intervals could be obtained from the posterior predictive distribution. This may be particularly valuable when reliable external evidence is available, for example from previous trials, registries, or available expert elicitation. However, in interim prediction settings, especially when follow-up is immature or there is limited knowledge on the disease areas, the construction of defensible priors for both the survival extrapolation and the censoring may itself be challenging. Developing Bayesian formulations in this settings is therefore a promising direction for future work.

\section*{Funding}

Edoardo Ratti is partially supported by the grant: Italian MUR
Dipartimenti di Eccellenza 2023--2027 (l.232/2016, art. 1, commi
314--337).

\section*{Conflicts of Interest}

The authors declare no conflicts of interest.

\section*{Data Availability Statement}

The data that support the findings of this study are available from
the corresponding author upon reasonable request.

\printbibliography

@article{Tian2022,
   author = {Qinglong Tian and Fanqi Meng and Daniel J. Nordman and William Q. Meeker},
   doi = {10.1080/01621459.2020.1850461},
   issn = {1537274X},
   number = {539},
   journal = {Journal of the American Statistical Association},
   pages = {1296-1310},
   publisher = {Taylor and Francis Ltd.},
   title = {Predicting the Number of Future Events},
   volume = {117},
   year = {2022}
}

@article{Bagiella2001,
   author = {Emilia Bagiella and Daniel F. Heitjan},
   doi = {10.1002/sim.843},
   issn = {02776715},
   number = {14},
   journal = {Statistics in Medicine},
   month = {7},
   pages = {2055-2063},
   pmid = {11439420},
   title = {Predicting analysis times in randomized clinical trials},
   volume = {20},
   year = {2001}
}

@article{YingHeitjanChen2004NonparametricPrediction,
  author  = {Ying, Gui-Shuang and Heitjan, Daniel F. and Chen, Tai-Tsang},
  title   = {Nonparametric prediction of event times in randomized clinical trials},
  journal = {Clinical Trials},
  year    = {2004},
  volume  = {1},
  number  = {4},
  pages   = {352--361},
  doi     = {10.1191/1740774504cn030oa}
}

@article{Donovan2006,
   author = {J. Mark Donovan and Michael R. Elliott and Daniel F. Heitjan},
   doi = {10.1080/10543400600609445},
   issn = {10543406},
   number = {3},
   journal = {Journal of Biopharmaceutical Statistics},
   month = {5},
   pages = {343-356},
   pmid = {16724489},
   title = {Predicting event times in clinical trials when treatment arm is masked},
   volume = {16},
   year = {2006}
}

@article{Ying2008,
   author = {Ying, Gui-Shuang and Heitjan, Daniel F.},
   doi = {10.1002/pst.271},
   issn = {15391604},
   number = {2},
   journal = {Pharmaceutical Statistics},
   month = {4},
   pages = {107-120},
   pmid = {17377932},
   title = {Weibull prediction of event times in clinical trials},
   volume = {7},
   year = {2008}
}

@article{Ying2013,
   author = {Ying, Gui-Shuang and Heitjan, Daniel F.},
   doi = {10.1177/1740774512470314},
   issn = {17407745},
   number = {2},
   journal = {Clinical Trials},
   month = {4},
   pages = {197-206},
   pmid = {23321264},
   title = {Prediction of event times in the {REMATCH} Trial},
   volume = {10},
   year = {2013}
}

@article{Ying2017,
   author = {Ying, Gui-Shuang and Zhang, Qiang and Lan, Yu and Li, Yimei and Heitjan, Daniel F.},
   doi = {10.1016/j.cct.2017.05.012},
   issn = {15592030},
   journal = {Contemporary Clinical Trials},
   month = {8},
   pages = {30-37},
   pmid = {28545934},
   publisher = {Elsevier Inc.},
   title = {Cure modeling in real-time prediction: How much does it help?},
   volume = {59},
   year = {2017}
}

@article{Lan2018,
   author = {Yu Lan and Daniel F. Heitjan},
   doi = {10.1177/1740774517750633},
   issn = {17407753},
   number = {2},
   journal = {Clinical Trials},
   month = {4},
   pages = {159-168},
   pmid = {29376735},
   publisher = {SAGE Publications Ltd},
   title = {Adaptive parametric prediction of event times in clinical trials},
   volume = {15},
   year = {2018}
}

@article{Ou2019,
   author = {Ou, Fang-Shu and Heller, Martin and Shi, Qian},
   doi = {10.1002/pst.1934},
   issn = {15391612},
   number = {4},
   journal = {Pharmaceutical Statistics},
   month = {7},
   pages = {433-446},
   pmid = {30806485},
   publisher = {John Wiley and Sons Ltd},
   title = {Milestone prediction for time-to-event endpoint monitoring in clinical trials},
   volume = {18},
   year = {2019}
}

@article{Anisimov2011,
   author = {Anisimov, Vladimir V.},
   doi = {10.1002/pst.525},
   issn = {15391604},
   number = {6},
   journal = {Pharmaceutical Statistics},
   month = {11},
   pages = {517-522},
   pmid = {22140055},
   title = {Predictive event modelling in multicentre clinical trials with waiting time to response},
   volume = {10},
   year = {2011}
}

@article{Zhang2012,
   author = {Xiaoxi Zhang and Qi Long},
   doi = {10.1002/bimj.201100180},
   issn = {15214036},
   number = {6},
   journal = {Biometrical Journal},
   pages = {735-749},
   pmid = {22907686},
   publisher = {Wiley-VCH Verlag},
   title = {Joint monitoring and prediction of accrual and event times in clinical trials},
   volume = {54},
   year = {2012}
}

@article{Heitjan2015,
   author = {Heitjan, Daniel F. and Ge, Zhiyun and Ying, Gui-Shuang},
   doi = {10.1016/j.cct.2015.07.010},
   issn = {15592030},
   journal = {Contemporary Clinical Trials},
   number = {Pt A},
   month = {10},
   pages = {26-33},
   pmid = {26188165},
   publisher = {Elsevier Inc.},
   title = {Real-time prediction of clinical trial enrollment and event counts: A review},
   volume = {45},
   year = {2015}
}

@article{Chen2016,
   author = {Chen, Tai-Tsang},
   doi = {10.1186/s12874-016-0117-3},
   issn = {14712288},
   number = {1},
   journal = {BMC Medical Research Methodology},
   month = {2},
   pages = {12},
   pmid = {26830911},
   publisher = {BioMed Central},
   title = {Predicting analysis times in randomized clinical trials with cancer immunotherapy},
   volume = {16},
   year = {2016}
}

@article{Wang2022RealTimeTTEM,
  author  = {Wang, Li and Liu, Yang and Chen, Xiaotian and Pulkstenis, Erik},
  title   = {Real time monitoring and prediction of time to endpoint maturation in clinical trials},
  journal = {Statistics in Medicine},
  year    = {2022},
  volume  = {41},
  number  = {18},
  pages   = {3596--3611},
  doi     = {10.1002/sim.9436},
  url     = {https://doi.org/10.1002/sim.9436},
  pmid    = {35587584}
}

@article{Machida2025DynamicPredictionTiming,
  author  = {Machida, Ryunosuke and Sakamaki, Kentaro and Ohigashi, Tomohiro and Sozu, Takashi},
  title   = {Dynamic Prediction of Analysis Timing in Clinical Trials Using Joint Models of Longitudinal and Time-to-Event Data},
  journal = {Statistics in Medicine},
  year    = {2025},
  volume  = {44},
  number  = {28-30},
  pages   = {e70312},
  doi     = {10.1002/sim.70312},
  url     = {https://doi.org/10.1002/sim.70312}
}

@article{Fu2025BayesianPredictionEventTimes,
  author  = {Fu, Jingyan and Zhao, Dan and Skanji, Donia and Liu, Hua and Tang, Rui Sammi and Yuan, Ying},
  title   = {{Bayesian} Prediction of Event Times Using Mixture Model for Blinded Randomized Controlled Trials},
  journal = {Statistics in Medicine},
  year    = {2025},
  volume  = {44},
  number  = {28-30},
  pages   = {e70310},
  doi     = {10.1002/sim.70310},
  url     = {https://doi.org/10.1002/sim.70310}
}

@article{Hong2013,
   author = {Yili Hong},
   doi = {10.1016/j.csda.2012.10.006},
   issn = {01679473},
   number = {1},
   journal = {Computational Statistics \& Data Analysis},
   month = {3},
   pages = {41-51},
   title = {On computing the distribution function for the {Poisson} binomial distribution},
   volume = {59},
   year = {2013}
}

@article{Tang2022,
   author = {Wenpin Tang and Fengmin Tang},
   doi = {10.1214/22-STS852},
   number = {1},
   journal = {Statistical Science},
   pages = {108--119},
   publisher = {Institute of Mathematical Statistics},
   title = {The {Poisson Binomial Distribution}---Old \& New},
   volume = {38},
   year = {2023}
}

@article{Efron1981CensoredBootstrap,
  author  = {Efron, Bradley},
  title   = {Censored Data and the Bootstrap},
  journal = {Journal of the American Statistical Association},
  year    = {1981},
  volume  = {76},
  number  = {374},
  pages   = {312--319},
  doi     = {10.1080/01621459.1981.10477650}
}

@techreport{Hjort1985BootstrappingCox,
  author      = {Hjort, Nils Lid},
  title       = {Bootstrapping {Cox}'s Regression Model},
  institution = {Department of Statistics, Stanford University},
  type        = {Technical Report},
  number      = {LCS 21 (also NSF-241)},
  year        = {1985},
  month       = nov,
  url         = {https://purl.stanford.edu/wv616sr1865}
}

@article{Akritas1986BootstrappingKM,
  author  = {Akritas, Michael G.},
  title   = {Bootstrapping the {Kaplan--Meier} Estimator},
  journal = {Journal of the American Statistical Association},
  year    = {1986},
  volume  = {81},
  number  = {396},
  pages   = {1032--1038},
  doi     = {10.1080/01621459.1986.10478369},
  url     = {https://doi.org/10.1080/01621459.1986.10478369}
}

@article{GrossLai1996BootstrapTruncatedCensored,
  author  = {Gross, Shulamith T. and Lai, Tze Leung},
  title   = {Bootstrap Methods for Truncated and Censored Data},
  journal = {Statistica Sinica},
  year    = {1996},
  volume  = {6},
  number  = {3},
  pages   = {509--530},
  url     = {https://www.jstor.org/stable/24305605}
}

@article{Bilker1997,
   author = {Warren B. Bilker and Mei Cheng Wang},
   doi = {10.1080/03610919708813372},
   issn = {03610918},
   number = {1},
   journal = {Communications in Statistics Part B: Simulation and Computation},
   pages = {141-171},
   publisher = {Marcel Dekker Inc.},
   title = {Bootstrapping left truncated and right censored data},
   volume = {26},
   year = {1997}
}

@book{MeekerHahnEscobar2017StatisticalIntervals,
  author    = {Meeker, William Q. and Hahn, Gerald J. and Escobar, Luis A.},
  title     = {Statistical Intervals: A Guide for Practitioners and Researchers},
  year      = {2017},
  edition   = {2},
  publisher = {John Wiley \& Sons},
  series    = {Wiley Series in Probability and Statistics},
  isbn      = {9780471687177},
  doi       = {10.1002/9781118594841},
  url       = {https://onlinelibrary.wiley.com/doi/book/10.1002/9781118594841}
}

@manual{flexsurv,
  title        = {{flexsurv}: Flexible Parametric Survival and Multi-State Models},
  author       = {Jackson, Christopher H.},
  year         = {2024},
  note         = {{R} package version 2.3.2},
  url          = {https://CRAN.R-project.org/package=flexsurv},
  urldate      = {2026-01-06}
}

@manual{eventTrack,
  title        = {{eventTrack}: Event Prediction for Time-to-Event Endpoints},
  author       = {Rufibach, Kaspar},
  year         = {2025},
  note         = {{R} package version 1.0.4},
  doi          = {10.32614/CRAN.package.eventTrack},
  url          = {https://CRAN.R-project.org/package=eventTrack},
  urldate      = {2026-01-06}
}

@manual{RCoreTeam2025,
  title        = {{R}: A Language and Environment for Statistical Computing},
  author       = {{R Core Team}},
  year         = {2025},
  organization = {R Foundation for Statistical Computing},
  address      = {Vienna, Austria},
  note         = {Version 4.5.1},
  url          = {https://www.R-project.org/}
}

@manual{rsimsum,
  title  = {{rsimsum}: Analysis of Simulation Studies Including {Monte Carlo} Error},
  author = {Gasparini, Alessandro and White, Ian R.},
  year   = {2024},
  note   = {{R} package version 0.13.0},
  doi    = {10.32614/CRAN.package.rsimsum},
  url    = {https://CRAN.R-project.org/package=rsimsum}
}

@article{Brilleman2020simsurv,
  author  = {Brilleman, Samuel L. and Wolfe, Rory and Moreno-Betancur, Margarita and Crowther, Michael J.},
  title   = {Simulating Survival Data Using the {simsurv} {R} Package},
  journal = {Journal of Statistical Software},
  year    = {2020},
  volume  = {97},
  number  = {3},
  pages   = {1--27},
  doi     = {10.18637/jss.v097.i03},
  url     = {https://doi.org/10.18637/jss.v097.i03}
}

@manual{survival,
  title  = {{survival}: Survival Analysis},
  author = {Therneau, Terry M.},
  year   = {2020},
  note   = {{R} package version 3.2-7},
  url    = {https://CRAN.R-project.org/package=survival}
}

@manual{muhaz,
  title  = {{muhaz}: Hazard Function Estimation in Survival Analysis},
  author = {Hess, Kenneth and Gentleman, Robert},
  year   = {2021},
  note   = {{R} package version 1.2.6.4},
  doi    = {10.32614/CRAN.package.muhaz},
  url    = {https://CRAN.R-project.org/package=muhaz}
}

@manual{poibin,
  title  = {{poibin}: The {Poisson Binomial Distribution}},
  author = {Hong, Yili and {R Core Team}},
  year   = {2024},
  note   = {{R} package version 1.6},
  doi    = {10.32614/CRAN.package.poibin},
  url    = {https://CRAN.R-project.org/package=poibin}
}

@article{Stacy1962,
   author = {E. W. Stacy},
   doi = {10.1214/aoms/1177704481},
   issn = {0003-4851},
   number = {3},
   journal = {The Annals of Mathematical Statistics},
   month = {9},
   pages = {1187-1192},
   publisher = {Institute of Mathematical Statistics},
   title = {A Generalization of the Gamma Distribution},
   volume = {33},
   year = {1962}
}

@article{Prentice1975,
   author = {R L Prentice},
   journal = {Biometrika},
   number = {3},
   pages = {607--614},
   title = {Discrimination among some parametric models},
   volume = {62},
   doi = {10.1093/biomet/62.3.607},
   year = {1975}
}

@article{Cox2007,
   author = {Christopher Cox and Haitao Chu and Michael F. Schneider and Alvaro Muñoz},
   doi = {10.1002/sim.2836},
   issn = {02776715},
   number = {23},
   journal = {Statistics in Medicine},
   month = {10},
   pages = {4352-4374},
   pmid = {17342754},
   title = {Parametric survival analysis and taxonomy of hazard functions for the generalized gamma distribution},
   volume = {26},
   year = {2007}
}

@article{BenderAugustinBlettner2005GeneratingSurvivalTimes,
  author  = {Bender, Ralf and Augustin, Thomas and Blettner, Maria},
  title   = {Generating survival times to simulate {Cox} proportional hazards models},
  journal = {Statistics in Medicine},
  year    = {2005},
  volume  = {24},
  number  = {11},
  pages   = {1713--1723},
  doi     = {10.1002/sim.2059},
  url     = {https://doi.org/10.1002/sim.2059}
}

@article{CrowtherLambert2013SimulatingComplexSurvival,
  author  = {Crowther, Michael J. and Lambert, Paul C.},
  title   = {Simulating biologically plausible complex survival data},
  journal = {Statistics in Medicine},
  year    = {2013},
  volume  = {32},
  number  = {23},
  pages   = {4118--4134},
  doi     = {10.1002/sim.5823},
  url     = {https://doi.org/10.1002/sim.5823}
}

@article{Gray2021,
   author = {Jodi Gray and Thomas Sullivan and Nicholas R. Latimer and Amy Salter and Michael J. Sorich and Robyn L. Ward and Jonathan Karnon},
   doi = {10.1177/0272989X20978958},
   issn = {1552681X},
   number = {2},
   journal = {Medical Decision Making},
   month = {2},
   pages = {179-193},
   pmid = {33349137},
   publisher = {SAGE Publications Inc.},
   title = {Extrapolation of Survival Curves Using Standard Parametric Models and Flexible Parametric Spline Models: Comparisons in Large Registry Cohorts with Advanced Cancer},
   volume = {41},
   year = {2021}
}

@article{Cooper2022,
   author = {Miranda Cooper and Sarah Smith and Troy Williams and Raquel Aguiar-Ibáñez},
   doi = {10.1080/13696998.2022.2030599},
   issn = {1941837X},
   number = {1},
   journal = {Journal of Medical Economics},
   pages = {260-273},
   pmid = {35060433},
   publisher = {Taylor and Francis Ltd.},
   title = {How accurate are the longer-term projections of overall survival for cancer immunotherapy for standard versus more flexible parametric extrapolation methods?},
   volume = {25},
   year = {2022}
}

@article{Latimer2022,
   author = {Nicholas R Latimer and Amanda I Adler},
   doi = {10.1136/bmjmed-2021-000094},
   number = {1},
   journal = {BMJ Medicine},
   month = {3},
   pages = {e000094},
   publisher = {BMJ},
   title = {Extrapolation beyond the end of trials to estimate long term survival and cost effectiveness},
   volume = {1},
   year = {2022}
}

@article{Sweeting2023,
   author = {Michael J. Sweeting and Mark J. Rutherford and Dan Jackson and Sangyu Lee and Nicholas R. Latimer and Robert Hettle and Paul C. Lambert},
   doi = {10.1177/0272989X231184247},
   issn = {1552681X},
   number = {6},
   journal = {Medical Decision Making},
   month = {8},
   pages = {737-748},
   pmid = {37448102},
   publisher = {SAGE Publications Inc.},
   title = {Survival Extrapolation Incorporating General Population Mortality Using Excess Hazard and Cure Models: A Tutorial},
   volume = {43},
   year = {2023}
}

@article{Jackson2023survextrap,
  author  = {Jackson, Christopher H.},
  title   = {{survextrap}: a package for flexible and transparent survival extrapolation},
  journal = {BMC Medical Research Methodology},
  year    = {2023},
  volume  = {23},
  pages   = {282},
  doi     = {10.1186/s12874-023-02094-1},
  url     = {https://doi.org/10.1186/s12874-023-02094-1}
}

@article{Palmer2023GuideFlexibleSurvival,
  author  = {Palmer, Stephen and Borget, Isabelle and Friede, Tim and Husereau, Don and Karnon, Jonathan and Kearns, Ben and Medin, Emma and Peterse, Elisabeth F. P. and Klijn, Sven L. and Verburg-Baltussen, Elisabeth J. M. and Fenwick, Elisabeth and Borrill, John},
  title   = {A Guide to Selecting Flexible Survival Models to Inform Economic Evaluations of Cancer Immunotherapies},
  journal = {Value in Health},
  year    = {2023},
  month   = feb,
  volume  = {26},
  number  = {2},
  pages   = {185--192},
  doi     = {10.1016/j.jval.2022.07.009},
  url     = {https://doi.org/10.1016/j.jval.2022.07.009}
}

@article{Chen2024,
   author = {Enoch Yi Tung Chen and Yuliya Leontyeva and Chia Ni Lin and Jung Der Wang and Mark S. Clements and Paul W. Dickman},
   doi = {10.1177/0272989X241227230},
   issn = {1552681X},
   number = {3},
   journal = {Medical Decision Making},
   month = {4},
   pages = {269-282},
   pmid = {38314657},
   publisher = {SAGE Publications Inc.},
   title = {Comparing Survival Extrapolation within All-Cause and Relative Survival Frameworks by Standard Parametric Models and Flexible Parametric Spline Models Using the {Swedish} Cancer Registry},
   volume = {44},
   year = {2024}
}

@article{Whitehead2001,
   author = {John Whitehead},
   doi = {10.1177/009286150103500435},
   number = {4},
   journal = {Drug Information Journal},
   pages = {1387-1400},
   title = {Predicting the duration of sequential survival studies},
   volume = {35},
   year = {2001}
}

@article{Hade2010,
   author = {Erinn M. Hade and David Jarjoura and Lai Wei},
   doi = {10.1177/1740774510367525},
   issn = {17407745},
   number = {3},
   journal = {Clinical Trials},
   month = {6},
   pages = {219-226},
   pmid = {20392786},
   title = {Sample size re-estimation in a breast cancer trial},
   volume = {7},
   year = {2010}
}

@article{Todd2012,
   author = {Susan Todd and Elsa Valdés-Márquez and Jodie West},
   doi = {10.1002/pst.516},
   issn = {15391604},
   number = {2},
   journal = {Pharmaceutical Statistics},
   month = {3},
   pages = {141-148},
   pmid = {22337635},
   title = {A practical comparison of blinded methods for sample size reviews in survival data clinical trials},
   volume = {11},
   year = {2012}
}

@article{Friede2019,
   author = {Tim Friede and Harald Pohlmann and Heinz Schmidli},
   doi = {10.1002/pst.1927},
   issn = {15391612},
   number = {3},
   journal = {Pharmaceutical Statistics},
   month = {5},
   pages = {351-365},
   pmid = {30652403},
   publisher = {John Wiley and Sons Ltd},
   title = {Blinded sample size reestimation in event-driven clinical trials: Methods and an application in multiple sclerosis},
   volume = {18},
   year = {2019}
}

@article{Mori2024,
   author = {Tim Mori and Sho Komukai and Satoshi Hattori and Tim Friede},
   doi = {10.1002/pst.2459},
   issn = {15391612},
   number = {2},
   journal = {Pharmaceutical Statistics},
   month = {3},
   pages = {e2459},
   publisher = {John Wiley and Sons Ltd},
   title = {Flexible Spline Models for Blinded Sample Size Reestimation in Event-Driven Clinical Trials},
   volume = {24},
   year = {2025}
}

@incollection{Kim2014MaximumDurationInformation,
  author    = {Kim, KyungMann},
  title     = {Maximum Duration and Information Trials},
  booktitle = {Methods and Applications of Statistics in Clinical Trials, Volume 1: Concepts, Principles, Trials, and Designs},
  editor    = {Balakrishnan, N.},
  publisher = {John Wiley \& Sons, Inc.},
  address   = {Hoboken, NJ},
  year      = {2014},
  pages     = {515--521},
  doi       = {10.1002/9781118596005.ch41}
}

@book{Wu,
  author    = {Wu, Jianrong},
  title     = {Statistical Methods for Survival Trial Design: With Applications to Cancer Clinical Trials Using {R}},
  year      = {2018},
  publisher = {CRC Press},
  series    = {Chapman \& Hall/CRC Biostatistics Series},
  doi       = {10.1201/9780429470172},
  isbn      = {9781138033221},
  url       = {https://www.crcpress.com/go/biostats}
}

@book{WassmerBrannath2025GroupSequential,
  author    = {Wassmer, Gernot and Brannath, Werner},
  title     = {Group Sequential and Confirmatory Adaptive Designs in Clinical Trials},
  year      = {2025},
  edition   = {2},
  publisher = {Springer},
  address   = {Cham},
  series    = {Springer Series in Pharmaceutical Statistics},
  doi       = {10.1007/978-3-031-89669-9},
  isbn      = {978-3-031-89668-2}
}

@book{JennisonTurnbull2025GroupSequentialAdaptive,
  author    = {Jennison, Christopher and Turnbull, Bruce W.},
  title     = {Group Sequential and Adaptive Methods for Clinical Trials},
  year      = {2026},
  edition   = {2},
  publisher = {Chapman \& Hall/CRC},
  address   = {New York},
  doi       = {10.1201/9781584888482},
  isbn      = {9781584888475},
  url       = {https://www.taylorfrancis.com/books/mono/10.1201/9781584888482/}
}

@article{Proschan1992,
   author = {Michael A Proschan and Dean A Follmann and Myron A Waclawiw},
   number = {4},
   journal = {Biometrics},
   pages = {1131-1143},
   title = {Effects of Assumption Violations on {Type I} Error Rate in Group Sequential Monitoring},
   volume = {48},
   doi = {10.2307/2532704},
   url = {https://www.jstor.org/stable/2532704},
   year = {1992}
}

@article{Lan1989,
   author = {Lan, K. K. Gordon and DeMets, David L.},
   number = {3},
   journal = {Biometrics},
   pages = {1017-1020},
   title = {Changing Frequency of Interim Analysis in Sequential Monitoring},
   volume = {45},
   doi = {10.2307/2531701},
   url = {https://www.jstor.org/stable/2531701},
   year = {1989}
}

@article{Whitehead1992Overrunning,
  author  = {Whitehead, John},
  title   = {Overrunning and underrunning in sequential clinical trials},
  journal = {Controlled Clinical Trials},
  year    = {1992},
  volume  = {13},
  number  = {2},
  pages   = {106--121},
  doi     = {10.1016/0197-2456(92)90017-T}
}

@article{Fang2011,
   author = {Liang Fang and Zheng Su},
   doi = {10.1016/j.cct.2011.05.013},
   issn = {15517144},
   number = {5},
   journal = {Contemporary Clinical Trials},
   month = {9},
   pages = {755-759},
   pmid = {21645644},
   title = {A hybrid approach to predicting events in clinical trials with time-to-event outcomes},
   volume = {32},
   year = {2011}
}

@misc{Anisimov2021AdvancedModelsArxiv,
  author        = {Anisimov, Vladimir V. and
                   Gormley, Stephen and
                   Baverstock, Rosalind and
                   Kineza, Cynthia},
  title         = {Advanced models for predicting event
                   occurrence in event-driven clinical trials
                   accounting for patient dropout, cure and
                   ongoing recruitment},
  year          = {2021},
  month         = aug,
  howpublished  = {arXiv preprint},
  eprint        = {2108.09196},
  archivePrefix = {arXiv},
  primaryClass  = {stat.ME},
  doi           = {10.48550/arXiv.2108.09196},
  url           = {https://arxiv.org/abs/2108.09196}
}

@article{Goodman2011,
   author = {Melody S. Goodman and Yi Li and Ram C. Tiwari},
   doi = {10.1080/02664763.2011.559209},
   issn = {02664763},
   number = {11},
   journal = {Journal of Applied Statistics},
   month = {11},
   pages = {2523-2532},
   title = {Detecting multiple change points in piecewise constant hazard functions},
   volume = {38},
   year = {2011}
}

@article{Beran1990,
   author = {Rudolf Beran},
   number = {411},
   journal = {Journal of the American Statistical Association},
   pages = {715--723},
   title = {Calibrating Prediction Regions},
   volume = {85},
   doi = {10.1080/01621459.1990.10474932},
   year = {1990}
}

@article{Harris1989,
   author = {Ian R Harris},
   number = {4},
   journal = {Biometrika},
   pages = {675-684},
   title = {Predictive Fit for Natural Exponential Families},
   volume = {76},
   doi = {10.1093/biomet/76.4.675},
   year = {1989}
}

@article{Shen2018,
   author = {Shen, Jieli and Liu, Regina Y. and Xie, Min-ge},
   doi = {10.1016/j.jspi.2017.09.012},
   issn = {03783758},
   journal = {Journal of Statistical Planning and Inference},
   month = {5},
   pages = {126-140},
   publisher = {Elsevier B.V.},
   title = {Prediction with confidence—A general framework for predictive inference},
   volume = {195},
   year = {2018}
}

@article{Hannig2006,
   author = {Jan Hannig and Hari Iyer and Paul Patterson},
   doi = {10.1198/016214505000000736},
   issn = {01621459},
   number = {473},
   journal = {Journal of the American Statistical Association},
   month = {3},
   pages = {254-269},
   title = {Fiducial generalized confidence intervals},
   volume = {101},
   year = {2006}
}

@article{Wang2012,
   author = {Wang, C.-M. and Hannig, Jan and Iyer, Hari K.},
   doi = {10.1016/j.jspi.2012.02.021},
   issn = {03783758},
   number = {7},
   journal = {Journal of Statistical Planning and Inference},
   month = {7},
   pages = {1980-1990},
   title = {Fiducial prediction intervals},
   volume = {142},
   year = {2012}
}

@techreport{Cario1997,
   author = {Cario, Marne C. and Nelson, Barry L.},
   institution = {Department of Industrial Engineering and Management Sciences, Northwestern University},
   address = {Evanston, IL},
   title = {Modeling and Generating Random Vectors with Arbitrary Marginal Distributions and Correlation Matrix},
   pages = {1--19},
   note = {pp. 1--19},
   year = {1997}
}

@article{Royston2002,
   author = {Royston, Patrick and Parmar, Mahesh K. B.},
   doi = {10.1002/sim.1203},
   issn = {02776715},
   number = {15},
   journal = {Statistics in Medicine},
   month = {8},
   pages = {2175-2197},
   pmid = {12210632},
   title = {Flexible parametric proportional-hazards and proportional-odds models for censored survival data, with application to prognostic modelling and estimation of treatment effects},
   volume = {21},
   year = {2002}
}

@book{KleinMoeschberger2003SurvivalAnalysis,
  author    = {Klein, John P. and Moeschberger, Melvin L.},
  title     = {Survival Analysis: Techniques for Censored and Truncated Data},
  year      = {2003},
  edition   = {2},
  publisher = {Springer},
  address   = {New York},
  series    = {Statistics for Biology and Health},
  isbn      = {9780387953991},
  doi       = {10.1007/b97377},
  url       = {https://link.springer.com/book/10.1007/b97377}
}

@article{Jiang2021,
   author = {Shu Jiang and David Swanson and Rebecca A. Betensky},
   doi = {10.1016/j.conctc.2021.100842},
   issn = {24518654},
   journal = {Contemporary Clinical Trials Communications},
   month = {9},
   pages = {100842},
   publisher = {Elsevier Inc.},
   title = {Estimation of the censoring distribution in clinical trials},
   volume = {23},
   year = {2021}
}

@article{Conter2024PEGAsparaginaseHRALL,
  author  = {Conter, Valentino and Valsecchi, Maria Grazia and Cario, Gunnar and Zimmermann, Martin and Attarbaschi, Andishe and Stary, Jan and Niggli, Felix and Dalla Pozza, Luciano and Elitzur, Sarah and Silvestri, Daniela and Locatelli, Franco and M{\"o}ricke, Anja and Engstler, Gernot and Smisek, Petr and Bodmer, Nicole and Barbaric, Draga and Izraeli, Shai and Rizzari, Carmelo and Boos, Joachim and Buldini, Barbara and Zucchetti, Massimo and von Stackelberg, Arend and Matteo, Cristina and Lehrnbecher, Thomas and Lanvers-Kaminsky, Claudia and Cazzaniga, Giovanni and Gruhn, Bernd and Biondi, Andrea and Schrappe, Martin},
  title   = {Four Additional Doses of {PEG-L-Asparaginase} During the Consolidation Phase in the {AIEOP-BFM ALL} 2009 Protocol Do Not Improve Outcome and Increase Toxicity in High-Risk {ALL}: Results of a Randomized Study},
  journal = {Journal of Clinical Oncology},
  year    = {2024},
  month   = mar,
  volume  = {42},
  number  = {8},
  pages   = {915--926},
  doi     = {10.1200/JCO.23.01388},
  pmid    = {38096462},
  url     = {https://doi.org/10.1200/JCO.23.01388}
}

@article{AubelAntignyFougerayDuboisSaintHilary2021,
  author  = {Aubel, P. and Antigny, M. and Fougeray, R. and Dubois, F. and Saint-Hilary, G.},
  title   = {A {Bayesian} approach for event predictions in clinical trials with time-to-event outcomes},
  journal = {Statistics in Medicine},
  year    = {2021},
  volume  = {40},
  number  = {28},
  pages   = {6344--6359},
  doi     = {10.1002/sim.9186}
}

@article{WangKeJiangZhangSnapinn2012,
  author  = {Wang, Jianming and Ke, Chunlei and Jiang, Qi and Zhang, Charlie and Snapinn, Steven},
  title   = {Predicting analysis time in event-driven clinical trials with event-reporting lag},
  journal = {Statistics in Medicine},
  year    = {2012},
  volume  = {31},
  number  = {9},
  pages   = {801--811},
  doi     = {10.1002/sim.4506}
}

@article{TsiatisDavidian2022,
  author  = {Tsiatis, Anastasios A. and Davidian, Marie},
  title   = {Group sequential methods for interim monitoring of randomized clinical trials with time-lagged outcome},
  journal = {Statistics in Medicine},
  year    = {2022},
  volume  = {41},
  number  = {28},
  pages   = {5517--5536},
  doi     = {10.1002/sim.9580}
}

@article{CrowtherLambert2014,
  author  = {Crowther, Michael J. and Lambert, Paul C.},
  title   = {A general framework for parametric survival analysis},
  journal = {Statistics in Medicine},
  year    = {2014},
  volume  = {33},
  number  = {30},
  pages   = {5280--5297},
  doi     = {10.1002/sim.6300}
}

@article{LiuPawitanClements2018,
  author  = {Liu, X.-R. and Pawitan, Y. and Clements, M.},
  title   = {Parametric and penalized generalized survival models},
  journal = {Statistical Methods in Medical Research},
  year    = {2018},
  volume  = {27},
  number  = {5},
  pages   = {1531--1546},
  doi     = {10.1177/0962280216664760}
}

@article{ZhangPuDengRoychoudhuryChuRobinson2025StudyDuration,
  author  = {Zhang, Hong and Pu, Jie and Deng, Shibing and Roychoudhury, Satrajit and Chu, Haitao and Robinson, Douglas},
  title   = {Study duration prediction for clinical trials with time-to-event endpoints accounting for heterogeneous population},
  journal = {Journal of Biopharmaceutical Statistics},
  year    = {2025},
  volume  = {35},
  number  = {6},
  pages   = {1255--1270},
  month   = oct,
  doi     = {10.1080/10543406.2025.2489294},
  url     = {https://doi.org/10.1080/10543406.2025.2489294}
}

@incollection{RenfroAlonzo,
  author    = {Renfro, Lindsay and Alonzo, Todd},
  title     = {Considerations for Pediatric Oncology Trials},
  booktitle = {Handbook of Statistics in Clinical Oncology},
  editor    = {Hoering, Antje and Othus, Megan and Crowley, John},
  publisher = {Chapman \& Hall/CRC},
  year      = {2026},
  edition   = {4},
  pages     = {247--259},
  chapter   = {18},
  doi       = {10.1201/9781003456247-21}
}

@incollection{Valsecchi,
  author    = {Valsecchi, Maria Grazia and De Lorenzo, Paola },
  title     = {Strategies for Trial Design and Analyses},
  booktitle = {New Agents for the Treatment of Acute Lymphoblastic Leukemia},
  editor    = {Saha, Vaskar and Kearns, Pamela},
  publisher = {Springer},
  year      = {2011},
  pages     = {83-104},
  chapter   = {5},
  doi       = {10.1007/978-1-4419-8459-3_5}
}

@article{Friede2009,
  author  = {Friede, Tim and Henderson, Robin},
  title   = {Exploring changes in treatment effects across design stages in adaptive trials},
  journal = {Pharmaceutical Statistics},
  year    = {2009},
  volume  = {8},
  number  = {1},
  pages   = {62--72},
  doi     = {10.1002/pst.332}
}

@article{SureshTaylorTsodikov2021,
  author  = {Suresh, Krithika and Taylor, Jeremy M. G. and Tsodikov, Alexander},
  title   = {A {Gaussian} copula approach for dynamic prediction of survival with a longitudinal biomarker},
  journal = {Biostatistics},
  year    = {2021},
  volume  = {22},
  number  = {3},
  pages   = {504--521},
  doi     = {10.1093/biostatistics/kxz049}
}

@article{ZhangCharalambousFoster2023,
  author  = {Zhang, Zili and Charalambous, Christiana and Foster, Peter},
  title   = {Joint modelling of longitudinal measurements and survival times via a multivariate copula approach},
  journal = {Journal of Applied Statistics},
  year    = {2023},
  volume  = {50},
  number  = {13},
  pages   = {2739--2759},
  doi     = {10.1080/02664763.2022.2081965}
}

@article{PanJoe2022,
  author  = {Pan, Shenyi and Joe, Harry},
  title   = {Predicting times to event based on vine copula models},
  journal = {Computational Statistics \& Data Analysis},
  year    = {2022},
  volume  = {175},
  pages   = {107546},
  doi     = {10.1016/j.csda.2022.107546}
}

\newpage
\section*{Supplementary materials}
\setcounter{figure}{0}
\appendix

\section{The Poisson--binomial distribution}\label{App:poibin}

The Poisson--binomial distribution is the distribution of a sum of independent Bernoulli random variables with possibly unequal success probabilities. The ordinary binomial distribution is recovered as the special case in which all success probabilities are equal.

Let $\zeta_i\sim\mathrm{Bernoulli}(\pi_i)$, $i=1,\ldots,r$, be independent indicators, and let
\[
Y_\Delta=\sum_{i=1}^{r}\zeta_i.
\]
Then $Y_\Delta$ takes values in $\{0,1,\ldots,r\}$ and follows a Poisson--binomial distribution with probability vector $\bm\pi=(\pi_1,\ldots,\pi_r)$. Its probability mass function and cumulative distribution function are
\[
f_{Y_\Delta}(y)
=
\sum_{A\in\Omega_y}
\prod_{i\in A}\pi_i
\prod_{i'\in A^c}(1-\pi_{i'}),
\qquad
F_{Y_\Delta}(y)
=
\sum_{\ell=0}^{y} f_{Y_\Delta}(\ell),
\]
where $\Omega_y$ is the collection of all subsets of $\{1,\ldots,r\}$ with cardinality $y$, and $A^c$ denotes the complement of $A$.

In the interim-monitoring setting considered in this paper, the Bernoulli indicators correspond to the patients in the interim risk set, and the success probabilities are the patient-specific conditional event probabilities $\pi_{i,\Delta}(\bm\eta)$ defined in Eq. 2, Section 2 of the manuscript. Direct evaluation of the probability mass function by enumeration is infeasible for moderate or large $r$, and several approximations and exact computational algorithms have therefore been proposed, including Poisson and normal approximations and recursive or Fourier-based methods \parencite{Hong2013}. See \textcite{Tang2022} for a detailed review.
\section{Derivation of $\pi_{i,\Delta}$ and special cases}
\label{App:different-pi}

For patient $i\in\mathcal R_c$, conditional on the realised follow-up time $\tau_i$ and covariates $\bm z_i$, the probability of contributing an event during the future window $(t_c,t_c+\Delta]$ is
\begin{align*}
\pi_{i,\Delta}(\bm\eta)
&=
\Pr\!\left\{
T_i\leq \min(\tau_i+\Delta,C_i)
\,\middle|\,
T_i>\tau_i,\ C_i>\tau_i,\ a_i \ \bm z_i
\right\} \\[2pt]
&=
\frac{
\Pr\!\left(
\tau_i<T_i\leq \tau_i+\Delta,\ T_i\leq C_i
\,\middle|\,
a_i \ \bm z_i
\right)
}{
\Pr\!\left(
T_i>\tau_i,\ C_i>\tau_i
\,\middle|\, a_i \
\bm z_i
\right)
}.
\end{align*}
Under conditional independence of event time and loss to follow-up,
$T_i\perp\!\!\!\perp C_i\mid\ (A_i ,\bm z_i)$, the numerator is
\[
\Pr\!\left(
\tau_i<T_i\leq \tau_i+\Delta,\ T_i\leq C_i
\,\middle|\, a_i \
\bm z_i
\right)
=
\int_{\tau_i}^{\tau_i+\Delta}
f_{\bm\theta}(u\mid\bm z_i)\,
\Pr(C_i\geq u)\,
\mathrm du
=
\int_{\tau_i}^{\tau_i+\Delta}
f_{\bm\theta}(u\mid\bm z_i)\,
G_{\bm\psi}(u)\,
\mathrm du,
\]
and the denominator is
\[
\Pr\!\left(
T_i>\tau_i,\ C_i>\tau_i
\,\middle|\, a_i \
\bm z_i
\right)
=
S_{\bm\theta}(\tau_i\mid\bm z_i)\,
G_{\bm\psi}(\tau_i).
\]
Combining the two terms gives
\begin{equation}\label{pi}
\pi_{i,\Delta}(\bm\eta)
=
\frac{
\displaystyle
\int_{\tau_i}^{\tau_i+\Delta}
f_{\bm\theta}(u\mid\bm z_i)\,
G_{\bm\psi}(u)\,
\mathrm du
}{
S_{\bm\theta}(\tau_i\mid\bm z_i)\,
G_{\bm\psi}(\tau_i)
},
\end{equation}

which is Eq. 2, Section 2 of the manuscript.

\subsection{Special cases and closed forms}
\label{pi_closed}

\paragraph{Administrative censoring only.}
In the absence of loss to follow-up throughout
the study, $G_\psi\equiv1$, and hence
\[
\pi_{i,\Delta}(\bm\theta)
=
1-
\frac{
S_{\bm\theta}(\tau_i+\Delta\mid\bm z_i)
}{
S_{\bm\theta}(\tau_i\mid\bm z_i)
}
=
1-\exp\!\left\{
H_{\bm\theta}(\tau_i\mid\bm z_i)
-
H_{\bm\theta}(\tau_i+\Delta\mid\bm z_i)
\right\},
\]
where $H_{\bm\theta}$ denotes the cumulative hazard. This is the form used in the case study.
\paragraph{Exponential event time and no loss to follow-up.}
If $T_i\mid\bm z_i\sim\mathrm{Exp}(\lambda_{\bm z_i})$ and
$G_{\bm\psi}\equiv 1$, then
\[
\pi_{i,\Delta}
=
1-\exp(-\lambda_{\bm z_i}\Delta).
\]
\section{Likelihood factorisation}
\label{App:Lik}

Throughout this Supplementary Materials, likelihoods are understood conditionally on the realised follow-up windows and covariates. With the notation used in the main text, $G_{\bm\psi}$ denotes the conditional loss-to-follow-up survival distribution entering this conditional likelihood.

Recall that
\[
\delta_i=\mathrm E \Longleftrightarrow T_i\leq \min(C_i,\tau_i),\qquad
\delta_i=\mathrm L \Longleftrightarrow C_i<\min(T_i,\tau_i),
\]
and
\[
\delta_i=\mathrm A \Longleftrightarrow \tau_i< \min(T_i,C_i).
\]
Under conditional independence of event time and loss to follow-up given covariates, the contribution of patient $i$ to the conditional likelihood is
\[
\begin{aligned}
\ell_i(\bm\theta,\bm\psi)
&=
\left[
f_{\bm\theta}(X_i\mid\bm z_i)\,
G_{\bm\psi}(X_i ;\widehat{\bm \xi})
\right]^{\mathbb I\{\delta_i=\mathrm E\}}  \\
&\quad\times
\left[
S_{\bm\theta}(X_i\mid\bm z_i)\,
g_{\bm\psi}(X_i ;\widehat{\bm \xi})
\right]^{\mathbb I\{\delta_i=\mathrm L\}}  \\
&\quad\times
\left[
S_{\bm\theta}(\tau_i\mid\bm z_i)\,
G_{\bm\psi }(\tau_i ;\widehat{\bm \xi})
\right]^{\mathbb I\{\delta_i=\mathrm A\}} .
\end{aligned}
\]
For administratively censored observations, $X_i=\tau_i$. Assuming $\Theta\times\Psi\times\Xi$,  this can be written compactly as the product of two right-censored likelihood contributions,
\[
\mathcal L(\bm\theta,\bm\psi)
\propto
\mathcal L_T(\bm\theta)\,
\mathcal L_C(\bm\psi),
\]
where
\begin{equation}\label{eq:lik-app}
\begin{aligned}
&\mathcal L_T(\bm\theta)
\propto
\prod_{i=1}^{n}
f_{\bm\theta}(X_i\mid\bm z_i)^{\mathbb I\{\delta_i=\mathrm E\}}
S_{\bm\theta}(X_i\mid\bm z_i)^{1-\mathbb I\{\delta_i=\mathrm E\}},
\\
&\mathcal L_C(\bm\psi)
\propto
\prod_{i=1}^{n}
g_{\bm\psi}(X_i ; {\widehat{\bm \xi}})^{\mathbb I\{\delta_i=\mathrm L\}}
 G_{\bm\psi}(X_i; {\widehat{\bm \xi}})^{1-\mathbb I\{\delta_i=\mathrm L\}}   
\end{aligned}
\end{equation}
This is the factorisation used in Eq.~4, Section 2 of the manuscript. Maximising the two factors for a fixed $\widehat{\bm\xi}$
gives the estimators $\widehat{\bm\theta}$ and $\widehat{\bm\psi}$.

Since prediction is conditional
on the realized interim follow-up window, conditioning on the realized
windows $\tau_i$ at a fixed interim calendar time $t_c$ is equivalent to conditioning on the observed entry
times $a_i$. The accrual CDF $Q_{\bm \xi}(a)$, which is defined on a known, fixed support \((0,a_{\max})\), enters in \eqref{eq:lik-app}
through $\Phi^{-1}\!\bigl\{Q_{\bm \xi}(a)\bigr\}$, which is is evaluated at the individual entry times observed. Thus, $\Phi^{-1}\!\bigl\{Q_{\bm \xi}(a)\bigr\}$ is treated as a conditioning constant and we estimate $\bm{\nu}$ and $\rho$ jointly by maximising
the loss-to-follow-up likelihood in \eqref{eq:lik-app} at the observed entry times as shown below. Nevertheless, when its parameter \(\bm \xi\) is unknown, it is first estimated by maximising the marginal accrual likelihood
\begin{equation}\label{app:likQ}
L_A(\bm{\xi})=\prod_{i=1}^n q_{\bm{\xi}}(a_i).
\end{equation}
The conditional loss-to-follow-up model is then fitted using the observed entry times, with \(\bm \xi\) fixed at its estimate in the latter likelihood. Furthermore, although prediction is conditional on the realised follow-up windows $\tau_i$, uncertainty in \(\widehat{\bm \xi}\) is propagated through the bootstrap procedure described in Supplementary materials \ref{App:Algo}. Note that, when \(Q_{\bm \xi}(a)\) is fully specified as in the case of the uniform distribution on the known support, no accrual parameter needs to be estimated.

\subsection{Gaussian copula model for loss to follow-up}
\label{app:gaussian-dropout}
Copula-based models have been used to derive survival
distributions conditional to a continuous covariate and predict time-to-event outcomes
\parencite{SureshTaylorTsodikov2021,ZhangCharalambousFoster2023,PanJoe2022}. In our setting, we use a Gaussian copula to model the dependence
between entry time and time to loss to follow-up.
Let $Q_{\bm \xi}(a)$ be the continuous marginal CDF for the entry time indexes by the parameter $\bm\xi$.
Let $F^{\mathrm m}_{\bm{\nu}}$ be the continuous marginal CDF for the
loss-to-follow-up time, with survival function
$G^{\mathrm m}_{\bm{\nu}}=1-F^{\mathrm m}_{\bm{\nu}}$ and density
$f^{\mathrm m}_{\bm{\nu}}$. Consider the Gaussian construction
\[
\begin{pmatrix}W_1\\W_2\end{pmatrix}
\sim \!\mathcal N\!\left[
\mathbf 0,
\begin{pmatrix}
1 & \rho\\
\rho & 1
\end{pmatrix}
\right],
\qquad |\rho|<1,
\]
\[
A=Q_{\bm \xi}^{-1}\!\bigl\{\Phi(W_1)\bigr\},
\qquad
C=(F^{\mathrm m}_{\bm{\nu}})^{-1}\!
   \bigl\{1-\Phi(W_2)\bigr\}.
\]
Here $\rho$ is the correlation on the latent Gaussian scale and
$\bm{\psi}=(\bm{\nu},\rho)$. For an observed entry time $A=a$, write
\[
h_{\bm \xi}(a)=\Phi^{-1}\!\bigl\{Q_{\bm \xi}(a)\bigr\},
\qquad
h_{\bm{\nu}}(t)=\Phi^{-1}\!\bigl\{G^{\mathrm m}_{\bm{\nu}}(t)\bigr\}.
\]
Under the Gaussian copula construction, the observed entry time
$A=a$ determines the first latent variable through
$W_1=\Phi^{-1}\{Q_{\bm\xi}(a)\}=h_{\bm\xi}(a)$.
Since $(W_1,W_2)$ is bivariate standard normal with correlation $\rho$,
the conditional mean of $W_2$ given $W_1=w_1$ is $\rho \, w_1$, and its
conditional variance is $1-\rho^2$. Substituting the latent value
associated with entry time $a$ therefore gives
\[
W_2\mid A=a
\sim \mathcal N\!\left(\rho\,h_{\bm\xi}(a),\,1-\rho^2\right).
\]

The loss-to-follow-up time is generated by applying the marginal
quantile function to $1-\Phi(W_2)$. Then we have, up to probability-zero
boundaries,
\[
C>t
\quad\Longleftrightarrow\quad
1-\Phi(W_2)>F^{\mathrm m}_{\bm\nu}(t)
\quad\Longleftrightarrow\quad
W_2<h_{\bm\nu}(t).
\]

Standardising $W_2$ under its conditional normal distribution gives
\begin{equation}
\begin{aligned}
G_{\bm{\psi}}(t)
&=\Phi\!\left(\frac{h_{\bm{\nu}}(t)-\rho \, h_{\bm \xi}(a)}{\sqrt{1- \rho ^2}}\right)\\[0.2cm]
& = \Phi\!\left(\frac{\Phi^{-1}\!\bigl\{G^{\mathrm m}_{\bm{\nu}}(t)\bigr\}-\rho \, \Phi^{-1}\!\bigl\{Q_{\bm \xi}(a)\bigr\}}{\sqrt{1-\rho^2}}\right)
\end{aligned}    
\end{equation}

Differentiating
$\Phi(h_{\bm{\nu}}(t))=G^{\mathrm m}_{\bm{\nu}}(t)$
gives
\[
h'_{\bm{\nu}}(t)
=-\frac{f^{\mathrm m}_{\bm{\nu}}(t)}
        {\phi(h_{\bm{\nu}}(t))},
\]
where $\phi$ is the standard normal density. The conditional density
on the interior of the support is therefore

\begin{equation}\label{eq:AppCondLossDist}
\begin{aligned}
g_\psi(t)
&= -\frac{\partial}{\partial t}G_\psi(t)\\
&= \frac{f_\nu^m(t)}{\sqrt{1-\rho^2}}
\frac{
\phi\left(
\frac{\Phi^{-1}\{G_\nu^m(t)\}
-\rho\Phi^{-1}\{Q_\xi(a)\}}
{\sqrt{1-\rho^2}}
\right)
}{
\phi\left[\Phi^{-1}\{G_\nu^m(t)\}\right]
}.
\end{aligned}
\end{equation}

\section{The CPB algorithm}
\label{App:Algo}

\begin{algorithm}[H]
\caption{Conditional parametric bootstrap (CPB) for estimating the conditional distribution of $Y_\Delta$.}
\label{Alg1}
\KwIn{Interim data $\mathcal D_c=\{(X_i,\delta_i,\tau_i,\bm z_i)\}_{i=1}^{n}$; prediction horizon $\Delta$; number of bootstrap replicates $B$.}
\KwOut{Bootstrap estimator $\widehat F^*_\Delta(\cdot\mid\mathcal D_c)$ of the conditional CDF of $Y_\Delta$.}
\BlankLine

\tcc{Step 1: fit the models on the observed interim data}
Compute $\widehat{\bm\xi}$ by maximizing the likelihood in \eqref{app:likQ}\;
Compute $\widehat{\bm\theta}$ by maximising the likelihood in \eqref{eq:lik-app}\; 
Compute $\widehat{\bm\psi}$ by maximising the likelihood in \eqref{eq:lik-app} conditioning on $\Phi^{-1}\!\bigl\{Q_{\widehat{\bm \xi}}(a_i)\bigr\}$\;
Set $\widehat{\bm\eta}=(\widehat{\bm\theta},\widehat{\bm\psi}, \widehat{\bm\xi})$\;
Set $\mathcal R_c=\{i:\delta_i=\mathrm A\}$ from the original interim data\;

\BlankLine
\tcc{Step 2: generate bootstrap datasets conditional on the observed follow-up windows}
Keep $(\tau_i,\bm z_i)$ fixed at their original values for all patients\;
\For{$b \leftarrow 1$ \KwTo $B$}{
  \For{$i \leftarrow 1$ \KwTo $n$}{
    Sample $T_i^{*(b)}$ from $f_{\widehat{\bm{\theta}}}
(\cdot\mid\bm{z}_i)$\;
    Sample $C_i^{*(b)}$ from from the loss-to-follow-up density
$g_{\widehat{\bm{\psi}}}(\cdot ;\widehat{\bm \xi})$ defined in  \eqref{eq:AppCondLossDist} conditioning on $\Phi^{-1}\!\bigl\{Q_{\widehat{\bm \xi}}(a_i)\bigr\}$ \;
Set $X_i^{*(b)}=\min\{T_i^{*(b)},C_i^{*(b)},\tau_i\}$\;
Set
$\displaystyle
\delta_i^{*(b)}=
\begin{cases}
\mathrm{E}, & T_i^{*(b)}\leq\min\{C_i^{*(b)},\tau_i\},\\
\mathrm{L}, & C_i^{*(b)}<\min\{T_i^{*(b)},\tau_i\},\\
\mathrm{A}, & \text{otherwise}.
\end{cases}
$\;
  }

  Construct $\mathcal D_c^{*(b)}=\{(X_i^{*(b)},\delta_i^{*(b)},\tau_i,\bm z_i)\}_{i=1}^{n}$\;

  \BlankLine
  \tcc{Step 3: refit and evaluate patient-specific event probabilities}
  Sample $n$ auxiliary entry times $A_i^{*(b)}$ from  $Q_{\widehat{\bm \xi}}(\cdot)$\;
Estimate $\widehat{\bm \xi}^{*(b)}$ from these auxiliary times by maximising the likelihood in \eqref{app:likQ}\;
  Compute $\widehat{\bm\theta}^{*(b)}$ by maximising the likelihood in \eqref{eq:lik-app} on $\mathcal D_c^{*(b)}$\;
  Compute $\widehat{\bm\psi}^{*(b)}$ by maximising the likelihood in \eqref{eq:lik-app} on $\mathcal D_c^{*(b)}$ conditioning on $\Phi^{-1}\!\bigl\{Q_{\widehat{\bm \xi}^{*(b)}}(a_i)\bigr\}$\;
  Set $\widehat{\bm\eta}^{*(b)}=(\widehat{\bm\theta}^{*(b)},\widehat{\bm\psi}^{*(b)}, \widehat{\bm \xi}^{*(b)})$\;
  Evaluate $\pi_{i,\Delta}(\widehat{\bm\eta}^{*(b)})$ for all $i\in\mathcal R_c$\;
}

\BlankLine
\tcc{Step 4: average the Poisson--binomial CDFs}
\For{$y\in\{0,1,\ldots,|\mathcal R_c|\}$}{
  Compute
  \[
  \widehat F^*_\Delta(y\mid\mathcal D_c)
  =
  \frac{1}{B}
  \sum_{b=1}^{B}
  F_{\mathrm{PB}}\{y;\bm\pi_\Delta(\widehat{\bm\eta}^{*(b)})\}.
  \]
}
\Return $\widehat F^*_\Delta(\cdot\mid\mathcal D_c)$\;
\end{algorithm}

\subsection{Sampling times in Step 2}
\label{app:bootstrap-sampling}

The event time can be generated by inverse transform sampling as
\[
T_i^{*(b)}
=F_{\widehat{\bm{\theta}}}^{-1}
  (U_i^{*(b)}\mid\bm{z}_i),
\qquad U_i^{*(b)}\sim\mathrm{Unif}(0,1),
\]
where $F_{\bm{\theta}}=1-S_{\bm{\theta}}$. For the loss-to-follow-up times, retain the observed $h_{\bm \xi}(a_i)$ and draw $V_{i}^{*(b)}\sim\mathcal N(0,1)$ independently
of the event-time draw and across patients. Set
\[
W_{2i}^{*(b)}
=\widehat \rho\,h_{\widehat{\bm \xi}}(a_i)+\sqrt{1-\widehat \rho^{\,2}}\,V_{i}^{*(b)},
\qquad
C_i^{*(b)}
=(F^{\mathrm m}_{\widehat{\bm{\nu}}})^{-1}
  \!\bigl\{1-\Phi(W_{2i}^{*(b)})\bigr\}.
\]
This draws $C_i^{*(b)}$ from the fitted conditional distribution
given $A_i$. Both $\bm{\nu}$ and $\rho$ are re-estimated in Step~3. The original entry times are retained throughout the bootstrap samples. When the accrual distribution is estimated, an independent auxiliary sample is generated from \(Q_{\widehat{\bm\xi}}(a)\) to obtain \(\widehat{\bm\xi}^{*(b)}\). The fitted \(\widehat{\bm\xi}\) is used to generate bootstrap loss-to-follow-up times, whereas \(\widehat{\bm\xi}^{*(b)}\) is used when refitting the conditional loss-to-follow-up model in each bootstrap samples and evaluating the predictive probabilities. For a fully specified accrual distribution, e.g. a uniform distribution, the auxiliary resample from \(Q_{\widehat{\bm\xi}}(a)\)  in Step~3 is omitted.

When censoring is exclusively administrative, set $C_i=+\infty$
in the model and $C_i^{*(b)}=+\infty$ in Step~2. Then
$G_{\bm \psi}(t)\equiv 1$. Steps~1 and~3 use only the
event-time likelihood, and the predictive parameter reduces to
$\bm{\theta}$. In each replicate,
$X_i^{*(b)}=\min\{T_i^{*(b)},\tau_i\}$, and the status is determined
by whether the event occurs by $\tau_i$.

To generate event times when covariates are included, it is sometimes useful to exploit the covariate-effect scale, i.e. the way the vector of covariates $\bm z$ acts on a baseline function of time. Here we review the strategy adopted for some specialised cases.

\paragraph{Proportional hazards (PH) models.}
Under the PH assumption,
\[
S_{\bm\theta}(t\mid\bm z)
=
\bigl[S_{\bm\theta,0}(t)\bigr]^{\exp(\bm\beta^\top\bm z)},
\]
where $S_{\bm\theta,0}(t)$ is the baseline survival function and $\bm\beta^\top\bm z$ is the covariate linear predictor. Since
\[
F_{\bm\theta}(t\mid\bm z)
=
1-\bigl[1-F_{\bm\theta,0}(t)\bigr]^{\exp(\bm\beta^\top\bm z)},
\]
we can sample from the baseline inverse CDF by setting
\[
T^\star
=
F_{\bm\theta,0}^{-1}(U^\dagger),
\qquad
U^\dagger
=
1-\bigl(1-U\bigr)^{\exp(-\bm\beta^\top\bm z_i)}, \qquad U\sim\mathrm{Unif}(0,1).
\]
This avoids drawing event times directly from
$F_{\bm\theta}(t\mid\bm z)$ when inverse sampling is not convenient, for example for a Royston--Parmar model with link function
$g\{S(t)\}=\log[-\log\{S(t)\}]$.

\paragraph{Proportional odds (PO) models.}
Under the PO assumption,
\[
\mathrm{logit}\{1-S_{\bm\theta}(t\mid\bm z)\}
=
\bm\beta^\top\bm z
+
\mathrm{logit}\{1-S_{\bm\theta,0}(t)\},
\]
where $1-S_{\bm\theta,0}(t)=F_{\bm\theta,0}(t)$ is the baseline incidence distribution function and $\bm\beta^\top\bm z$ is the covariate linear predictor on the log-odds scale. Then times can be sampled from
\[
T^\star
=
F_{\bm\theta,0}^{-1}(U^\dagger),
\qquad
U^\dagger
=
\mathrm{expit}
\!\left[
\mathrm{logit}(U)-\bm\beta^\top\bm z_i
\right], \qquad U\sim\mathrm{Unif}(0,1),
\]
where $\mathrm{expit}(a)=1/\{1+\exp(-a)\}$ and
$\mathrm{logit}(a)=\mathrm{expit}^{-1}(a)$. This avoids drawing event times directly from
$F_{\bm\theta}(t\mid\bm z)$ when inverse sampling is not convenient, for example for a Royston--Parmar model with
link function
$g\{S(t)\}=\operatorname{logit}\{1-S(t)\}$,
corresponding to a logit link for the event
distribution.

\paragraph{Linear probit (LP) models.}
Under the LP assumption,
\[
\Phi^{-1}\{1-S_{\bm\theta}(t\mid\bm z)\}
=
\bm\beta^\top\bm z
+
\Phi^{-1}\{1-S_{\bm\theta,0}(t)\},
\]
where $\Phi^{-1}(\cdot)$ is the quantile function of the standard normal distribution and $\bm\beta^\top\bm z$ is the linear predictor on the probit scale. Then times can be sampled from
\[
T^\star
=
F_{\bm\theta,0}^{-1}(U^\dagger),
\qquad
U^\dagger
=
\Phi
\!\left[
\Phi^{-1}(U)-\bm\beta^\top\bm z_i
\right], \qquad U\sim\mathrm{Unif}(0,1).
\]
This avoids drawing event times directly from
$F_{\bm\theta}(t\mid\bm z)$ when inverse sampling is not convenient, for example for a Royston--Parmar model with
link function
$g\{S(t)\}=\Phi^{-1}\{1-S(t)\}=-\Phi^{-1}\{S(t)\}$,
corresponding to a probit link for the event
distribution.

These three schemes cover the Royston--Parmar specifications used in the simulations and case study
\parencite{BenderAugustinBlettner2005GeneratingSurvivalTimes,CrowtherLambert2013SimulatingComplexSurvival}.

\section{Asymptotic nominal coverage of the CPB }
\label{App:Cov}

This section connects the conditional parametric bootstrap proposed in Section 3 of the manuscript with the direct-bootstrap theory of \textcite{Tian2022}. The purpose is to show how the argument used to establish asymptotic nominal coverage for the original direct-bootstrap multiple-cohort method extends to the present patient-level setting.

\textcite{Tian2022} establish their coverage result in two steps. They first prove the key statements in the single-cohort case - all the items are put on test at the same time-, where the conditional predictive distribution of the future failure count is binomial. They then extend the same lemmas and theorems to the multiple-cohort case - different groups of items are put on test at staggered time-, where the conditional predictive distribution is Poisson--binomial with grouped probabilities. In Section D of their Supplementary Materials, they note that the multiple-cohort proof follows the same method as the binomial case once the standardised binomial predictand is replaced by the corresponding Poisson--binomial predictand.

Our setting is the patient-level analogue of this multiple-cohort setting. Each patient still event-free at the interim analysis can be represented as a singleton cohort, with a patient-specific conditional event probability. Conditional on the observed interim data,
\[
Y_\Delta\mid \mathcal D_c,\bm\eta_0
\sim
\mathrm{PoiBin}\{\bm\pi_\Delta(\bm\eta_0)\},
\qquad
\bm\pi_\Delta(\bm\eta_0)
=
\{\pi_{i,\Delta}(\bm\eta_0):i\in\mathcal R_c\},
\]
where $\bm\eta_0=(\bm\theta_0,\bm\psi_0, \bm \xi_0)$ is the true parameter and $\pi_{i,\Delta}$ is given by \eqref{eq:piDelta}. Compared with the original multiple-cohort framework, the grouped cohort probabilities are replaced by subject-specific probabilities, and the probability map also incorporates loss to follow-up. The structure of the predictive distribution remains Poisson--binomial.

In Theorem~3, \textcite{Tian2022} show that, under the assumptions of their Theorems~1--2 and Lemmas~1--3, the $p$-quantile of the ideal direct-bootstrap predictive distribution has asymptotically correct coverage:
\[
\lim_{n\to\infty}\Pr(Y\leq Y_p^*)=p,
\qquad
\lim_{n\to\infty}\Pr(Y\geq Y_p^*)=1-p.
\]
We therefore impose the corresponding conditions in the present patient-level notation and describe the adaptations needed for the argument to apply. In particular, differentiability of the patient-specific probabilities is combined with conditions on their aggregate mean and variance over the original interim risk set. Define
\[
\mu_n(\bm\eta)=\sum_{i\in\mathcal R_c}\pi_{i,\Delta}(\bm\eta),
\qquad
s_n^2(\bm\eta)=\sum_{i\in\mathcal R_c}
\pi_{i,\Delta}(\bm\eta)\{1-\pi_{i,\Delta}(\bm\eta)\}.
\]
The original risk set $\mathcal R_c$ is held fixed when evaluating these quantities, including across bootstrap replicates. Throughout this section, the event-time, conditional loss-to-follow-up and accrual models are assumed as correctly specified, and the accrual distribution has known, fixed support \((0,a_{\max})\). The observed entry times and covariates are retained in the bootstrap data generation and predictive evaluation, while uncertainty in estimated accrual parameters is propagated through auxiliary bootstrap samples. The parameter dimension and prediction horizon \(\Delta>0\) are fixed as \(n\to\infty\). When the accrual distribution is fully specified, as for a uniform distribution on the known support, \(\bm \xi\) is omitted from \(\bm \eta\), and the procedure and conditions reduce to the fixed-accrual case.

\subsection{Conditions corresponding to Theorems~1--2 of \textcite{Tian2022}}
\label{App:Cov:Conds}

The following conditions are the patient-level counterparts of the assumptions used in Theorems~1--2 of \textcite{Tian2022}.

\begin{enumerate}
\item[(C1)] \emph{Asymptotic normality of the parameter estimator.}
Based on the interim data $\mathcal D_c$, the estimator
\(\widehat{\bm\eta}=(\widehat{\bm\theta},\widehat{\bm \psi},\widehat{\bm \xi})\) satisfies
\[
\sqrt n\bigl(\widehat{\bm\eta}-\bm\eta_0\bigr)
\xrightarrow{d}
\mathrm{MVN}(\bm 0,\bm V_0),
\]
where $\bm V_0$ is positive definite. The estimator \(\widehat{\bm\eta}=(\widehat{\bm\theta},\widehat{\bm \psi},\widehat{\bm \xi})\) is obtained by two-stage maximum likelihood estimation: \(\bm \xi\) is estimated from the marginal accrual likelihood, and \(\bm \psi\) is subsequently estimated from the conditional loss-to-follow-up likelihood evaluated at \(\widehat{\bm\xi}\). Under standard regularity conditions and the conditional likelihood factorisation in Supplementary materials ~\ref{App:Lik}, the asymptotic covariance matrix \(\bm V_0\) accounts for the dependence between \(\hat\psi\) and \(\hat\xi\) induced by the two-stage estimation procedure.

\item[(C2)] \emph{Bootstrap approximation of the estimator distribution.}
Conditionally on $\mathcal D_c$, the distribution of
$\sqrt n(\widehat{\bm\eta}^{*}-\widehat{\bm\eta})$
converges weakly, in probability, to $\mathrm{MVN}(\bm 0,\bm V_0)$.
This is the estimator-distribution consistency requirement used in Lemma~1 of \textcite{Tian2022}. The bootstrap convergence concerns the joint
distribution of all estimated parameters, including
the accrual parameter. Bootstrap samples retain the original follow-up windows and covariates, while event and censoring indicators are regenerated as in Supplementary materials~\ref{App:Algo}.

\item[(C3)] \emph{Continuity and non-degeneracy of the conditional probabilities.}

For each patient $i\in\mathcal R_c$, the survival functions entering \eqref{eq:piDelta} are continuous at $\bm\eta_0$ for $t\in[\tau_i,\tau_i+\Delta]$, with
\[
S_{\bm\theta_0}(\tau_i\mid\bm z_i)G_{\bm\psi_0}(\tau_i\mid a_i)>0.
\]
Here the conditioning on entry time is written explicitly. In the absence of loss to follow-up, set $G\equiv1$ and omit $\bm\psi$ from the parameter vector.
The risk set satisfies $|\mathcal R_c|/n\xrightarrow{p}c\in(0,1]$, and its empirical distribution of follow-up windows, entry times and covariates converges to a limit. For a constant $s_0^2>0$, assume
\[
\frac{s_n^2(\bm\eta_0)}n\xrightarrow{p}s_0^2,
\qquad
\frac{s_n^2(\widehat{\bm\eta})}n\xrightarrow{p}s_0^2,
\qquad
\frac{s_n^2(\widehat{\bm\eta}^{*})}n
\xrightarrow{p^*}s_0^2\quad\text{in probability}.
\]
These are the aggregate variance counterparts of the limiting cohort proportions and non-degeneracy assumptions in Section D of the Supplementary Materials of \textcite{Tian2022}.

\item[(C4)] \emph{Differentiability of the probability map.}
The maps $\bm\eta\mapsto\pi_{i,\Delta}(\bm\eta)$ are continuously differentiable in a neighbourhood of $\bm\eta_0$. Their average gradient converges locally uniformly at $\bm\eta_0$ to a finite vector $\bm d_0$: for every deterministic sequence $\epsilon_n\downarrow0$,
\[
\sup_{\|\bm\eta-\bm\eta_0\|\leq\epsilon_n}
\left\|\frac1n\nabla_{\bm\eta}\mu_n(\bm\eta)-\bm d_0\right\|
\xrightarrow{p}0.
\]
This condition permits the delta-method expansion of the aggregate predictive mean. A non-zero $\bm d_0$ yields a non-zero first-order contribution from parameter estimation; the coverage argument also allows $\bm d_0=\bm0$.
\end{enumerate}

Under (C1)--(C4), the argument of Lemmas~1--3 of \textcite{Tian2022} carries over as follows. Lemma~1 applies to the joint parameter estimator. For the counterpart of Lemma~2, a Taylor expansion gives
\[
\frac{\mu_n(\widehat{\bm\eta})-\mu_n(\bm\eta_0)}{\sqrt n}
=\bm d_0^\top\sqrt n(\widehat{\bm\eta}-\bm\eta_0)+o_p(1),
\]
and the bootstrap expansion is
\[
\frac{\mu_n(\widehat{\bm\eta}^{*})-\mu_n(\widehat{\bm\eta})}{\sqrt n}
=\bm d_0^\top\sqrt n(\widehat{\bm\eta}^{*}-\widehat{\bm\eta})
+o_{p^*}(1)\quad\text{in probability}.
\]
This replaces the finite vector of cohort probabilities by their aggregate mean, without requiring a fixed number of distinct patient profiles.

For the counterpart of Lemma~3, write
\[
\frac{Y_\Delta-\mu_n(\widehat{\bm\eta})}{s_n(\widehat{\bm\eta})}
=
\frac{Y_\Delta-\mu_n(\bm\eta_0)}{s_n(\bm\eta_0)}
\frac{s_n(\bm\eta_0)}{s_n(\widehat{\bm\eta})}
-
\frac{\mu_n(\widehat{\bm\eta})-\mu_n(\bm\eta_0)}{s_n(\widehat{\bm\eta})}.
\]
Conditionally on $\mathcal D_c$, the first standardised term is a sum of independent Bernoulli variables. Its Berry--Esseen error is bounded by a constant times $1/s_n(\bm\eta_0)$ and hence tends to zero by (C3). Combining this conditional approximation with the Taylor expansion and taking expectation over the interim data, as in the proof of Lemma~3 and its extension in Section D of Tian et al.'s Supplementary Materials, gives
\[
\frac{Y_\Delta-\mu_n(\widehat{\bm\eta})}{s_n(\widehat{\bm\eta})}
\xrightarrow{d}N(0,1+\kappa),
\qquad
\kappa=\frac{\bm d_0^\top\bm V_0\bm d_0}{s_0^2}.
\]
No finite-sample independence between the future count and the fitted estimator is required.

Let $Y_\Delta^*$ be an auxiliary variable with conditional law
$Y_\Delta^*\mid\widehat{\bm\eta}^{*},\mathcal D_c
\sim\mathrm{PoiBin}\{\bm\pi_\Delta(\widehat{\bm\eta}^{*})\}$.
Its bootstrap marginal CDF is the ideal mixture approximated by \eqref{eq:FhatStar}.
The same conditional normal approximation and the bootstrap Taylor expansion yield the counterpart of Lemma~3, Part~2:
\[
\sup_{z\in\mathbb R}\left|
\Pr^*\!\left\{
\frac{Y_\Delta^*-\mu_n(\widehat{\bm\eta})}{s_n(\widehat{\bm\eta})}\leq z
\right\}
-\Phi\!\left(\frac{z}{\sqrt{1+\kappa}}\right)
\right|\xrightarrow{p}0.
\]
The estimation terms enter the original prediction error and its bootstrap counterpart with opposite signs; their centred normal limits are symmetric, giving the same limiting distribution.

The quantile argument in the proof of Theorem~3 of \textcite{Tian2022} then applies. Writing $q_{p,n}^*$ for the $p$-quantile of the ideal bootstrap predictive distribution, for $p\in(0,1)$,
\[
\Pr_{\bm\eta_0}(Y_\Delta\leq q_{p,n}^*)\to p,
\qquad
\Pr_{\bm\eta_0}(Y_\Delta\geq q_{p,n}^*)\to1-p.
\]
Consequently, the equal-tailed prediction interval has asymptotically nominal repeated-sampling coverage. Discreteness can affect finite-sample coverage but vanishes on the diverging scale $s_n(\widehat{\bm\eta})$. The estimator in \eqref{eq:FhatStar} approximates the ideal bootstrap CDF by Monte Carlo averaging. The same asymptotic conclusion holds for the quantiles in \eqref{eq:PIstar} when the number of bootstrap replicates $B=B_n\to\infty$; a fixed finite $B$ provides a numerical approximation to the theoretical procedure.

\subsection{Remarks on the conditions}

In the multiple-cohort framework of Tian et al. (2022), the convergence of cohort proportions ensures the existence of deterministic limits for the aggregate predictive variance and the parameter-estimation contribution to the predictive mean. In our patient-level setting, the number of distinct follow-up and covariate profiles may increase with sample size, making assumptions on a fixed collection of cohort proportions unnecessarily restrictive.

Conditions (C1)--(C2) are the asymptotic-normality and bootstrap-consistency assumptions on the fitted estimator used in the direct-bootstrap argument of \textcite{Tian2022}. Conditions (C3) and (C4) replace these finite-cohort requirements by directly imposing the corresponding aggregate limits on the predictive variance and the gradient of the predictive mean. When the patients are partitioned into a fixed number of cohorts, these limits reduce to those obtained under the cohort-proportion assumptions of Tian et al.

Moreover, the bootstrap consistency requirement for censoring probabilities in Tian et al. is used to control the bootstrap risk-set sizes, particularly in their calibration-bootstrap construction. In the proposed conditional direct-bootstrap procedure, the original interim risk set is retained throughout the predictive calculation, while bootstrap samples are generated only to reproduce parameter-estimation variability. Consequently, consistency of the bootstrap estimator distribution (C2), together with the aggregate variance and smoothness conditions (C3)–(C4), provides the corresponding ingredients needed for the direct-bootstrap coverage argument without requiring a separate assumption on bootstrap risk-set proportions.

For the pointwise differentiability in (C4), the additional feature is that $\pi_{i,\Delta}(\bm\eta)$ contains an integral involving both the event-time and loss-to-follow-up models. Writing
\[
N_i(\bm\eta)
=
\int_{\tau_i}^{\tau_i+\Delta}
f_{\bm\theta}(u\mid\bm z_i)
G_{\bm\psi}(u \mid a_i)\,
\mathrm du,
\qquad
D_i(\bm\eta)
=
S_{\bm\theta}(\tau_i\mid\bm z_i)
G_{\bm\psi}(\tau_i \mid a_i),
\]
we have
\[
\pi_{i,\Delta}(\bm\eta)
=
\frac{N_i(\bm\eta)}{D_i(\bm\eta)}.
\]
The quotient rule gives
\[
\nabla_{\bm\eta}\pi_{i,\Delta}(\bm\eta)
=
\frac{
D_i(\bm\eta)\nabla_{\bm\eta}N_i(\bm\eta)
-
N_i(\bm\eta)\nabla_{\bm\eta}D_i(\bm\eta)
}{
D_i(\bm\eta)^2
}.
\]
Thus continuous differentiability of $\pi_{i,\Delta}$ follows from continuous differentiability of $N_i$ and $D_i$, and from $D_i(\bm\eta_0)\neq0$. Differentiation under the integral sign for $N_i$ requires the usual dominated-convergence conditions on
the parameter derivatives of $f_{\bm\theta}(u\mid\bm z_i)G_{\bm\psi}(u\mid a_i)$. This establishes pointwise smoothness, while the local uniform control of the aggregate gradient is stated in (C4).

In particular, writing
$h_{\bm\xi}(a)=\Phi^{-1}\{Q_{\bm\xi}(a)\}$,
\[
\nabla_{\bm\xi} h_{\bm\xi}(a)
=
\frac{\nabla_{\bm\xi} Q_{\bm\xi}(a)}
{\varphi\{h_{\bm\xi}(a)\}}.
\]
The fixed support of $Q_{\bm\xi}(a)$ ensures that the
support does not vary with the estimated parameter,
but does not by itself imply the locally uniform
aggregate gradient convergence required by (C4).

\begin{figure}[!t]
  \centering
  \begin{subfigure}[t]{0.49\linewidth}
    \centering
    \includegraphics[width=\linewidth]{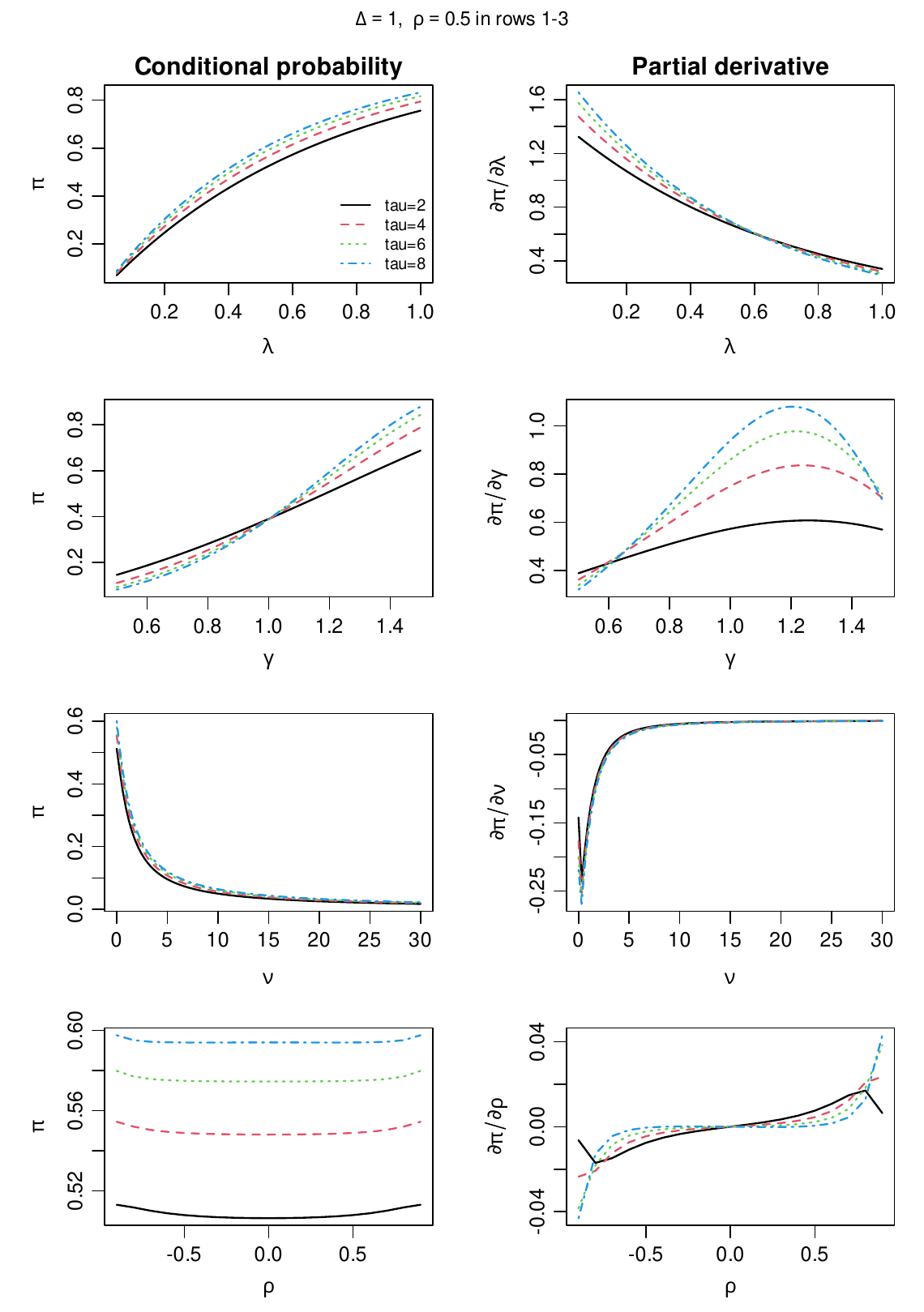}
    \caption{$\Delta=1$.}
    \label{fig:pi_sens_dt1}
  \end{subfigure}\hfill
  \begin{subfigure}[t]{0.49\linewidth}
    \centering
    \includegraphics[width=\linewidth]{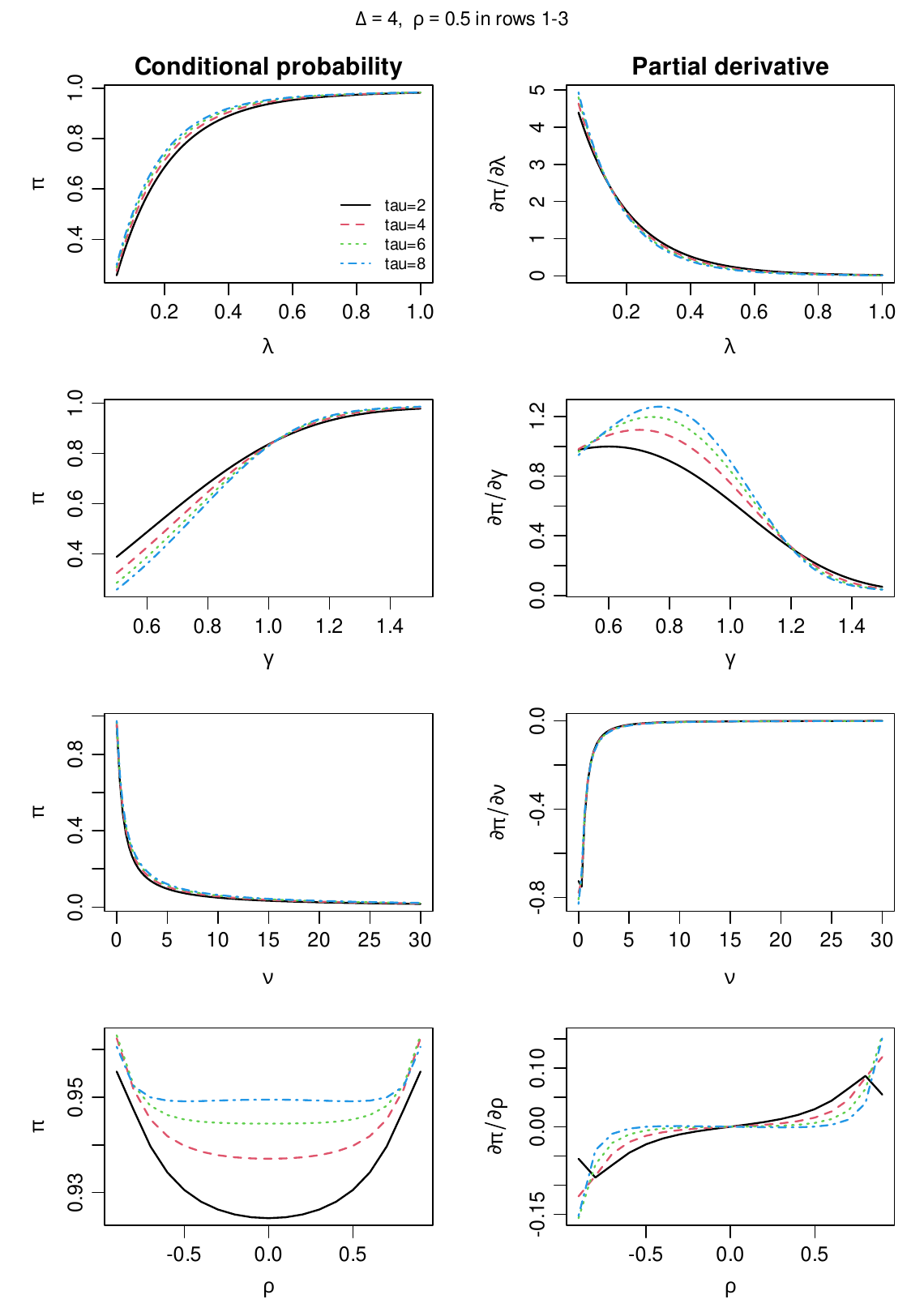}
    \caption{$\Delta=4$.}
    \label{fig:pi_sens_dt4}
  \end{subfigure}

  \caption{Sensitivity of the conditional event probability $\pi_{i,\Delta}(\bm\eta)$ to model parameters, shown for prediction horizons $\Delta=1$ and $\Delta=4$ and interim follow-up times $\tau\in\{2,4,6,8\}$. For each horizon, the four rows correspond to the Weibull parameters $\lambda$ and $\gamma$, the marginal exponential loss-to-follow-up rate $\nu$, and the latent Gaussian correlation parameter $\rho$, respectively. Their dependence is specified through a Gaussian copula. The left column shows $\pi_{i,\Delta}(\bm\eta)$ as a function of each parameter, and the right column shows its corresponding numerical partial derivative. Each parameter is varied in turn, holding the others at their reference values $\lambda=0.5$, $\gamma=1.2$, $\nu=0.03$, and $\rho=0.5$, with $\mathrm{HR}=1$. Entry time is fixed at the median of the accrual distribution. Correlation curves are evaluated at increments of $0.1$. Colours and line styles identify the values of $\tau$ consistently across all plots.}
  \label{App:Figu:pi_sensitivity}
\end{figure}

For completeness, Figure~\ref{App:Figu:pi_sensitivity} provides a numerical sensitivity check of $\pi_{i,\Delta}(\bm\eta)$ and its finite-difference partial derivatives under a Weibull event-time distribution, an exponential loss-to-follow-up distribution and uniform entry dates, over a representative range of parameter values and follow-up windows. For this numerical illustration, the accrual distribution is fully specified, so that \(\bm \xi\) is omitted from the parameter vector. This diagnostic illustrates point-wise smoothness over the displayed parameter values and follow-up windows and it does not by itself verify the aggregate asymptotic conditions.

\FloatBarrier

\newpage

\section{Additional details for the simulation study}\label{App:Simu}

\subsection{General simulation setup}\label{App:SimuSetup}

Let $Z$ denote the binary treatment indicator, with
$Z\sim\mathrm{Bernoulli}(0.5)$, corresponding to equal allocation between control
($Z=0$) and treatment ($Z=1$). Conditional on $Z$, event times were generated from a
Weibull proportional hazards model with hazard
\[
h(t\mid Z)=h_0(t)\exp(\beta Z),
\qquad
h_0(t)=\gamma\lambda_0 t^{\gamma-1},
\qquad t>0.
\]
Thus the cumulative hazard is
\[
H(t\mid Z)=\lambda_0\exp(\beta Z)t^\gamma.
\]
Throughout the simulations, the Weibull shape parameter was fixed at $\gamma=0.6$.
Event times were generated by inverse transform sampling as
\[
T=\left\{\frac{-\log(U)}{\lambda_0\exp(\beta Z)}\right\}^{1/\gamma},
\qquad U\sim\mathrm{Unif}(0,1),
\]
following standard simulation methods for survival data
\parencite{BenderAugustinBlettner2005GeneratingSurvivalTimes}. For each scenario, the baseline rate $\lambda_0$ was calibrated by
numerically solving for a prespecified probability of administrative
censoring at interim. The probability was evaluated by numerical
integration, as detailed in the following subsections.

Entry times were generated over a fixed accrual period of length $a_{\max}=3$ years.
In the simulation factors, $t_c$ denotes the time from accrual closure to the interim
analysis, considered as a measure of follow-up maturity. Hence, the corresponding calendar time of the interim analysis is
$a_{\max}+t_c$, and the available follow-up for a patient entering at calendar time
$A$ is
\[
\tau=a_{\max}+t_c-A.
\]

The interim analyses were scheduled at $t_c\in\{1,3,5\}$ years after
accrual closure, and predictions were evaluated at horizons
$\Delta t\in\{1,4\}$ years. 

In study $\mathcal S_1$, we used the Gaussian copula construction
described in Supplementary materials~\ref{app:gaussian-dropout} to generate dependent
entry and loss-to-follow-up times with marginal distributions
\[
A=a_{\max}\Phi(W_1)\sim\mathrm{Unif}(0,a_{\max}),
\qquad
C=-\frac{\log\Phi(W_2)}{\nu}\sim\mathrm{Exp}(\nu),
\qquad \nu=k\lambda_0.
\]

For these marginals, the conditional survival function of the
loss-to-follow-up time is
\[
G_{\bm \psi}(t)
=
\Phi\!\left(
  \frac{\Phi^{-1}(e^{-\nu t})
        -\rho \, \Phi^{-1}(a/a_{\max})}
       {\sqrt{1-(\rho)^2}}
\right),
\qquad t>0,\quad 0<a<a_{\max}.
\]

We then jointly estimated the marginal
loss-to-follow-up rate $\nu$ and the latent Gaussian correlation
$\rho$ by minimizing the negative conditional log-likelihood using
\texttt{nlminb()} in R with an analytical gradient. Optimisation was
performed on the scales $\log\nu$ and $\operatorname{atanh}(\rho)$.
The fit to the original interim data was initialised at $\rho=0$
and at a rate equal to the number of observed losses to follow-up
divided by the total observed follow-up time.

Because the marginal transformations generally change the
correlation \parencite{Cario1997}, we calibrated the latent Gaussian correlation $\rho$
to achieve the target population correlation $\rho^{\mathrm{{T}}}$ between $A$ and
$C$. For a candidate value of $\rho$, this correlation was evaluated as
\[
\operatorname{Corr}_{\rho}(A,C)
=\sqrt{12}\int_{-\infty}^{\infty}
\left[
  \Phi\!\left(\frac{\rho w}{\sqrt{2-(\rho)^2}}\right)
  -\frac12
\right]
[-\log\Phi(w)]\,\phi(w)\,\mathrm{d}w,
\]
where $\phi$ is the standard normal density. We evaluated this integral
numerically and used a root-finding algorithm to solve
$\operatorname{Corr}_{\rho}(A,C)=\rho^{\mathrm{{T}}}$.

For each simulated patient, the observed time in $\mathcal S_1$ was
\[
X=\min(T,C,\tau),
\]
with the corresponding event, loss-to-follow-up, or administrative
censoring indicator determined by the minimum. In $\mathcal S_2$, loss to follow-up was not simulated, so the observed time was $X=\min(T,\tau)$.

\subsection{\texorpdfstring{Data-generating mechanism for $\mathcal S_1$}{Data-generating mechanism for S1}}
\label{app:data_mec1}

In study $\mathcal{S}_{1}$, each data-generating scenario was defined
by the interim follow-up maturity $t_c$, the treatment effect
$\mathrm{HR}\in\{0.2,0.8\}$, the target correlation
$\rho^{\mathrm{T}}\in\{0.1,0.5\}$ between entry and loss-to-follow-up times, and the
loss-to-follow-up multiplier $k\in\{0.1,0.2,0.3\}$. For each scenario,
the baseline rate $\lambda_0$ was calibrated to yield an administrative
censoring probability $p_{\text{cens}}$ of $0.5$ at interim. This is the probability
that a patient was still under follow-up and event-free at interim.

The calibration proceeded as follows:
\begin{enumerate}
  \item Fix $t_c$, $\mathrm{HR}$, $\rho^{\mathrm{{T}}}$, and $k$, and obtain the
  latent correlation $\rho$ corresponding to $\rho^{\mathrm{{T}}}$ as
  described in Section~F.1.

  \item For a candidate value of $\lambda_0$, set
  $\nu=k\lambda_0$ and write $\tau=a_{\max}+t_c-a$.

  \item Compute the probability of administrative censoring $p_{\text{cens}}$ at
  interim by averaging over entry times and treatment arms.
  Conditional on entry and treatment, this probability is the
  product of the event-free and loss-to-follow-up survival
  probabilities at $\tau$. Hence
  \[
  p_{\text{cens}}(\lambda_0)
  =\frac{1}{2a_{\max}}\int_0^{a_{\max}}
  \left[
    e^{-\lambda_0\tau^\gamma}
    +e^{-\lambda_0\,\mathrm{HR}\,\tau^\gamma}
  \right]
  G_{\bm \psi}\!\left(\tau\right)
  \,\mathrm da.
  \]

  \item Evaluate the integral using adaptive numerical quadrature.

  \item Use a numerical root-finding algorithm to solve
  $p_{\text{cens}}(\lambda_0)=0.5$.
\end{enumerate}
After calibration, $N=1000$ Monte Carlo datasets of size $n=1000$
were generated for each scenario using the calibrated $\lambda_0$,
the corresponding $\nu=k\lambda_0$, and the calibrated latent $\rho$.
Each interim dataset was used for both prediction horizons.

\subsection{\texorpdfstring{Data-generating mechanism for $\mathcal S_2$}{Data-generating mechanism for S2}}
\label{app:data_mec2}

In study $\mathcal{S}_{2}$, event times were generated from the Weibull
proportional hazards model described in Section~F.1, with administrative
censoring at interim as the sole source of censoring. Each
data-generating scenario was defined by the interim follow-up maturity
$t_c$, the treatment effect $\mathrm{HR}\in\{0.2,0.8\}$, and the target
proportion of administratively censored patients at interim,
$p_{\text{cens}}\in\{0.2,0.8\}$.

The calibration proceeded as follows.
\begin{enumerate}
  \item Fix the scenario values of $t_c$, $\mathrm{HR}$, and $p_{\text{cens}}$.

  \item Choose a candidate value of $\lambda_0$ and write
  $\tau=a_{\max}+t_c-a$ for the available follow-up of a patient
  entering at calendar time $a$.

  \item Compute the probability of administrative censoring $p_{\text{cens}}$ at interim
  by averaging the probability of remaining event-free up to $\tau(a)$
  over the two treatment arms and the uniform distribution of entry
  times:
  \[
  p_{\text{cens}}(\lambda_0)
  = \frac{1}{2a_{\max}}\int_0^{a_{\max}}
  \left[
    e^{-\lambda_0\tau^\gamma}
    + e^{-\lambda_0\,\mathrm{HR}\,\tau^\gamma}
  \right]\,\mathrm{d}a.
  \]

  \item Evaluate the integral using adaptive numerical quadrature.

  \item Use a numerical root-finding algorithm to solve
  $p_{\text{cens}} = 0.2$ and $p_{\text{cens}} = 0.8$ .
\end{enumerate}

After calibration, $N=1000$ Monte Carlo datasets of size $n=1000$
were generated for each scenario using the calibrated value of
$\lambda_0$. Each interim dataset was used for both prediction horizons.

\subsection{Software}\label{App:Software}

All simulations were implemented in R 4.5.1 \parencite{RCoreTeam2025}. Simulation
summaries were produced using the \texttt{rsimsum} package \parencite{rsimsum}.
Event times from the Weibull proportional hazards model were generated using inverse
transform sampling; equivalent survival-data simulation routines are available in
the \texttt{simsurv} package \parencite{Brilleman2020simsurv}. Generalized
gamma and Royston--Parmar models were fitted with \texttt{flexsurv}
\parencite{flexsurv}. Random variates from fitted parametric survival models were
obtained using the corresponding quantile functions in \texttt{flexsurv}. Maximization of the conditonal loss-to-follow-up likelihood was performed with \texttt{nlminb()}
\parencite{RCoreTeam2025}. Evaluation of the Poisson--binomial distribution was
carried out using the \texttt{poibin} package \parencite{poibin}.

For the case study, nonparametric and semiparametric survival analyses were conducted
using \texttt{survival} and \texttt{muhaz} \parencite{muhaz}. Weibull, log-normal,
and log-logistic regression models were fitted with \texttt{survival}, while
generalized gamma and Royston--Parmar models were fitted with \texttt{flexsurv}.

\subsection{\texorpdfstring{Additional results for $\mathcal S_1$}{Additional results for S1}}
\label{App:Simu1}

\begingroup
\footnotesize
\setlength{\tabcolsep}{3pt}
\setlength{\LTcapwidth}{\linewidth}
\begin{longtable}{@{}*{13}{r}@{}}
\caption{Generating parameters, ranges of fitted values, and mean oracle prediction limits in study $\mathcal{S}_1$.}\label{tab:lambda_mec1}\\
\toprule
$t_c$ & $k$ & $\Delta t$ & $HR$ & $\rho$ & $\lambda_0$ & $\min(\hat\lambda_0)$ & $\max(\hat\lambda_0)$ & $\nu$ & $\min(\hat\nu)$ & $\max(\hat\nu)$ & $\bar L_{\mathrm{true}}$ & $\bar U_{\mathrm{true}}$ \\
\midrule
\endfirsthead
\toprule
$t_c$ & $k$ & $\Delta t$ & $HR$ & $\rho$ & $\lambda_0$ & $\min(\hat\lambda_0)$ & $\max(\hat\lambda_0)$ & $\nu$ & $\min(\hat\nu)$ & $\max(\hat\nu)$ & $\bar L_{\mathrm{true}}$ & $\bar U_{\mathrm{true}}$ \\
\midrule
\endhead
\midrule
\endfoot
\bottomrule
\endlastfoot
1 & 0.1 & 1 & 0.2 & 0.1 & 0.622 & 0.517 & 0.730 & 0.062 & 0.045 & 0.085 & 38.2 & 63.5 \\
1 & 0.1 & 1 & 0.2 & 0.5 & 0.607 & 0.487 & 0.753 & 0.061 & 0.045 & 0.082 & 38.4 & 63.8 \\
1 & 0.1 & 1 & 0.8 & 0.1 & 0.396 & 0.318 & 0.483 & 0.040 & 0.025 & 0.058 & 51.4 & 81.0 \\
1 & 0.1 & 1 & 0.8 & 0.5 & 0.390 & 0.306 & 0.484 & 0.039 & 0.025 & 0.060 & 51.3 & 80.8 \\
1 & 0.1 & 4 & 0.2 & 0.1 & 0.622 & 0.517 & 0.730 & 0.062 & 0.045 & 0.085 & 113.2 & 148.3 \\
1 & 0.1 & 4 & 0.2 & 0.5 & 0.607 & 0.487 & 0.753 & 0.061 & 0.045 & 0.082 & 115.3 & 150.3 \\
1 & 0.1 & 4 & 0.8 & 0.1 & 0.396 & 0.318 & 0.483 & 0.040 & 0.025 & 0.058 & 160.2 & 202.1 \\
1 & 0.1 & 4 & 0.8 & 0.5 & 0.390 & 0.306 & 0.484 & 0.039 & 0.025 & 0.060 & 160.9 & 202.8 \\
1 & 0.2 & 1 & 0.2 & 0.1 & 0.501 & 0.376 & 0.600 & 0.100 & 0.078 & 0.125 & 31.7 & 55.6 \\
1 & 0.2 & 1 & 0.2 & 0.5 & 0.487 & 0.388 & 0.604 & 0.097 & 0.072 & 0.119 & 32.2 & 56.2 \\
1 & 0.2 & 1 & 0.8 & 0.1 & 0.347 & 0.275 & 0.436 & 0.069 & 0.051 & 0.087 & 44.1 & 72.0 \\
1 & 0.2 & 1 & 0.8 & 0.5 & 0.340 & 0.274 & 0.415 & 0.068 & 0.051 & 0.085 & 43.9 & 71.8 \\
1 & 0.2 & 4 & 0.2 & 0.1 & 0.501 & 0.376 & 0.600 & 0.100 & 0.078 & 0.125 & 93.6 & 127.3 \\
1 & 0.2 & 4 & 0.2 & 0.5 & 0.487 & 0.388 & 0.604 & 0.097 & 0.072 & 0.119 & 96.8 & 130.6 \\
1 & 0.2 & 4 & 0.8 & 0.1 & 0.347 & 0.275 & 0.436 & 0.069 & 0.051 & 0.087 & 136.0 & 176.4 \\
1 & 0.2 & 4 & 0.8 & 0.5 & 0.340 & 0.274 & 0.415 & 0.068 & 0.051 & 0.085 & 137.1 & 177.5 \\
1 & 0.3 & 1 & 0.2 & 0.1 & 0.422 & 0.345 & 0.514 & 0.127 & 0.100 & 0.156 & 27.0 & 49.7 \\
1 & 0.3 & 1 & 0.2 & 0.5 & 0.410 & 0.337 & 0.534 & 0.123 & 0.101 & 0.155 & 27.5 & 50.2 \\
1 & 0.3 & 1 & 0.8 & 0.1 & 0.310 & 0.237 & 0.387 & 0.093 & 0.066 & 0.118 & 38.5 & 65.0 \\
1 & 0.3 & 1 & 0.8 & 0.5 & 0.302 & 0.234 & 0.392 & 0.091 & 0.063 & 0.109 & 38.5 & 65.0 \\
1 & 0.3 & 4 & 0.2 & 0.1 & 0.422 & 0.345 & 0.514 & 0.127 & 0.100 & 0.156 & 79.9 & 112.3 \\
1 & 0.3 & 4 & 0.2 & 0.5 & 0.410 & 0.337 & 0.534 & 0.123 & 0.101 & 0.155 & 83.0 & 115.5 \\
1 & 0.3 & 4 & 0.8 & 0.1 & 0.310 & 0.237 & 0.387 & 0.093 & 0.066 & 0.118 & 117.9 & 156.9 \\
1 & 0.3 & 4 & 0.8 & 0.5 & 0.302 & 0.234 & 0.392 & 0.091 & 0.063 & 0.109 & 119.8 & 158.7 \\
3 & 0.1 & 1 & 0.2 & 0.1 & 0.401 & 0.328 & 0.488 & 0.040 & 0.030 & 0.053 & 18.0 & 37.4 \\
3 & 0.1 & 1 & 0.2 & 0.5 & 0.395 & 0.321 & 0.502 & 0.040 & 0.029 & 0.050 & 18.0 & 37.5 \\
3 & 0.1 & 1 & 0.8 & 0.1 & 0.262 & 0.205 & 0.331 & 0.026 & 0.017 & 0.037 & 24.9 & 47.4 \\
3 & 0.1 & 1 & 0.8 & 0.5 & 0.260 & 0.206 & 0.333 & 0.026 & 0.017 & 0.035 & 24.9 & 47.3 \\
3 & 0.1 & 4 & 0.2 & 0.1 & 0.401 & 0.328 & 0.488 & 0.040 & 0.030 & 0.053 & 68.2 & 98.9 \\
3 & 0.1 & 4 & 0.2 & 0.5 & 0.395 & 0.321 & 0.502 & 0.040 & 0.029 & 0.050 & 68.7 & 99.5 \\
3 & 0.1 & 4 & 0.8 & 0.1 & 0.262 & 0.205 & 0.331 & 0.026 & 0.017 & 0.037 & 95.4 & 132.0 \\
3 & 0.1 & 4 & 0.8 & 0.5 & 0.260 & 0.206 & 0.333 & 0.026 & 0.017 & 0.035 & 95.5 & 132.1 \\
3 & 0.2 & 1 & 0.2 & 0.1 & 0.313 & 0.251 & 0.380 & 0.063 & 0.049 & 0.077 & 14.2 & 32.1 \\
3 & 0.2 & 1 & 0.2 & 0.5 & 0.308 & 0.228 & 0.388 & 0.062 & 0.049 & 0.076 & 14.3 & 32.3 \\
3 & 0.2 & 1 & 0.8 & 0.1 & 0.224 & 0.169 & 0.302 & 0.045 & 0.034 & 0.059 & 20.5 & 41.4 \\
3 & 0.2 & 1 & 0.8 & 0.5 & 0.222 & 0.171 & 0.272 & 0.044 & 0.035 & 0.055 & 20.5 & 41.5 \\
3 & 0.2 & 4 & 0.2 & 0.1 & 0.313 & 0.251 & 0.380 & 0.063 & 0.049 & 0.077 & 54.6 & 83.4 \\
3 & 0.2 & 4 & 0.2 & 0.5 & 0.308 & 0.228 & 0.388 & 0.062 & 0.049 & 0.076 & 55.6 & 84.5 \\
3 & 0.2 & 4 & 0.8 & 0.1 & 0.224 & 0.169 & 0.302 & 0.045 & 0.034 & 0.059 & 78.9 & 113.2 \\
3 & 0.2 & 4 & 0.8 & 0.5 & 0.222 & 0.171 & 0.272 & 0.044 & 0.035 & 0.055 & 79.4 & 113.8 \\
3 & 0.3 & 1 & 0.2 & 0.1 & 0.258 & 0.197 & 0.317 & 0.077 & 0.063 & 0.096 & 11.6 & 28.2 \\
3 & 0.3 & 1 & 0.2 & 0.5 & 0.254 & 0.173 & 0.340 & 0.076 & 0.061 & 0.094 & 11.8 & 28.5 \\
3 & 0.3 & 1 & 0.8 & 0.1 & 0.196 & 0.146 & 0.256 & 0.059 & 0.045 & 0.073 & 17.3 & 36.9 \\
3 & 0.3 & 1 & 0.8 & 0.5 & 0.193 & 0.148 & 0.247 & 0.058 & 0.047 & 0.073 & 17.3 & 36.9 \\
3 & 0.3 & 4 & 0.2 & 0.1 & 0.258 & 0.197 & 0.317 & 0.077 & 0.063 & 0.096 & 45.4 & 72.5 \\
3 & 0.3 & 4 & 0.2 & 0.5 & 0.254 & 0.173 & 0.340 & 0.076 & 0.061 & 0.094 & 46.5 & 73.8 \\
3 & 0.3 & 4 & 0.8 & 0.1 & 0.196 & 0.146 & 0.256 & 0.059 & 0.045 & 0.073 & 67.2 & 99.7 \\
3 & 0.3 & 4 & 0.8 & 0.5 & 0.193 & 0.148 & 0.247 & 0.058 & 0.047 & 0.073 & 67.6 & 100.2 \\
5 & 0.1 & 1 & 0.2 & 0.1 & 0.306 & 0.233 & 0.377 & 0.031 & 0.023 & 0.041 & 10.9 & 27.2 \\
5 & 0.1 & 1 & 0.2 & 0.5 & 0.303 & 0.233 & 0.393 & 0.030 & 0.023 & 0.039 & 10.9 & 27.3 \\
5 & 0.1 & 1 & 0.8 & 0.1 & 0.204 & 0.156 & 0.258 & 0.020 & 0.015 & 0.030 & 15.5 & 34.3 \\
5 & 0.1 & 1 & 0.8 & 0.5 & 0.203 & 0.157 & 0.259 & 0.020 & 0.014 & 0.028 & 15.5 & 34.3 \\
5 & 0.1 & 4 & 0.2 & 0.1 & 0.306 & 0.233 & 0.377 & 0.031 & 0.023 & 0.041 & 47.6 & 75.0 \\
5 & 0.1 & 4 & 0.2 & 0.5 & 0.303 & 0.233 & 0.393 & 0.030 & 0.023 & 0.039 & 47.9 & 75.4 \\
5 & 0.1 & 4 & 0.8 & 0.1 & 0.204 & 0.156 & 0.258 & 0.020 & 0.015 & 0.030 & 66.6 & 99.1 \\
5 & 0.1 & 4 & 0.8 & 0.5 & 0.203 & 0.157 & 0.259 & 0.020 & 0.014 & 0.028 & 66.7 & 99.2 \\
5 & 0.2 & 1 & 0.2 & 0.1 & 0.234 & 0.170 & 0.302 & 0.047 & 0.038 & 0.059 & 8.2 & 23.0 \\
5 & 0.2 & 1 & 0.2 & 0.5 & 0.231 & 0.174 & 0.287 & 0.046 & 0.038 & 0.056 & 8.3 & 23.2 \\
5 & 0.2 & 1 & 0.8 & 0.1 & 0.171 & 0.117 & 0.220 & 0.034 & 0.026 & 0.043 & 12.3 & 29.6 \\
5 & 0.2 & 1 & 0.8 & 0.5 & 0.170 & 0.125 & 0.241 & 0.034 & 0.027 & 0.043 & 12.3 & 29.7 \\
5 & 0.2 & 4 & 0.2 & 0.1 & 0.234 & 0.170 & 0.302 & 0.047 & 0.038 & 0.059 & 37.1 & 62.4 \\
5 & 0.2 & 4 & 0.2 & 0.5 & 0.231 & 0.174 & 0.287 & 0.046 & 0.038 & 0.056 & 37.7 & 63.1 \\
5 & 0.2 & 4 & 0.8 & 0.1 & 0.171 & 0.117 & 0.220 & 0.034 & 0.026 & 0.043 & 53.9 & 83.9 \\
5 & 0.2 & 4 & 0.8 & 0.5 & 0.170 & 0.125 & 0.241 & 0.034 & 0.027 & 0.043 & 54.2 & 84.3 \\
5 & 0.3 & 1 & 0.2 & 0.1 & 0.190 & 0.140 & 0.242 & 0.057 & 0.046 & 0.066 & 6.5 & 20.2 \\
5 & 0.3 & 1 & 0.2 & 0.5 & 0.188 & 0.134 & 0.244 & 0.057 & 0.047 & 0.070 & 6.6 & 20.3 \\
5 & 0.3 & 1 & 0.8 & 0.1 & 0.148 & 0.101 & 0.189 & 0.044 & 0.035 & 0.055 & 10.1 & 26.2 \\
5 & 0.3 & 1 & 0.8 & 0.5 & 0.146 & 0.106 & 0.190 & 0.044 & 0.035 & 0.054 & 10.1 & 26.2 \\
5 & 0.3 & 4 & 0.2 & 0.1 & 0.190 & 0.140 & 0.242 & 0.057 & 0.046 & 0.066 & 30.4 & 54.0 \\
5 & 0.3 & 4 & 0.2 & 0.5 & 0.188 & 0.134 & 0.244 & 0.057 & 0.047 & 0.070 & 30.8 & 54.6 \\
5 & 0.3 & 4 & 0.8 & 0.1 & 0.148 & 0.101 & 0.189 & 0.044 & 0.035 & 0.055 & 45.1 & 73.3 \\
5 & 0.3 & 4 & 0.8 & 0.5 & 0.146 & 0.106 & 0.190 & 0.044 & 0.035 & 0.054 & 45.4 & 73.6 \\
\end{longtable}
%\noindent\textit{Note.} Oracle bounds are averaged over 1,000 Monte Carlo replicates. The minima and maxima are computed across the interim fits within each scenario.\par
\endgroup

\begin{figure}[H]
  \centering
  \includegraphics[width=0.6\textwidth]{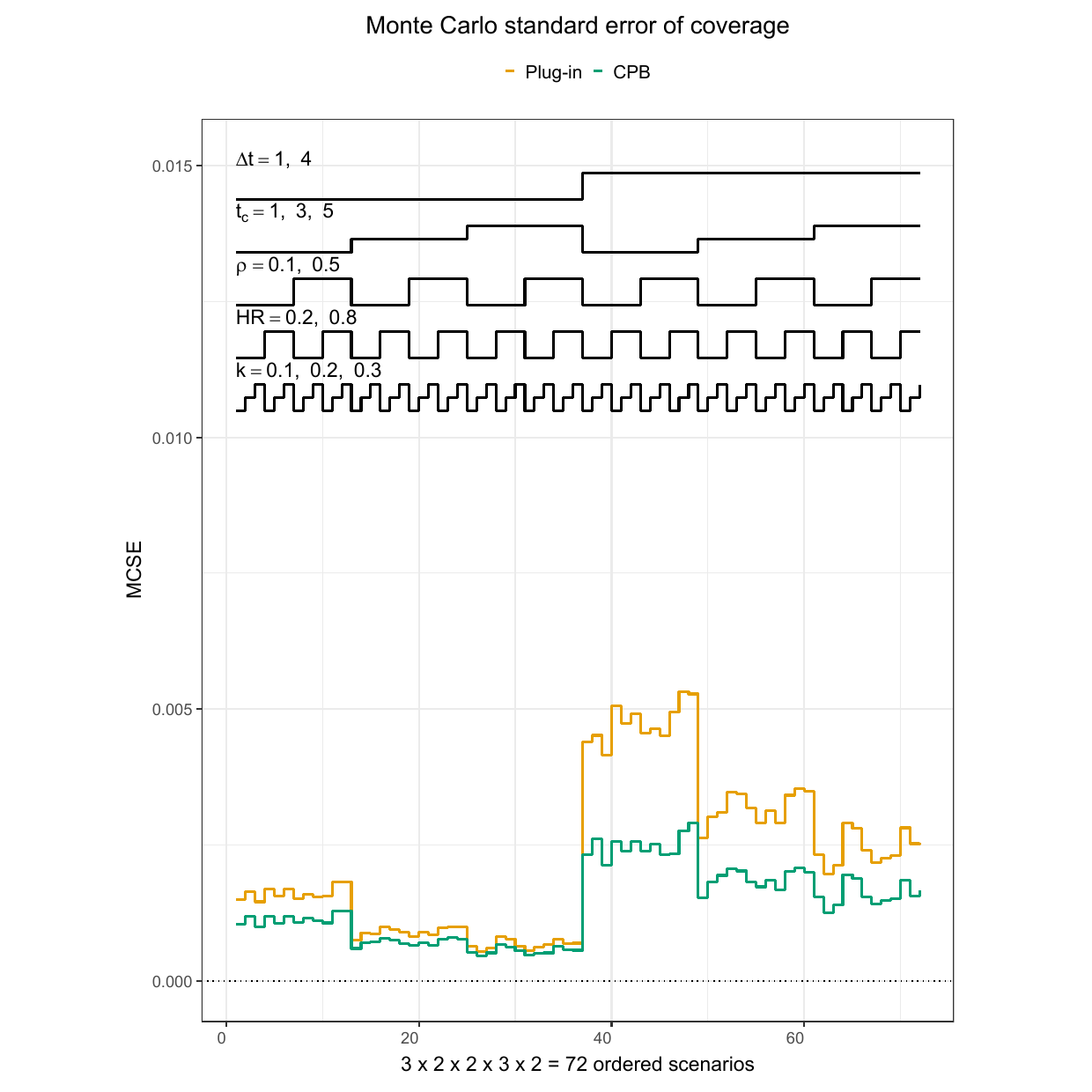}
  \caption{Study $\mathcal S_1$: Monte Carlo standard errors of the estimated coverage probabilities for the conditional parametric bootstrap (CPB) and the plug-in approach.}
  \label{App:Fig:se-drop}
\end{figure}

Table~\ref{tab:lambda_mec1} reports the calibrated values of $\lambda_0$ and $\nu$
for $\mathcal S_1$, together with the mean lower and upper oracle prediction bounds
computed from the conditional predictive distribution evaluated at the true
data-generating parameters.  Figure~\ref{App:Fig:se-drop} reports the Monte Carlo standard errors of the estimated
coverage probabilities for $\mathcal S_1$. These standard errors provide a check on
the simulation variability of the coverage estimates reported in the main paper.

\FloatBarrier

\FloatBarrier
\newpage
\subsection{\texorpdfstring{Additional results for $\mathcal S_2$}{Additional results for S2}}
\label{App:Simu2}

\begingroup
\footnotesize
\setlength{\tabcolsep}{3pt}
\setlength{\LTcapwidth}{\linewidth}
\begin{longtable}{@{}*{7}{r}@{}}
\caption{Generating Weibull parameter and mean oracle prediction limits in study $\mathcal{S}_2$.}\label{tab:lambda_mec2}\\
\toprule
$t_c$ & $p_{\text{cens}}$ & $\Delta t$ & $HR$ & $\lambda$ & $\bar L_{\mathrm{true}}$ & $\bar U_{\mathrm{true}}$ \\
\midrule
\endfirsthead
\toprule
$t_c$ & $p_{\text{cens}}$ & $\Delta t$ & $HR$ & $\lambda_0$ & $\bar L_{\mathrm{true}}$ & $\bar U_{\mathrm{true}}$ \\
\midrule
\endhead
\midrule
\endfoot
\bottomrule
\endlastfoot
1 & 0.2 & 1 & 0.2 & 2.849 & 34.1 & 56.5 \\
1 & 0.2 & 1 & 0.8 & 1.102 & 53.8 & 79.8 \\
1 & 0.2 & 4 & 0.2 & 2.849 & 98.6 & 125.7 \\
1 & 0.2 & 4 & 0.8 & 1.102 & 136.3 & 160.4 \\
1 & 0.8 & 1 & 0.2 & 0.232 & 27.3 & 50.7 \\
1 & 0.8 & 1 & 0.8 & 0.146 & 29.2 & 53.6 \\
1 & 0.8 & 4 & 0.2 & 0.232 & 101.8 & 139.7 \\
1 & 0.8 & 4 & 0.8 & 0.146 & 111.2 & 152.1 \\
3 & 0.2 & 1 & 0.2 & 1.932 & 16.5 & 34.5 \\
3 & 0.2 & 1 & 0.8 & 0.744 & 27.6 & 49.3 \\
3 & 0.2 & 4 & 0.2 & 1.932 & 62.0 & 88.4 \\
3 & 0.2 & 4 & 0.8 & 0.744 & 92.8 & 120.2 \\
3 & 0.8 & 1 & 0.2 & 0.160 & 13.1 & 30.8 \\
3 & 0.8 & 1 & 0.8 & 0.101 & 14.1 & 32.5 \\
3 & 0.8 & 4 & 0.2 & 0.160 & 58.6 & 89.8 \\
3 & 0.8 & 4 & 0.8 & 0.101 & 63.4 & 96.6 \\
5 & 0.2 & 1 & 0.2 & 1.538 & 10.3 & 25.8 \\
5 & 0.2 & 1 & 0.8 & 0.592 & 18.0 & 36.9 \\
5 & 0.2 & 4 & 0.2 & 1.538 & 44.9 & 69.4 \\
5 & 0.2 & 4 & 0.8 & 0.592 & 69.7 & 96.9 \\
5 & 0.8 & 1 & 0.2 & 0.128 & 8.0 & 22.9 \\
5 & 0.8 & 1 & 0.8 & 0.081 & 8.8 & 24.1 \\
5 & 0.8 & 4 & 0.2 & 0.128 & 40.5 & 67.6 \\
5 & 0.8 & 4 & 0.8 & 0.081 & 43.7 & 72.3 \\
\end{longtable}
\endgroup

Table~\ref{tab:lambda_mec2} reports the calibrated values of $\lambda_0$ for
$\mathcal S_2$, together with the mean lower and upper oracle prediction bounds
computed from the conditional predictive distribution evaluated at the true
data-generating parameters. Since $\mathcal S_2$ includes only administrative
censoring, this table describes the calibration of the event-time model across
scenarios with different censoring proportions and follow-up maturities.

\begin{figure}[H]
  \centering
  \includegraphics[width=0.6\textwidth]{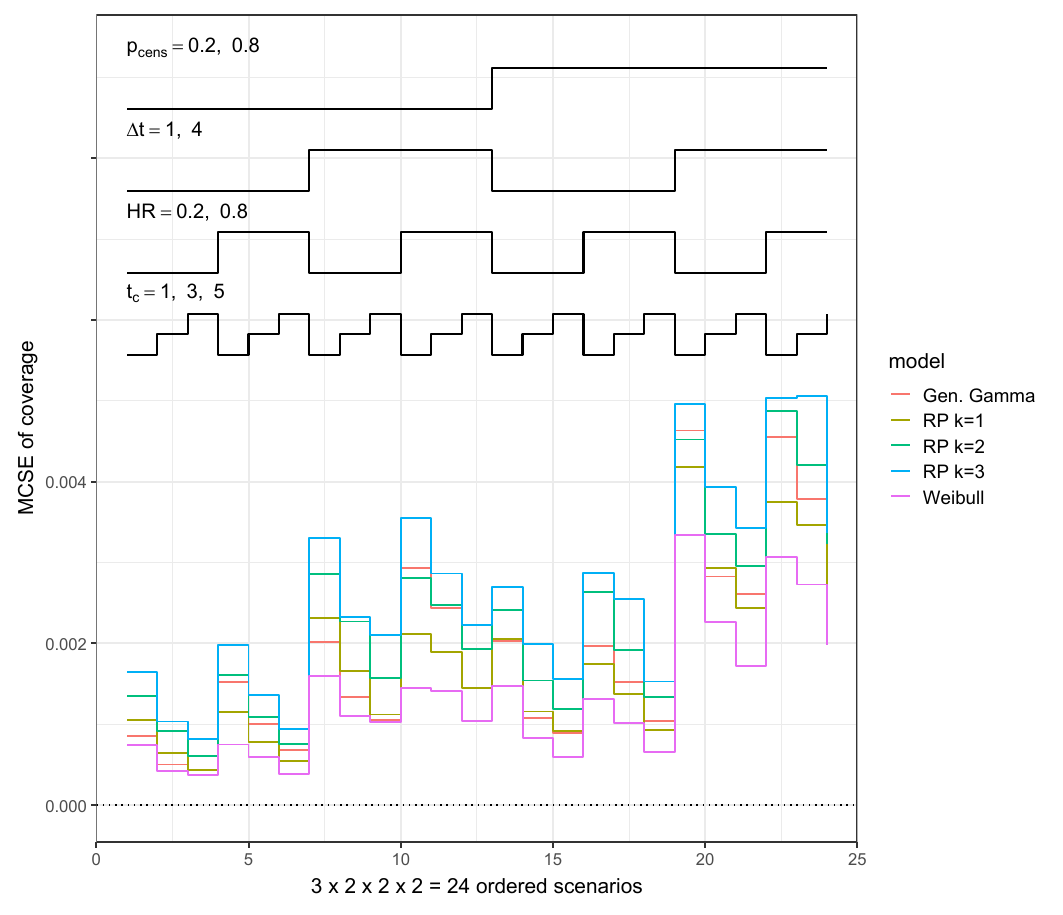}
  \caption{Study $\mathcal S_2$: Monte Carlo standard errors of the estimated coverage probabilities.}
  \label{App:Fig:se-s2}
\end{figure}

\FloatBarrier

\newpage
\section{Further case study results}\label{App:Case}
\FloatBarrier
 
\begin{figure}[H]
  \centering
  \includegraphics[width=0.8\textwidth]{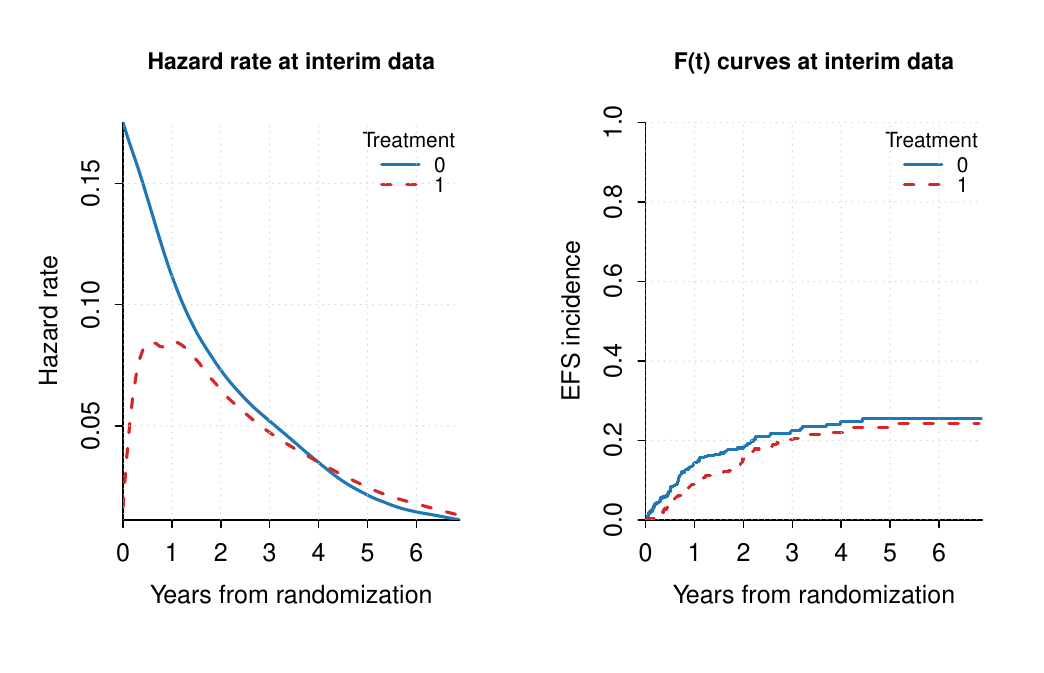}
  \caption{Smoothed hazard function and Incidence function for event survival probability (EFS) at 1-06-2017. 1 = Active treatment, 0 = standard of care.}
  \label{App:Fig:hazard}
\end{figure}

\begin{figure}[H]
  \centering
\includegraphics[width=1\textwidth]{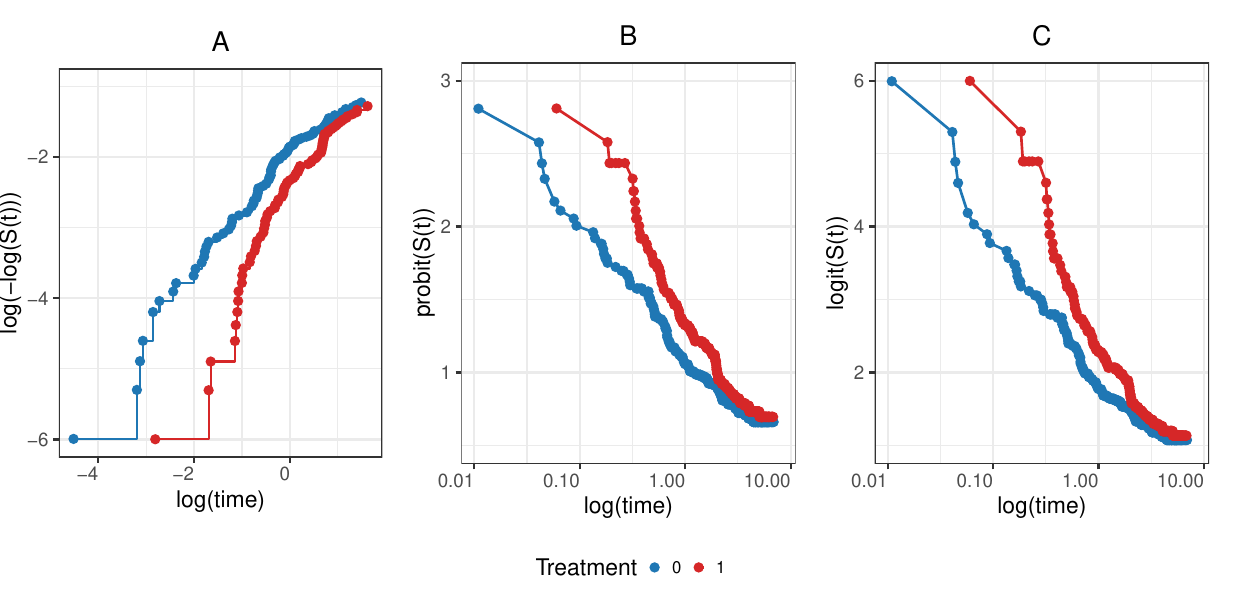}
\caption{Survival function transformations versus log-time
by treatment arm (1 = active treatment, 0 = standard of care).
Panel A shows the log cumulative hazard, $\log\{-\log S(t)\}$;
Panel B shows the probit transformation of survival,
$\Phi^{-1}\{S(t)\}$, where $\Phi^{-1}$ is the standard normal
quantile function; Panel C shows the logit transformation
of survival, $\log\{S(t)/(1-S(t))\}$.
Approximately straight and parallel lines suggest the
appropriateness of the corresponding standard parametric
model and covariate-effect scale, respectively.}
\end{figure}

% ===== Parametric analysis =====
\begin{figure}[H]
  \centering
\includegraphics[width=0.8\textwidth]{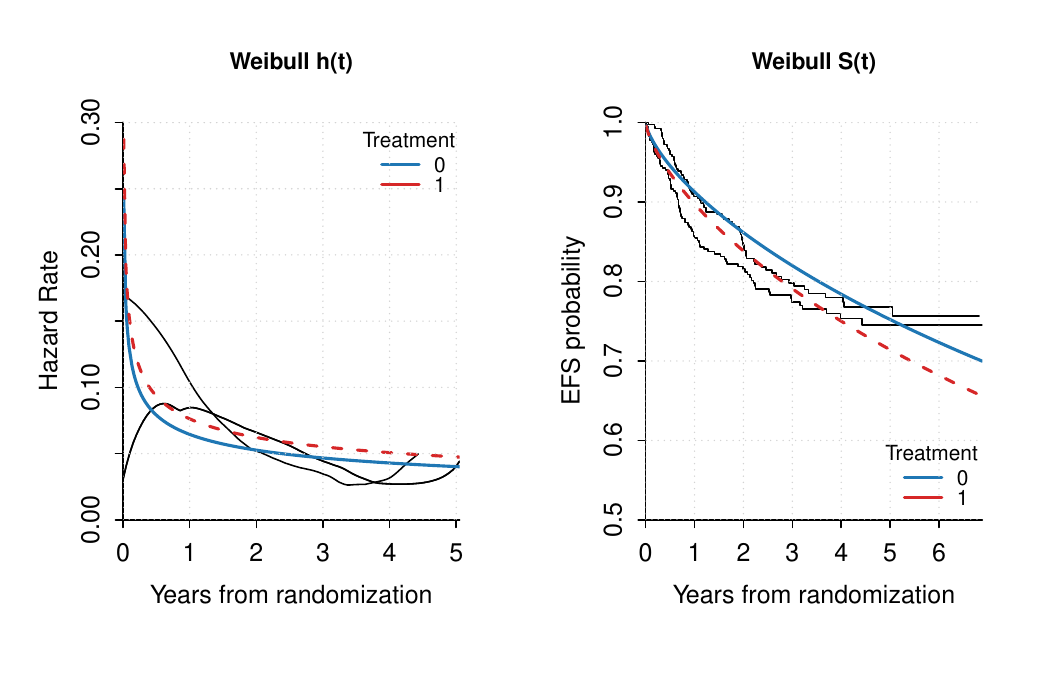}
  \caption{Observed (black lines) vs predicted hazard curves (left) and survival curves (right) under Weibull models at 1-06-2017. Solid lines refer to active treatment, dashed lines to standard of care.}
  \label{App:Fig:weibull}
\end{figure}

\begin{figure}[H]
  \centering
\includegraphics[width=0.8\textwidth]{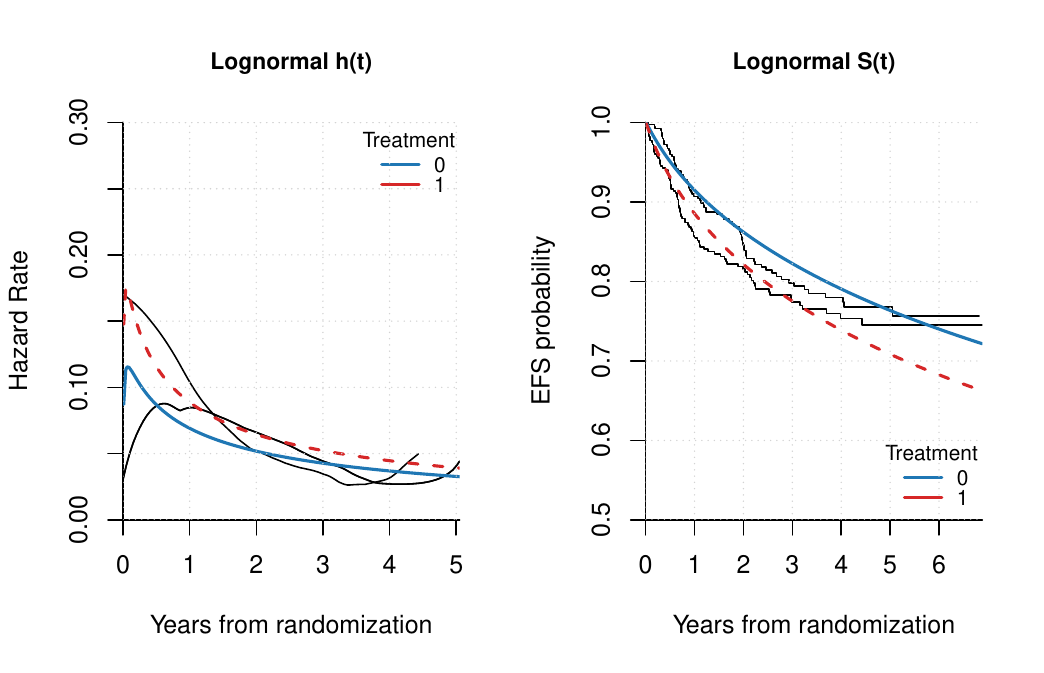}
\caption{Observed vs predicted hazard curves (left) and survival curves (right) under Lognormal models at 1-06-2017. Solid lines refer to active treatment, dashed lines to standard of care.}
  \label{App:Fig:lognormal}

\end{figure}

\begin{figure}[H]
  \centering
\includegraphics[width=0.8\textwidth]{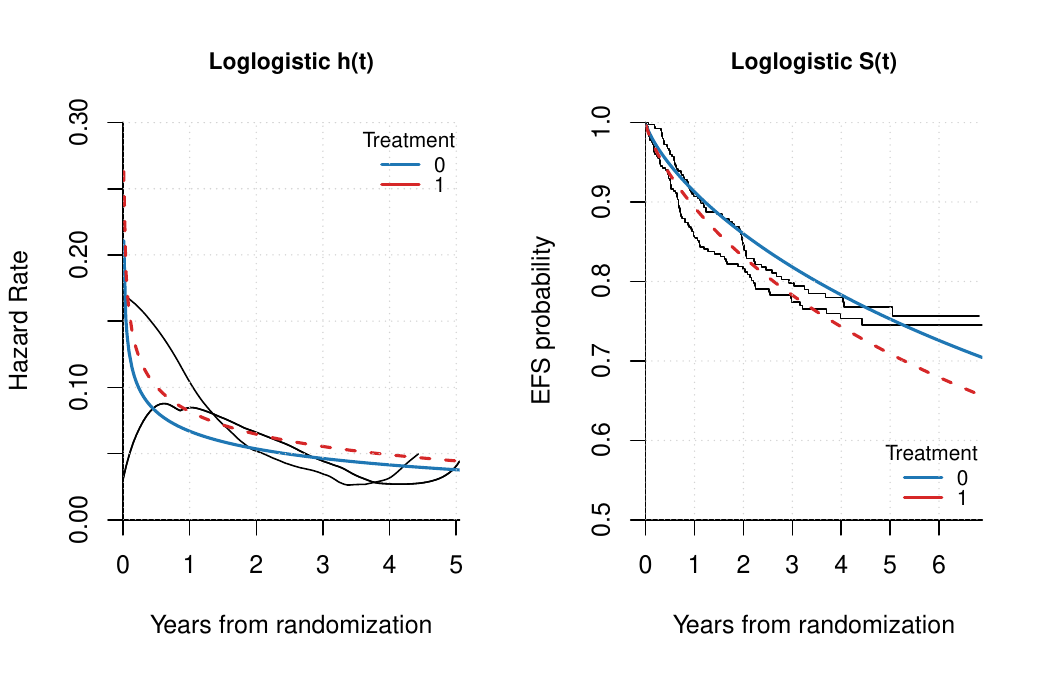}
\caption{Observed vs predicted hazard curves (left) and survival curves (right) under Loglogistic models at 1-06-2017. Solid lines refer to active treatment, dashed lines to standard of care.}
\label{App:Fig:loglogis}
\end{figure}

\begin{figure}[H]
  \centering  \includegraphics[width=0.8\textwidth]{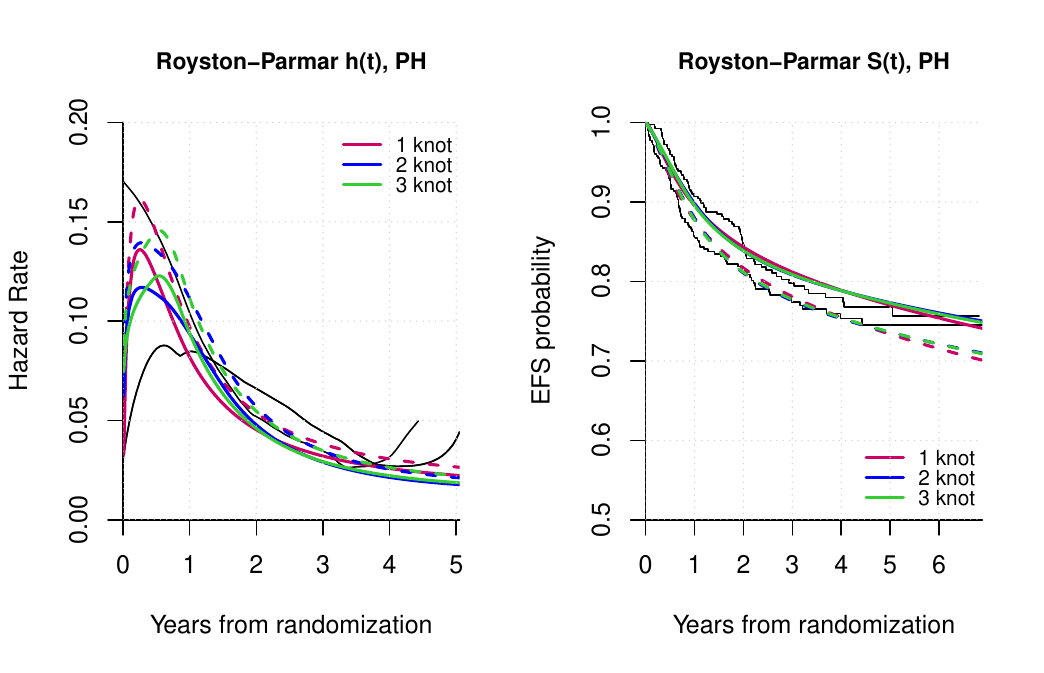}
  \caption{Observed vs predicted hazard curves (left) and survival curves (right) under Royston-Parmar models on the proportional hazard scale at 1-06-2017. Solid lines refer to active treatment, dashed lines to standard of care.}
  \label{App:Fig:rp-ph}
\end{figure}

\begin{figure}[H]
  \centering  \includegraphics[width=0.8\textwidth]{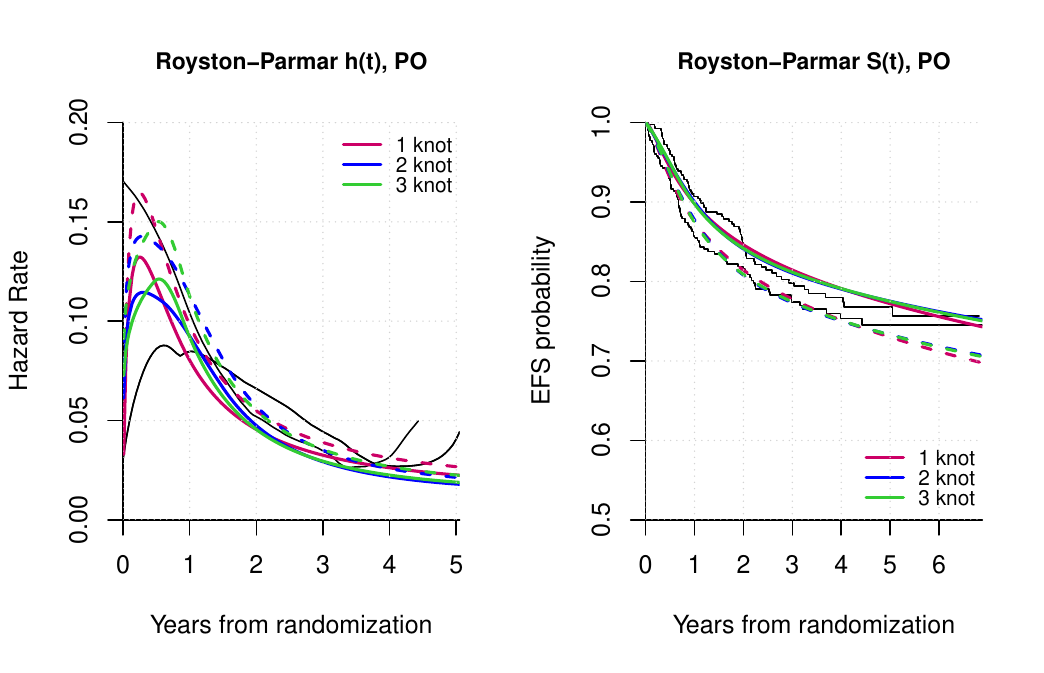}
  \caption{Observed vs predicted hazard curves (left) and survival curves (right) under Royston-Parmar models on the proportional odds scale at 1-06-2017. Solid lines refer to active treatment, dashed lines to standard of care.}
  \label{App:Fig:rp-po}
\end{figure}

\begin{figure}[H]
  \centering  \includegraphics[width=0.8\textwidth]{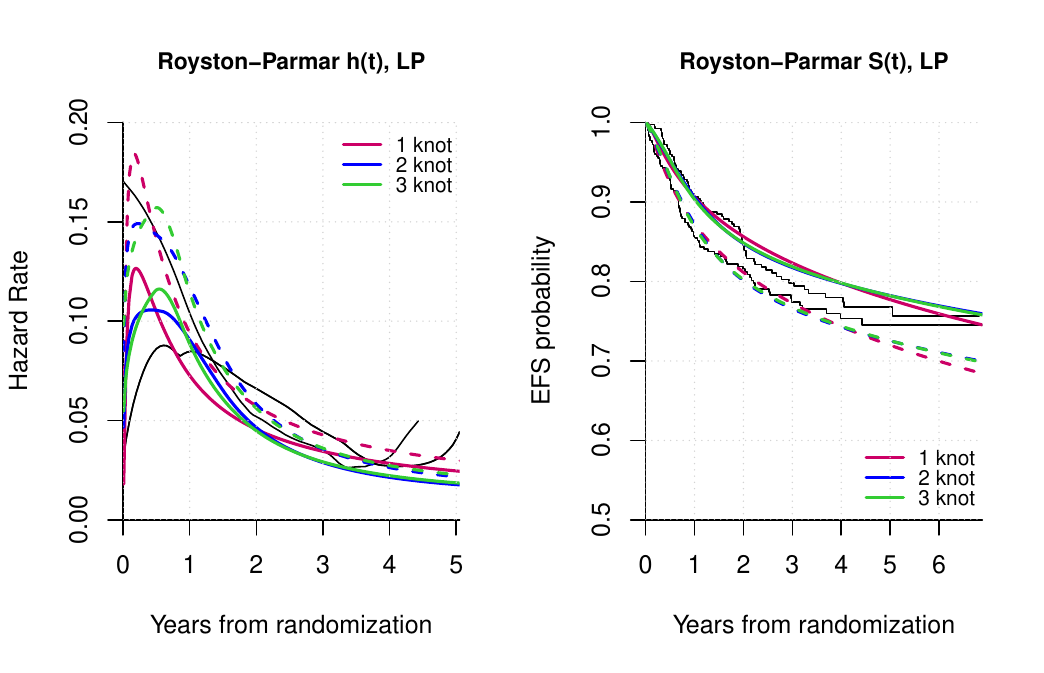}
  \caption{Observed vs predicted hazard curves (left) and survival curves (right) under Royston-Parmar models on the linear probit scale at 1-06-2017. Solid lines refer to active treatment, dashed lines to standard of care.}
  \label{App:Fig:rp-pp}
\end{figure}

\begin{figure}[H]
  \centering
\includegraphics[width=0.8\textwidth]{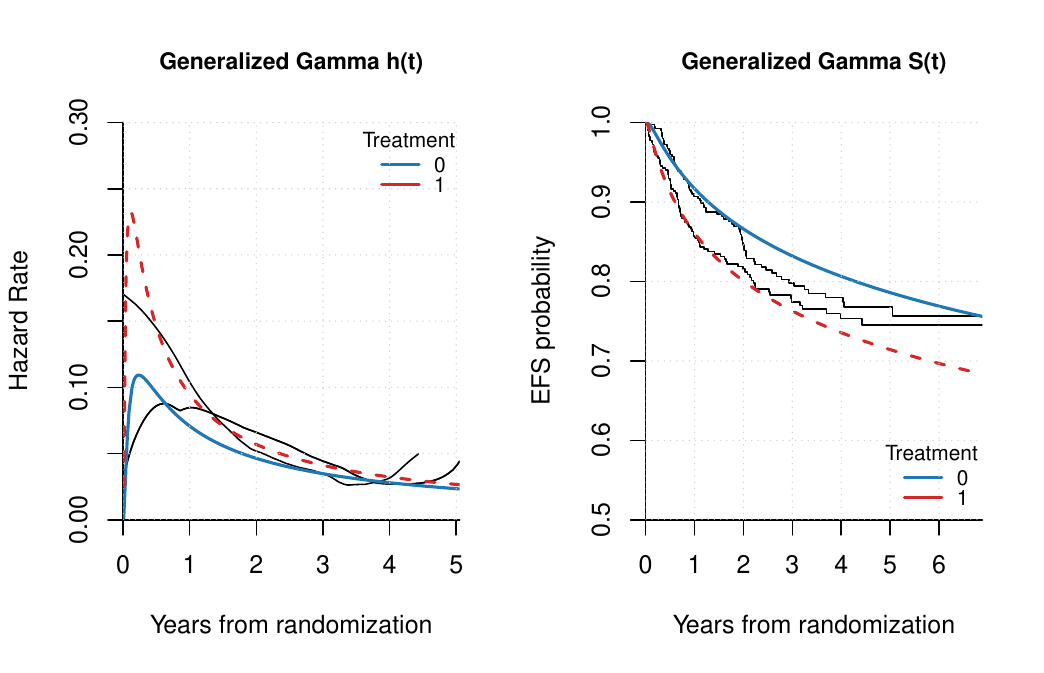}
  \caption{Observed vs predicted hazard curves (left) and survival curves (right) under Generalized Gamma models at 1-06-2017. Solid lines refer to active treatment, dashed lines to standard of care.}
\label{App:Fig:gengamma}
\end{figure}

\begin{figure}[H]
  \centering
\includegraphics[width=0.8\textwidth]{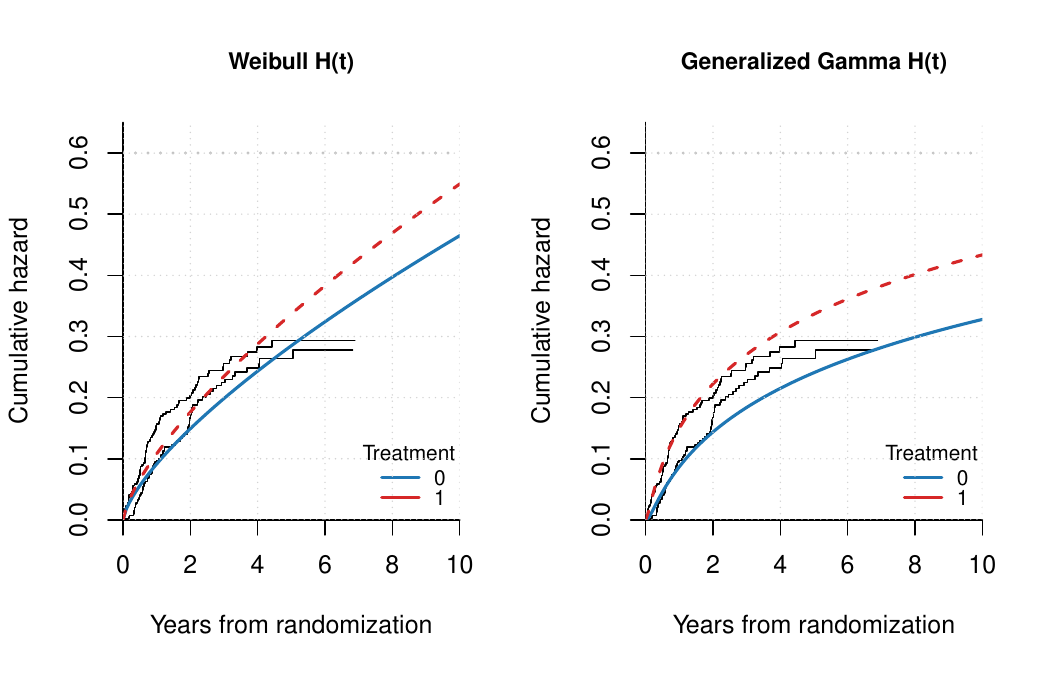}
  \caption{Observed vs predicted cumulative hazard curves under the Weibull and the Generalized Gamma model at 1-06-2017. Solid lines refer to active treatment, dashed lines to standard of care. A grey dashed line is added for reference in comparing the plots}
\label{App:Fig:weibull-gg-Ht}
\end{figure}

\begin{figure}[H]
  \centering
\includegraphics[width=0.8\textwidth]{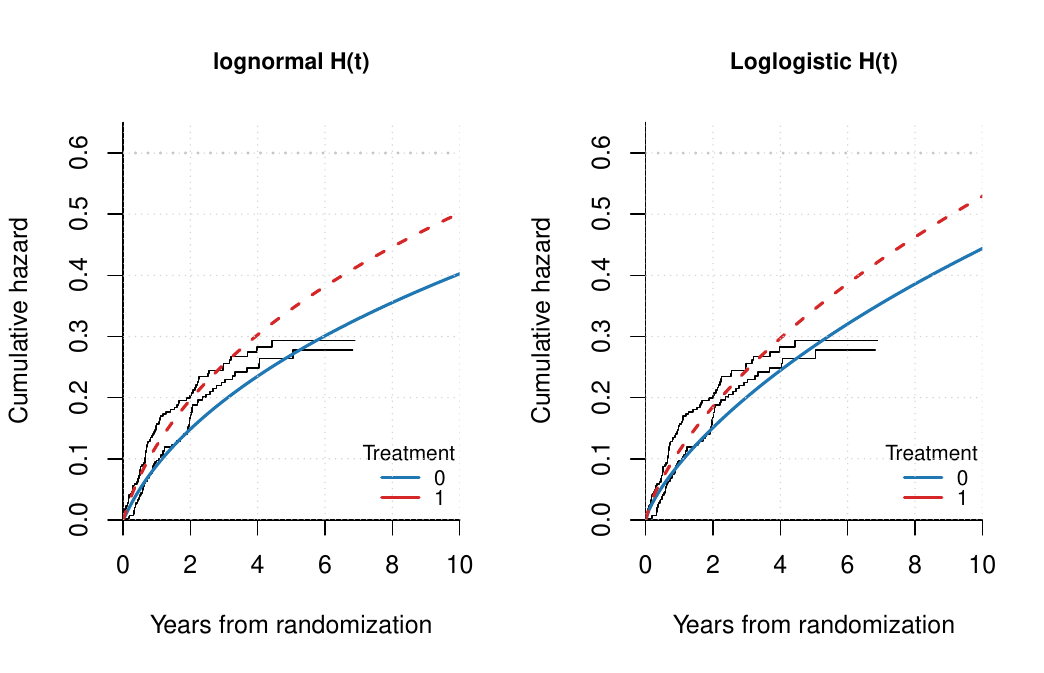}
  \caption{Observed vs predicted cumulative hazard curves under the Lognormal and the Loglogistic model at 1-06-2017. Solid lines refer to active treatment, dashed lines to standard of care. A grey dashed line is added for reference in comparing the plots}
\label{App:Fig:log-Ht}
\end{figure}

\begin{figure}[H]
  \centering
\includegraphics[width=0.8\textwidth]{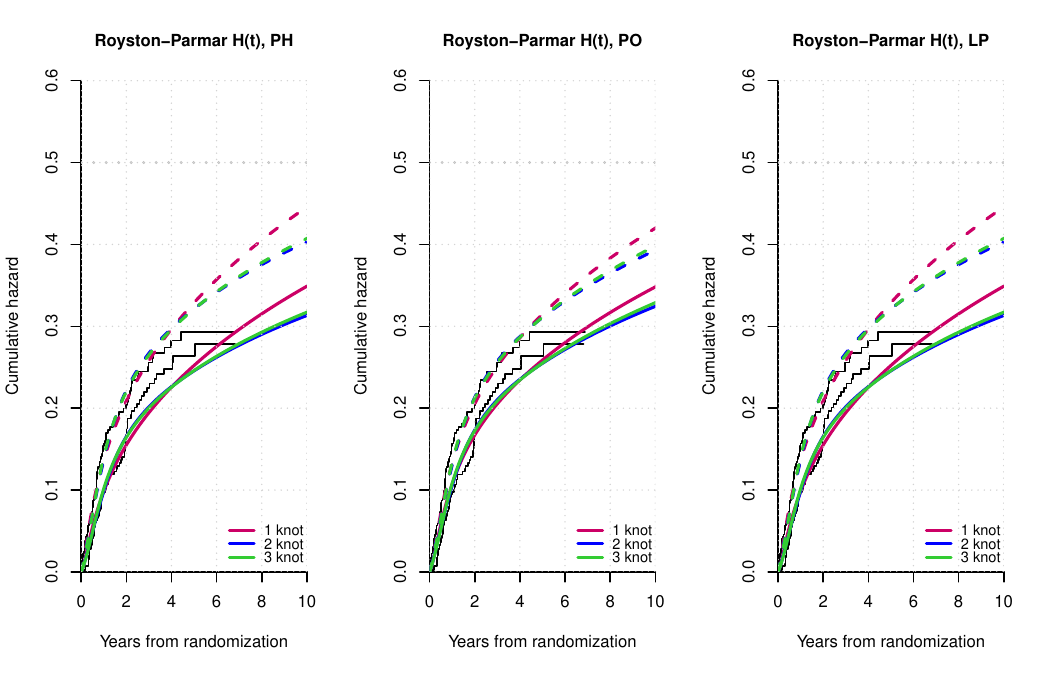}
  \caption{Observed vs predicted cumulative hazard curves under the Royston - Parmar model at 1-06-2017. Solid lines refer to active treatment, dashed lines to standard of care. PH = proportional hazards, PO = proportional odds, LP = linear probit. A grey dashed line is added for reference in comparing the plots}
\label{App:Fig:rp-Ht}
\end{figure}

\FloatBarrier
\begin{table}[H]
\centering
\small
\caption{Cumulative number of additional events by semester and corresponding 95\% prediction intervals (lower, upper) — Weibull, Royston--Parmar PH extensions and Generalized Gamma}
\label{tab:pi-weib}
\begin{tabular}{lcccccc}
\hline
Date & Additional Events & Weibull & RP, 1 knot & RP, 2 knots & RP, 3 knots & GG \\
\hline
1/12/2017 & 17 & (9, 26)   & (6, 22)  & (6, 21)  & (6, 21)  & (6, 21) \\
1/06/2018 & 29 & (20, 45)  & (14, 36) & (13, 34) & (13, 34) & (14, 36) \\
1/12/2018 & 43 & (31, 62)  & (20, 48) & (18, 45) & (19, 45) & (21, 48) \\
1/06/2019 & 50 & (41, 79)  & (26, 59) & (23, 55) & (24, 54) & (28, 58) \\
1/12/2019 & 53 & (51, 94)  & (31, 68) & (27, 63) & (28, 64) & (33, 68) \\
1/06/2020 & 55 & (61, 109) & (35, 78) & (30, 71) & (31, 73) & (38, 77) \\
1/12/2020 & 60 & (70, 123) & (39, 86) & (33, 78) & (35, 81) & (43, 85) \\
1/06/2021 & 62 & (79, 136) & (43, 94) & (36, 85) & (38, 89) & (47, 92) \\
\hline
\end{tabular}

\par\smallskip
\begin{minipage}{\linewidth}
\centering
\footnotesize\emph{Note:} RP = Royston--Parmar model; GG = Generalized Gamma model.
\end{minipage}
\end{table}

% ===== Table 3: Lognormal =====
\begin{table}[H]
\centering
\small
\caption{Cumulative number of additional events by semester and corresponding 95\% prediction intervals (lower, upper) — Lognormal and Royston--Parmar LP extensions}
\label{tab:pi-logn}
\begin{tabular}{lccccc}
\hline
Date & Additional Events & Lognormal & RP, 1 knot & RP, 2 knots & RP, 3 knots \\
\hline
1/12/2017 & 17 & (8, 24)   & (7, 22)  & (6, 21)  & (6, 21) \\
1/06/2018 & 29 & (18, 42)  & (15, 37) & (13, 35) & (14, 35) \\
1/12/2018 & 43 & (28, 57)  & (22, 50) & (19, 46) & (20, 46) \\
1/06/2019 & 50 & (37, 71)  & (29, 61) & (24, 55) & (25, 55) \\
1/12/2019 & 53 & (45, 84)  & (35, 72) & (28, 64) & (29, 64) \\
1/06/2020 & 55 & (53, 95)  & (40, 81) & (31, 71) & (33, 72) \\
1/12/2020 & 60 & (61, 106) & (45, 90) & (35, 79) & (36, 79) \\
1/06/2021 & 62 & (67, 116) & (50, 99) & (38, 85) & (39, 86) \\
\hline
\end{tabular}

\par\smallskip
\begin{minipage}{\linewidth}
\centering
\footnotesize\emph{Note:} RP = Royston--Parmar model.
\end{minipage}
\end{table}

\begin{table}[H]
\centering
\small
\caption{Cumulative number of additional events by semester and corresponding 95\% prediction intervals (lower, upper) — Loglogistic and Royston--Parmar PO extensions}
\label{tab:pi-loglog}
\begin{tabular}{lccccc}
\hline
Date & Additional Events & Loglogistic & RP, 1 knot & RP, 2 knots & RP, 3 knots \\
\hline
1/12/2017 & 17 & (9, 25)   & (6, 22)  & (6, 21)  & (6, 21) \\
1/06/2018 & 29 & (20, 44)  & (14, 36) & (13, 35) & (13, 35) \\
1/12/2018 & 43 & (31, 61)  & (21, 48) & (19, 46) & (18, 47) \\
1/06/2019 & 50 & (41, 76)  & (27, 59) & (23, 55) & (23, 57) \\
1/12/2019 & 53 & (50, 91)  & (32, 68) & (27, 64) & (27, 66) \\
1/06/2020 & 55 & (59, 104) & (36, 77) & (31, 72) & (30, 75) \\
1/12/2020 & 60 & (68, 117) & (41, 85) & (34, 79) & (33, 83) \\
1/06/2021 & 62 & (76, 129) & (45, 93) & (37, 86) & (36, 91) \\
\hline
\end{tabular}

\par\smallskip
\begin{minipage}{\linewidth}
\centering
\footnotesize\emph{Note:} RP = Royston--Parmar model.
\end{minipage}
\end{table}

\end{document}